\documentclass[12pt]{article}

\usepackage{amsmath}
\usepackage{graphicx}
\usepackage{enumerate}
\usepackage{natbib}
\usepackage{url}
\expandafter\def\expandafter\UrlBreaks\expandafter{\UrlBreaks
  \do\a\do\b\do\c\do\d\do\e\do\f\do\g\do\h\do\i\do\j%
  \do\k\do\l\do\m\do\n\do\o\do\p\do\q\do\r\do\s\do\t%
  \do\u\do\v\do\w\do\x\do\y\do\z\do\A\do\B\do\C\do\D%
  \do\E\do\F\do\G\do\H\do\I\do\J\do\K\do\L\do\M\do\N%
  \do\O\do\P\do\Q\do\R\do\S\do\T\do\U\do\V\do\W\do\X%
  \do\Y\do\Z}

\newcommand{\blind}{1}
\usepackage{amsfonts,amssymb,bbm}
\usepackage{enumitem}
\usepackage{multirow}
\usepackage{makecell}
\usepackage{xcolor}
\usepackage{hyperref}
\newtheorem{theorem}{Theorem}

\allowdisplaybreaks
\begin{document}
\def\spacingset#1{\renewcommand{\baselinestretch}%
{#1}\small\normalsize} \spacingset{1}


\if1\blind
{
  \title{\bf \textit{smoothEM}: a new approach for the simultaneous assessment of smooth patterns and spikes}
  \author{Huy Dang \thanks{
    The authors gratefully acknowledge the Irish Social Science Data Archive for having generously provided access to the Smart Meter Electricity data. The extreme temperature data used in this application was obtained from the National Oceanic and Atmospheric Administration (NOAA) via the Environmental Protection Agency (APE) website. M.A. Cremona acknowledges the support of the Natural Sciences and Engineering Research Council of Canada (NSERC) and of the Faculty of Business Administration of Université Laval. F.~Chiaromonte acknowledges the support of the Huck Institutes of the Life Sciences of the Pennsylvania State University. }\hspace{.2cm}\\
    Dept.~of Statistics, Pennsylvania State University, USA\\
    and \\
    Marzia A. Cremona \\
    Dept.~of Operations and Decision Systems, Université Laval, Canada\\
    CHU de Québec – Université Laval Research Center, Canada\\
    and \\
    Francesca Chiaromonte\\
    Dept.~of Statistics, Pennsylvania State University, USA\\
    Inst.~of Economics and L'EMbeDS, Sant'Anna School of Advanced Studies, Italy}
  \maketitle
} \fi

\if0\blind
{
  \bigskip
  \bigskip
  \bigskip
  \begin{center}
    {\LARGE\bf Title}
\end{center}
  \medskip
} \fi

\bigskip
\begin{abstract}
We consider functional data where an underlying smooth curve is composed not just with errors, but also with irregular spikes. We propose an approach that, combining regularized spline smoothing and an Expectation-Maximization (\textsc{EM}) algorithm, allows one to both identify spikes and estimate the smooth component. Imposing some assumptions on the error distribution, we prove consistency of EM estimates. Next, we demonstrate the performance of our proposal on finite samples and its robustness to assumptions violations through simulations. Finally, we apply our proposal to data on the annual heatwaves index in the US and on weekly electricity consumption in Ireland. In both data sets, we are able to characterize underlying smooth trends and to pinpoint irregular/extreme behaviors.
\end{abstract}

\noindent%

\newpage
\spacingset{1.9} 

\section{Introduction and Motivation}

%
%
%
The past two decades have witnessed an increasing interest in functional data, where one or more variables are data varying over a continuum and often possessing additional structures of interest. The vast majority of past and current literature focuses on functional data obeying certain smoothness conditions \citep[see, e.g.,][]{ramsaysilverman07, kozoszka17}, which can be effectively represented in low dimension through basis functions (e.g., Fourier basis, spline basis or polynomial basis). In the majority of applications, a penalty is employed to ensure such representation -- an approach commonly referred to as penalized/regularized smoothing \citep{yaolee06, goldsmith11}.

Real-world data often contain, on top of underlying smooth trends, sporadic discontinuities that can be interesting in and of themselves. Meaningful examples include data on extreme temperatures and electricity consumption, to name a few. As we shall see later, in both cases one can observe discontinuous spikes occurring on top of smooth underlying trends, due, e.g., to sporadic abnormalities in weather conditions or occasional over-consumption of electricity. For these discontinuous data, a simple application of penalized smoothing that does not account for the spikes
can generate an inaccurate approximation of the underlying curve. In this article, we seek to produce reliable estimates of the underlying trends and identify the spikes simultaneously, the latter of which can be analyzed 
to gain insight into their frequency, location, magnitude and spread. 

In existing literature, wavelet representation is often used to handle discontinuities due to its multi-scale nature and ability to adapt to local features in the data
\citep{nason08}. However, wavelet representation often suffers from boundary effects and does not present an intuitive way to incorporate smoothness information of the underlying trend. 
Another approach is provided by \citet{descary}, who consider functions with rough components that are assumed smooth at local scales. More specifically, these authors consider functions expressed as the sum of two uncorrelated components, $X(t) = Y(t) + W(t)$, where $Y(t)$ is taken to be of finite rank and of smoothness class $C^k$ ($k \geq 2)$, and $W(t)$ is locally highly variable but continuous at a shorter time scale. While interesting, this set-up is only an approximation for functions that contain true discontinuities, and are thus non-smooth at any scale. Another possible candidate is adaptive smoothing \citep{luo97,pintore06}, whose adaptive nature might be useful for increasing the local penalty where there are discontinuities, thus reducing their influence. However, by nature of their objective functions, adaptive smoothing methods tend to do the opposite: they decrease the local penalty to accommodate discontinuities, thus making the estimated smooth component even more wiggly. On a separate front, the field of spectroscopic analysis has produced several baseline correction methods to treat signals (e.g., raw Raman spectra or standard chromatograms) consisting of a smooth baseline with occasional positive spikes -- which are not of interest and are treated as contamination to be eliminated. Many of these methods employ iterative penalized smoothing, with adaptive weights informed by the sign and magnitude of residuals \citep{wei22, zhang10}. While of great interest, these methods are generally limited to simple and slowly varying smooth baselines, and are therefore ineffective when more complex smooth structures exist. 
In this article, we propose a novel framework that 
represents discontinuities explicitly. Broad distributional assumptions about the data allow us to exploit both the magnitude and the variance of the spikes, and to simultaneously perform smooth curve estimation and spike identification. In symbols, we are interested in data $(x_i, y_i)_{i = 1}^n$ of the form 
\begin{equation}
    \label{eq:1}
    y_i = f(x_i) + \mu^*\cdot \mathbbm{1}(x_i \in   \mathbb{S}) + \epsilon_i
\end{equation}
where $x_i$, $i=1\ldots, n$ are  locations in a domain (which we map in $[0,1]$ without loss of generality); $f \in C^p[0,1]$
is a smooth function with $p$ continuous derivatives; $\mathbb{S}$ is a random collection of intervals on $[0,1]$ affected by spikes of size $\mu^*$ such that $\mathbb{S}$ has probability measure $1-\alpha^*$;
$\mathbbm{1}(x_i \in \mathbb{S})$, $i=1\ldots, n$ is an indicator of spike occurrences;  
and $\epsilon_i$, $i=1\ldots, n$ are independent random errors. 
Assuming a fixed design, our main goal is to capture the underlying  curve and to identify  the spikes, i.e.~to estimate $f$ and the spike classification vector $\mathbbm{1}_{\mathbb{S}} = (\mathbbm{1}(x_i \in   \mathbb{S}))_{i = 1}^n$. 

With the additional assumption that errors are Gaussian, we rewrite Equation (\ref{eq:1}) as $y_i = f(x_i) + \xi_i$, where $\xi_i :=\mu^*\cdot \mathbbm{1}(x_i \in   \mathbb{S}) + \epsilon_i$ can be modeled as a mixture of Gaussians:
\begin{equation}
    \label{model}
    \xi_i \sim \alpha^*N(0, \sigma^{*2}) + (1-\alpha^*)N(\mu^*, \sigma^{*2}).
\end{equation}
Thus, with probability $\alpha^* \in (0.5,1)$, the departure from $f$ is Gaussian with mean $0$ and variance $\sigma^{*2}$, but with probability $(1-\alpha^*)$ it is ``spiked'' by a scalar amount $\mu^*$. Equation (\ref{model}) can be extended to accommodate size-varying spikes as
\begin{equation}
    \label{MSIV_model}
    \xi_i \sim \alpha^*N(0, \sigma^{*2}) + (1-\alpha^*)N(\mu^*, \sigma^{*2} + \sigma_h^{*2}).
\end{equation}
We deal with size-varying spikes in Section \ref{simulations}.  
Here we focus attention 
on the model in Equation \ref{model}. In this setting, maximum likelihood estimates (MLE) of $\alpha^*, \mu^*$ and $\sigma^*$ could be obtained through the {\it Expectation-Maximization} (EM) algorithm if the $\xi_i$'s were observable.
Previous work on EM convergence often assumed that the only parameter to be estimated is $\mu^*$, with both $\sigma^{*2}$ and $\alpha^*$ taken as known; see for instance \citep{wu17} and \citep{balakrishnan}. Drawing inspiration from the latter, we study convergence when all parameters $\mu^*, \sigma^{*2}$ and $\alpha^*$ are unknown -- proving that $\nu-$strong concavity, Lipschitz smoothness and Gradient smoothness conditions hold for our Gaussian mixture model. In such a case, guarantees on the convergence rate are harder to establish, but can in fact be provided if the ``contamination'' level $\alpha^*$ is  taken as known. We use simulations to demonstrate the practical effectiveness of our approach notwithstanding this shortcoming in theoretical guarantees. 

The remainder of this article is organized as follows. Section~2 provides background on smoothing splines and EM algorithm. Section~3 details our approach and the conditions under which it performs well. 
Section~4 provides convergence guarantees for the EM algorithm. Sections~5 and 6 demonstrate the performance of our proposal through simulations, comparisons with existing methods and real data analyses. Section~7 contains final remarks.

\section{Technical background}

    \subsection{Penalized smoothing splines}
    \label{sec:spline}
    Suppose that the data $(x_i, y_i)_{i=1}^n$ are generated according to
    \begin{equation}
        \label{huy12}
        y_i = f(x_i) + \epsilon_i
    \end{equation}
    where $x_i \in [0,1], i = 1, \ldots n$ are either fixed or random, and the $\epsilon_i$'s represent white noise (independent and Gaussian random errors). 
    Assuming that $f$ has $p$ continuous derivatives on $[0,1]$, i.e.~that $f \in C^p[0,1]$, it is often of interest to approximate $f$ from $(x_i, y_i)_{i = 1}^n$. 
    In a spline approximation \citep{deboor, xiao19} the estimator $\hat{f}$ is restricted to lie in the space of spline functions of order $m$. 
    Functions in this space have the representation  $\sum_k a_k N_k(x)$ where $N_k(x)$
    is the $k^{th}$ B-spline function. 
    Depending on the number of basis functions 
    $\hat{f}$ can either underfit or overfit the data. 
To prevent this, $\hat{f}$ is regularized by placing a penalty on its higher-order derivatives \citep[see, e.g.,][]{osullivan, ramsaysilverman07, xiao19};
    that is, to minimize
    \begin{equation}
        \label{eq:3}
        \sum_{i = 1}^{n} \bigg\{y_i - \sum_{k = 1}^{K} a_k N_k(x_i)\bigg\}^2 + \lambda \int_{0}^{1} \bigg \{\sum_{k = 1}^{K}a_kN_k^{(q)}(x)\bigg\}^2 dx
    \end{equation}
    where $\lambda$ is a tuning parameter whose optimal value can be found using cross validation.
    Equation (\ref{eq:3}) has a vectorized representation
    \begin{equation}
        \label{eq:4}
        \min_{\boldsymbol{a}} 
        \bigg(
        \frac{1}{n}\left\Vert  y-Na\right\Vert_2^2 + \lambda a^\top P_qa 
        \bigg)
    \end{equation}
    where $y = (y_1, \ldots, y_n)^\top$, $N = (N_1(x), \ldots, N_K(x))^\top$, $a = (a_1, \ldots, a_K)^\top$ are column vectors and $P_q$ is the penalization matrix. The explicit form of $P_q$ is not needed for 
    the purposes of this article; interested readers can find more details in \cite{xiao19}. 
    Equation (\ref{eq:4}) can be solved explicitly; indeed, if we set $H_n := N^\top N/n+\lambda P_q$, then the solutions are
    \begin{align*}
        \hat{a} = H_n^{-1} (N^\top y/n), \ \        \hat{f}(x) = N^\top(x)H_n^{-1}(N^\top y/n).
    \end{align*}

    \subsection{EM algorithm for Gaussian mixtures}
    \label{EMbackground}
    
    Let $\xi \in \Xi$ and $z \in Z$ be random variables whose joint density function is $\phi_{\theta^*}$, where $\theta^*$ belongs to a (non-empty) convex parameter space $\Omega$. 
    Suppose we can observe data $(\xi_i)_{i = 1}^n$, while the $(z_i)_{i = 1}^n$ are unobservable, and that $(\xi_i|z_i = j) \overset{iid}{\sim}G_j$ where the $G_j$'s are Gaussian distributions.
    Our goal is to estimate the unknown $\theta^*$ using Maximum Likelihood; that is, to find $\hat{\theta}$ that maximizes
    $$
        \ell_n(\theta) = \frac{1}{n} \sum_{i=1}^{n} \log\bigg(\int_Z \phi_\theta(\xi_i,z_i) dz_i\bigg) \ .
    $$
    In practice, the function $\ell_n$ is usually hard to optimize. The EM algorithm provides a way of searching for such maximum indirectly through the maximization of another function $Q_n: \Omega \times \Omega \rightarrow \mathbb{R}$ defined as
    $$
        Q_n(\theta|\theta') = \frac{1}{n}\sum_{i = 1}^{n}\bigg(\int_Z k_{\theta'}(z|\xi_i)\log \phi_\theta(\xi_i,z) dz\bigg)
    $$
    where $k_{\theta'}(z|\xi)$ is the conditional density of $z$ given $\xi$. Given this function and a current estimate $\theta_{n,t}$, the sample EM update is defined as
    $$
        \theta_{n,t+1} = \theta_{n,t} +s\nabla Q_n(\theta|\theta_{n,t})\big|_{\theta = \theta_{n,t}}, \qquad t= 0,1, \ldots \ .
    $$
    where $s$ is the step size. To study convergence of the EM to a (neighborhood of) the global optimum, \cite{balakrishnan} define the population level versions $\ell$ of $\ell_n$ and $Q$ of $Q_n$ as
    \begin{align*}
        \ell(\theta) = & \  \int_\Xi log\bigg(\int_Z \phi_\theta(\xi_i,z_i) dz_i\bigg)g_{\theta^*}(\xi)d\xi  \\
            Q(\theta|\theta') = & \ \int_\xi\bigg(\int_Z k_{\theta'}(z|\xi_i)\log \phi_\theta(\xi_i,z) dz\bigg)g_{\theta^*}(\xi)d\xi \ .
    \end{align*}
    Correspondingly, one has a population version of the EM update
    $$
        \theta_{t+1} = \theta_t + s \nabla Q(\theta|\theta_t)\big|_{\theta = \theta_t}, \qquad t= 0,1, \ldots \ .
    $$
    Based on this, since $\theta^*$ maximizes $\ell(\theta)$, to prove that the sample EM update converges to (a neighborhood of) $\theta^*$, one needs to prove that
    (i) the population EM  update converges to (a neighborhood of) $\theta^*$; and
    (ii) the sample EM update tracks closely the population update
    (this is precisely what we do in Theorem \ref{thm1} and Theorem  \ref{thm2} below).
    
   \section{The \emph{smoothEM} approach}
{\label{c2sec:approach}}

Let us consider again data as in Equations~(\ref{eq:1}) and (\ref{model}); that is 
\begin{align*}
    \xi_i &:= y_i - f(x_i) = \begin{cases}
    \mu^* + \epsilon_i, \text{ if } x_i \in   \mathbb{S}\\
    \epsilon_i, \text{ otherwise}
    \end{cases}  \\
    &\sim \alpha^*N(0, \sigma^{*2}) + (1-\alpha^*)N(\mu^*,\sigma^{*2})
\end{align*}
where the design (the $x_i$'s) is taken as fixed.
If we knew $f$, 
the EM algorithm could be used to search for the MLE of the mixture parameters $\theta^* = (\alpha^*, \mu^*, \sigma^{*2})$ and to estimate membership (i.e.~posterior) probabilities for each point, and thus the classification vector $\mathbbm{1}_{\mathbb{S}}$.
In reality, we do not know $f$, so we use the EM with an estimate $\hat{f}$ of $f$.
The traditional penalized smoothing technique in Equation (\ref{eq:3}) is ill-fitted for the above purpose. Indeed, it declares as optimal an $\hat{f}$ that minimizes a combination of sum of squared errors and degree of roughness. The tuning parameter $\lambda$, generally chosen by cross validation, determines the balance between these two competing criteria. In the case of noisy curves with spikes as defined in Equation (\ref{eq:1}), a sufficiently large $\mu^*$ causes the sum of squared errors term to dominate the roughness criterion. This in turns causes the cross validation procedure to be biased towards small values of $\lambda$, i.e. towards under-smoothed $\hat{f}$. 
This is the key observation that gives rise to our iterated penalized smoothing procedure, which we illustrate through two simple examples.
\begin{figure}[!tbh]
   \centering
    \includegraphics[width=\textwidth]{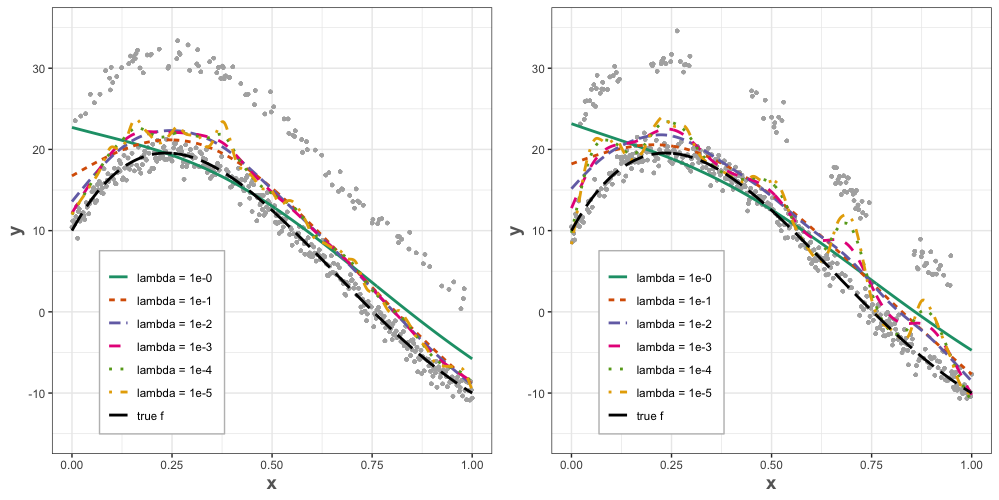}
    \caption{Simulated data and smoothing spline fit for $f$ with different values of the tuning parameter $\lambda$ (curves of different types and colors). The dashed black curve represents the true $f$. Left: independently and uniformly distributed spikes. Right: spikes clustered into ``hovering clouds''.}
    \label{fig:2} 
\end{figure}
Figure \ref{fig:2} plots $n = 500$ data points simulated across the $[0,1]$ domain.
The majority of such points are scattered about a fourth degree polynomial, the curve $f$, with independent errors $\epsilon_i \sim N(0,1)$. In the left panel, spikes occur uniformly across the domain and independently from other data points, while in the right panel, spikes form ``hovering clouds'' of different denseness at different locations along the domain. In both scenarios, spikes are shifted vertically by $\mu^* = 12$. 
We note that the clustered spikes (right panel) are correlated, and thus represent a departure from our assumed mixture Gaussian model. Even if our theoretical results require independence, we show with simulations in Section~\ref{simulations} that our algorithm is robust to this departure from the theoretical assumptions, maintaining its ability to identify spikes and recover the true $f$ also in this case. The true $f$ is plotted in black, and the estimates $\hat{f}$, obtained through 300 cubic spline basis functions with equally spaced knots and penalty of order 1, are plotted in different colors depending on the values of $\lambda$. At first glance, none of the $\hat{f}$'s approximates $f$ well, though some give better fits than others. In particular, the presence of spikes, especially when coupled with larger $\lambda$, affects estimation at both spike and non-spike locations. 
Also, for the correlated case (right panel), ``denser'' spikes sites distort $\hat{f}$ to a higher degree -- especially for small $\lambda$ values. Generalized cross-validation here selects $\lambda = 10^{-4}$, which still results in a highly distorted $\hat{f}$. 
Suppose now we were to identify spikes and do inference on the parameters $\alpha^*, \mu^*, \sigma^{*2}$ through the residuals from $\hat{f}$. Such a distorted estimate would generate misleading residuals.
On the other hand, just from visual inspection, $\lambda =  10^{-1}$ and $\lambda =  10^{-2}$ produce much more reasonable $\hat{f}$'s and, correspondingly, residuals which are a much better approximation of the underlying $\xi_i$'s, giving some hope that one may be able to identify spikes based on their magnitudes. If we identify and filter out the spikes, and then repeat regularized smoothing, we can obtain a much improved fit to $f$. 
\subsection{smoothEM: the algorithm}
Based on the above reasoning, we propose our \textbf{\textit{\MakeLowercase{smooth}EM}} procedure, which comprises the following steps:
\begin{enumerate}[label = S\arabic*]
    \label{fullalgorithm}
    \item
        Fit 
        penalized smoothing 
        splines over a grid of $\lambda$'s
        and obtain 
        residuals $\xi_\lambda = \{\xi_i(\lambda)\}_{i = 1}^n$.
    \item
        For each pair $(\lambda,\xi_\lambda)$, classify as ``spikes'' the set of largest $\xi_i(\lambda)$'s 
        (with a bound on its cardinality, e.g., $\leq 50\%$ of the observations),
        thus producing binary 
        memberships  $M_\lambda=\{M_i(\lambda)\}_{i = 1}^n$ 
        ($M_i(\lambda) = 0,1$ 
        if 
        $i$ is classified as ``smooth'' or ``spike'', respectively).
    \item
        For each pair $(\lambda, M_\lambda)$, fit a second
        penalized smoothing spline using only observations with $M_i(\lambda) = 0$, and compute updated residuals $\xi'_\lambda$ for all observations. 
        Here we also 
        compute an ``overfit'' score $F(\lambda)$
        based on how much the fit $\hat{f}$ 
        changes 
        when 
        smooth observations ($M_i(\lambda) = 0$) are perturbed by a small amount (see below).
    \item
        For each $\lambda$, 
        run the EM algorithm on $\xi'_\lambda$
        with $M_\lambda$ as initialization,
        to produce parameter estimates $(\hat{\alpha}, \hat{\mu}, \hat{\sigma}^2)_{\lambda}$ and new 
        memberships 
        $M'_\lambda$, obtained by thresholding 
        posterior probabilities
        (our implementation automatically 
        selects the threshold 
        between 0.5 and 1 such that the
        classification maximizes the log-likelihood).
    \item
        Considering a combination of log-likelihood and ``overfit'' scores,
        select $\lambda^* = \arg \max_\lambda \allowbreak \left[\ell(\lambda;\hat{\alpha}, \hat{\mu}, \hat{\sigma}^2) + F(\lambda) \right]$
        and thus parameter estimates
        $(\hat{\alpha}, \hat{\mu}, \hat{\sigma}^2)_{\lambda^*}$ and 
        memberships $M'_{\lambda^*}$.
    \item 
        Refit a 
        penalized smoothing spline 
        with $\lambda = \lambda^*$, using only observations 
        that satisfy  $M'_i(\lambda^*) = 0$. 
\end{enumerate}

\subsection{Some observations and remarks}
The performance of 
our procedure depends critically on the grid of $\lambda$ values chosen for its implementation. A relatively small grid suffices 
if one has prior knowledge about the smoothness of $f$. 
Otherwise, we recommend exploring a 
dense grid, 
likely 
to include a good $\lambda$ value
,  
withstanding the greater computational cost. 
%
Also, the initial residual magnitude-based classification in Step~2 can be 
implemented in 
different ways, with different computational costs (e.g., running a 1-dimensional $K$-means,
or pinpointing
the largest 
difference in 
the ordered 
sequence). 

In our experience, the overfit score $F$ in Step~3 is especially helpful when there is a limited number of observations.
Without it, the algorithm might favor very small values of $\lambda$ and
overfit most/all 
observations, leading to near zero residuals (almost) everywhere and producing very high likelihood values in later steps. 
The overfit score $F$ is calculated as
follows.
Let $\xi$ be 
the collection of residuals 
and $\xi^{(0)}$ the sub-collection corresponding to 
``smooth'' observations
($M_i = 0$). 
Let $\xi_p^{(0)}$ be a perturbed version of $\xi^{(0)}$ obtained as $\xi_p^{(0)} = \xi^{(0)} + \tau$, where $\tau$ is a vector of independent variables from a $N(0, \sigma^2_{\tau})$. 
Let $\hat{f}(\xi^{(0)})$ and $\hat{f}(\xi_p^{(0)})$ be the curves fitted to $\xi^{(0)}$ and $\xi_p^{(0)}$, respectively, and compute the score as $F = \small\|\hat{f}(\xi^{(0)}) - \hat{f}(\xi_p^{(0)})\small\|_2$.
Here, the amount of perturbation can be chosen to match the overall level of noise in the original data. We recommend using a robust estimation of standard deviation, e.g., $\sigma_{\tau} = median(|\hat{\xi}_i - median(\hat{\xi})|)$, where $\hat{\xi}$ are rough residual estimates obtained from fitting a loess curve to the raw data.

An \textsf{R} implementation of \emph{smoothEM} and some examples are provided at \url{https://github.com/hqd1/smoothEM}

\section{Theoretical remarks and guarantees}
\label{sec:theoretical_guarantees}

    \subsection{Remarks for iterated smoothing}
    
    Here we discuss in more detail the interplay between the smoothness of $f$, $\lambda$, $\mu^*/\sigma^*$, $n$, and the denseness of spikes, in determining the effectiveness of iterated smoothing.
    We 
    assume that $(x_i, y_i)_{i = 1}^n$ are generated as at the beginning of Section~\ref{c2sec:approach}, but the remarks in this subsection also extend to the case of correlated spikes.
    For a given realization of $\mathbbm{1}(x_i \in \mathbb{S})$ (i.e.~with 
    spike locations fixed), the only randomness is from 
    $(\epsilon_i)_{i=1}^n$.
    Assuming without loss of generality that $\mu^* >0$, let $r(x_p)$ and $r(x_s)$ be residuals from Step~1 at spike location $x_p$ and smooth location $x_s$. 
    The difference 
    \begin{align*}
        r(x_p) - r(x_s) 
        &= 
        y_p-\hat{f}(x_p) -  (y_s - \hat{f}(x_s)) \\
        &= f(x_p) + \mu^* + \epsilon_p - N(x_p)H^{-1}_nN^\top(f(x)+ \mu^* \cdot \mathbbm{1}_{\mathbb{S}} + \epsilon)/n \\
        & \quad - [f(x_s) + \epsilon_s - N(x_s)H^{-1}_nN^\top(f(x)+ \mu^* \cdot \mathbbm{1}_{\mathbb{S}} + \epsilon)/n] 
    \end{align*}
    (see Section~\ref{sec:spline}) can be decomposed as $r(x_p) - r(x_s) = R_1 + R_2 + R_3$, where
    \begin{align*}
        &R_1 = 
        f(x_p) - N(x_p)H^{-1}_nN^\top(f(x) + \epsilon)/n 
        - [f(x_s)- N(x_s)H^{-1}_nN^\top(f(x)+ \epsilon)/n],\\ 
        &R_2 = 
        \mu^*  - N(x_p)H^{-1}_nN^\top\mu^* \cdot \mathbbm{1}_{\mathbb{S}}/n 
        - [0 - N(x_s)H^{-1}_nN^\top\mu^* \cdot \mathbbm{1}_{\mathbb{S}}/n] ,\\  
        &R_3 = 
        \epsilon_p-\epsilon_s.
    \end{align*}
    The first component is $R_1 = r'(x_p) -r'(x_s)$, the difference
    of the residuals from fitting a spline to the noisy curve only. 
    As long as the underlying true $f(x)$ is $\in C^p[0,1]$ for $p\leq m$, where $m$ is the order of the smoothing spline, \citet{xiao19} guarantees that under appropriate conditions $\max_x r'(x)$ is of order $o\left\{\left(\log{n}/n\right)^{-m/(2m+1)}\right\}$. 
    The second component is $R_2 = r''(x_p) - r''(x_s)$,
    the difference of the residuals from fitting a spline to the $n$-discretized piecewise constant $\mu^* \cdot \mathbbm{1}_{\mathbb{S}}$.
    As spline smoothing is highly localized by nature, intuitively, if the number of spikes in a neighborhood of $x_p$ is sufficiently small and $\lambda$ is sufficiently large, the smoothing spline will prioritize approximation of the constant line $g(\cdot) = 0$, causing $r''(x_p) - r''(x_s)$ to be near $\mu^*$. As more spikes gather around $x_p$, this magnitude decreases, making spikes less distinguishable.
    It is important to note that the same $\lambda$ is used to fit the noisy curve in $R_1$ and $\mu^* \cdot \mathbbm{1}_{\mathbb{S}}$ in $R_2$. 
    As a consequence, if $f(x)$ 
    is rather jagged and thus the optimal $\lambda$ to fit it is small, one may easily overfit $\mu^* \cdot \mathbbm{1}_{\mathbb{S}}$ -- especially when spikes are dense in a neighborhood of $x_p$. 
    The third component is simply $R_3 = \epsilon_p - \epsilon_s$. 
    Under ideal conditions, 
    $R_1$ will be close to 0, 
    $R_2$ will be close to $\mu^*$, and as long as $\mu^*/\sigma^{*2}$ is large, the effect of 
    $R_3$ will be negligible.

    \subsection{Results for EM}
    \label{sec:EMtheory}
    
    In the following, let $q(\theta) = Q(\theta|\theta^*)$ (see Section~\ref{EMbackground}), $\omega$ is an arbitrarily small, positive number, and let $\mathbb{B}_2(r;\theta^*)$ denote an $L_2$ ball centered at $\theta^*$ with radius $r$. 
   \begin{theorem}[\textbf{Population level guarantees for the Gaussian mixture model}]
   \label{thm1}
        Consider the Gaussian mixture model in Equation (\ref{model}) with unknown parameters $\theta^* = (\alpha^*, \mu^*, \sigma^{*2}) \in (0.5,1) \times \mathbb{R} \times \mathbb{R}^+$. Given any initialization $\theta_0 \in  \mathbb{B}_2(r;\theta^*)$,  the population first order EM iterates satisfy the bound
        $$
            \left\Vert  \theta_k - \theta^*\right\Vert_2 \leq \bigg(1-\frac{2\nu-\gamma}{L+\nu}\bigg)^k\left\Vert  \theta_0 - \theta^*\right\Vert_2 \quad \text{ for all } k = 1,2, \ldots
        $$
        where $0 \leq \gamma < \nu \leq L$ and \begin{itemize}
            \item
                $\nu = \min \left\{\left(\frac{1}{(\alpha^*+r)^2} \vee 1 \right),  \frac{\sigma^{*2} - r}{2(\sigma^{*2}+r)^3}-\frac{(1-\alpha^*)r}{(\sigma^{*2} -r)^2}, \frac{1-\alpha^*}{\sigma^{*2}+r}-\frac{(1-\alpha^*)r}{(\sigma^{*2} -r)^2}\right\}$;
            \item
                $L = \max\left\{\frac{\alpha^*}{(\alpha^*-r \ \vee \ 0.5)^2} + \frac{1-\alpha^*}{(1-\alpha^*-r \ \vee \ \omega)^2},\frac{(1-\alpha^*)\sigma^{*2}}{(\sigma^{*2} -r)^2},\frac{\sigma^{*2}+r}{2(\sigma^{*2}-r)^3} +  \frac{1-\alpha^*}{\sigma^{*2} -r}\right\}$;
            \item
                $\gamma(\alpha^*,\mu^*, \sigma^{*2}) \sim O\left(\frac{\mu^{*5}}{\sigma^{*8}}\exp(-\frac{\mu^* -r}{\sigma^{*2}+r})\right)$, which decays exponentially with large $\frac{\mu^*}{\sigma^{*2}}$.
        \end{itemize} 
    \end{theorem}
    Before proceeding, some remarks are in order on the use of Theorem \ref{thm1}. 
    First, \cite{balakrishnan} 
    proved exponential convergence rate for the same model, but assuming known 
    $\alpha^*$ and $\sigma^{*2}$.
    Such exponential rate is 
    achievable because, 
    if $\alpha^*$ and $\sigma^{*2}$ are known, $\nu = L$ and $\left[1-\left(2\nu-\gamma\right)/\left(L+\nu\right)\right]$ is reduced to just $\gamma/(2L)$. Theorem \ref{thm1} provides a slower convergence rate, but considers all parameters as unknown (a proof is provided in the Supplementary Material).
    
    Second, $\nu$ decreases as $r$ increases; this creates an undesirable trade off, as ideally we would want both to be large. A larger $r$ allows for a larger basin of attraction for convergence but slows convergence, whereas a larger $\nu$ hastens the convergence rate. Choosing $r = \sigma^{*2}/5$ ensures that $\nu >0$.
 
    Third, even with a positive $\nu$, the convergence rate can be slowed by a large Lipschitz smoothness constant $L$.
    In particular, an arbitrarily large $ (1-\alpha^*)/(1-\alpha^*-r \ \vee \ \omega)^2$ makes the rate $\left[1-\left(2\nu-\gamma\right)/\left(L+\nu\right)\right]^k$ unacceptably slow. A workaround is to require that $\alpha_0 \in \mathbb{B}_2(r_\alpha;\alpha^*)$, where $r_\alpha < r$ is a small constant, so that $L$ is bounded by $ (1-\alpha^*)/(1-\alpha^*-r_\alpha)^2$ instead. 
    As an example, a reasonable choice for $r_\alpha$ is $0.25 - \alpha^*/4$.
    
    \begin{table}[b!]
\caption{\label{convergencetable} Convergence rate $CR=\left[1-(2\nu-\gamma)/(L+\nu)\right]$ for different values of $\alpha^*$, $\sigma^{*2}$, and $r = \sigma^{*2}/5$.
Rates are calculated 
assuming a sufficiently large signal to noise ratio $\mu^*/\sigma^{*2}$.}

        \centering
        \small
        \fbox{%
        \begin{tabular}{ c|c|c|c|c|c|c| } 
            \multicolumn{2}{c|}{\multirow{2}*{
            $CR$}}&
            \multicolumn{5}{c|}{ $(\sigma^{*2},r)$} \\
            \cline{3-7}
            \multicolumn{2}{c|}{}&(1, 0.2) & (2, 0.4) & (3, 0.6) & (4, 0.8) & (5, 1)\\
            \hline
            	&.6&0.932&0.969&0.979&0.985&0.988\\
            \cline{2-7}
	&.7&0.957&0.979&0.986&0.989&0.992\\
            \cline{2-7}
            &.8&0.966&0.989&0.993&0.995&0.996\\
            \cline{2-7}
            \multirow{-4}*{$\alpha^{*}$}&.9&0.979&0.996&0.998&0.999&0.999\\
            \hline
        \end{tabular}} 
    \end{table}

    Table~\ref{convergencetable} provides the convergence rates for various parameter settings, assuming sufficiently large signal to noise ratio $\mu^*/\sigma^{*2}$. 
    As to be expected, convergence rates improve with lower
    noise levels and more balanced proportions 
    between spike and smooth components. 
    Of course convergence
    with unknown parameters is 
    slower than that in the case of known $\alpha^*$ and $\sigma^{*2}$, but
    we still observe very reasonable convergence rates in simulations (see Section~\ref{simulations}).


    The next theorem concerns the convergence of the sample EM updates (a proof is provided in the Supplementary Material).
    
    \begin{theorem}[\textbf{Sample level guarantees for the Gaussian mixture model}]
        \label{thm2}
        Consider the Gaussian mixture model in Equation (\ref{model}) with unknown parameters $\theta^* = (\alpha^*, \mu^*, \sigma^{*2}) \in (0.5,1) \times \mathbb{R} \times \mathbb{R}^+$ and let n 
        be the sample size.  Given any 
        initialization $\theta_0 \in \mathbb{B}_2(r;\theta^*)$, with probability at least $1- \delta$
        the finite sample EM iterates $\{\theta_k\}_{k=0}^\infty$ satisfy the bound
        $$
            \left\Vert  \theta_t - \theta^*\right\Vert_2 \leq \left(1- \frac{2\nu-2\gamma}{L + \nu}\right)^t\left\Vert  \theta_0- \theta^*\right\Vert_2 + \epsilon_n \, 
        $$        
where $\epsilon_n \rightarrow 0$ almost surely.
    \end{theorem}
    Combined with Theorem \ref{thm1}, Theorem \ref{thm2} 
    shows that sample EM updates have the same convergence rate as the population updates, asymptotically.


\section{Simulations}
\label{simulations}
\subsection{Uniformly distributed spikes}
In our simulations $(x_i)_{i=1}^n$ are equispaced along the interval $[0,1]$. The true smooth component is a polynomial of degree 4 (or, equivalently, of order 5).
Different distributions of spike locations affect traditional spline smoothing methods differently, due to their local nature. The results in this section concerns spikes that are uniformly distributed across the domain. 
In the following, we demonstrate in detail the performance of \emph{smoothEM} 
considering different settings, namely: 
$n= 2000, 1000, 500, 200$, $\alpha^*$ 
between $0.8$ 
and $0.98$ 
(i.e.~spike contamination levels between $0.2$ and $0.02$), and ``signal to noise''
(STN) ratios $\mu^*/(6\sigma^*)$
between $0.2$ 
and $2$ -- these are implemented fixing $\sigma^*=1$ and changing the spike size
$\mu^*$.
Note that 
$6\sigma^*$ represents the width of the $95\%$ Gaussian noise interval at any given location. Thus, for instance, if $(x_i,y_i)$ lies $3\sigma^*$ \textit{below} $f(x_i)$
and 
$x_i \in \mathbb{S}$, then $\mu^* = 2 \cdot 6\sigma^*$ brings $(x_i,y_i)$ to a height well separated from the smooth underlying pattern, $3\sigma^*$ \textit{above} the graph of $f$. 

\begin{figure}[!tbh]
        \centering
        \begin{tabular}{cc}
        \includegraphics[width=.45\textwidth]{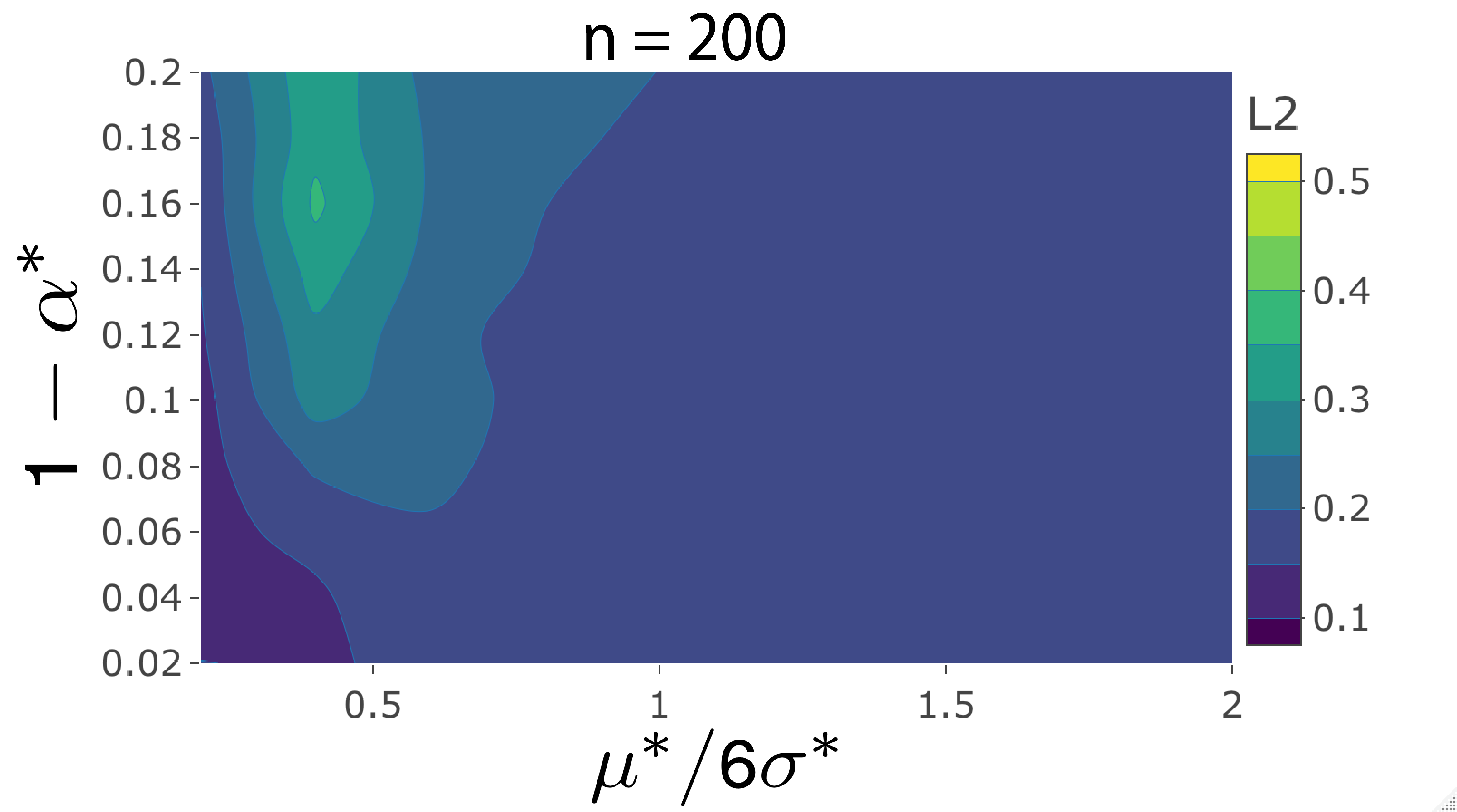}&
        \includegraphics[width=.45\textwidth]{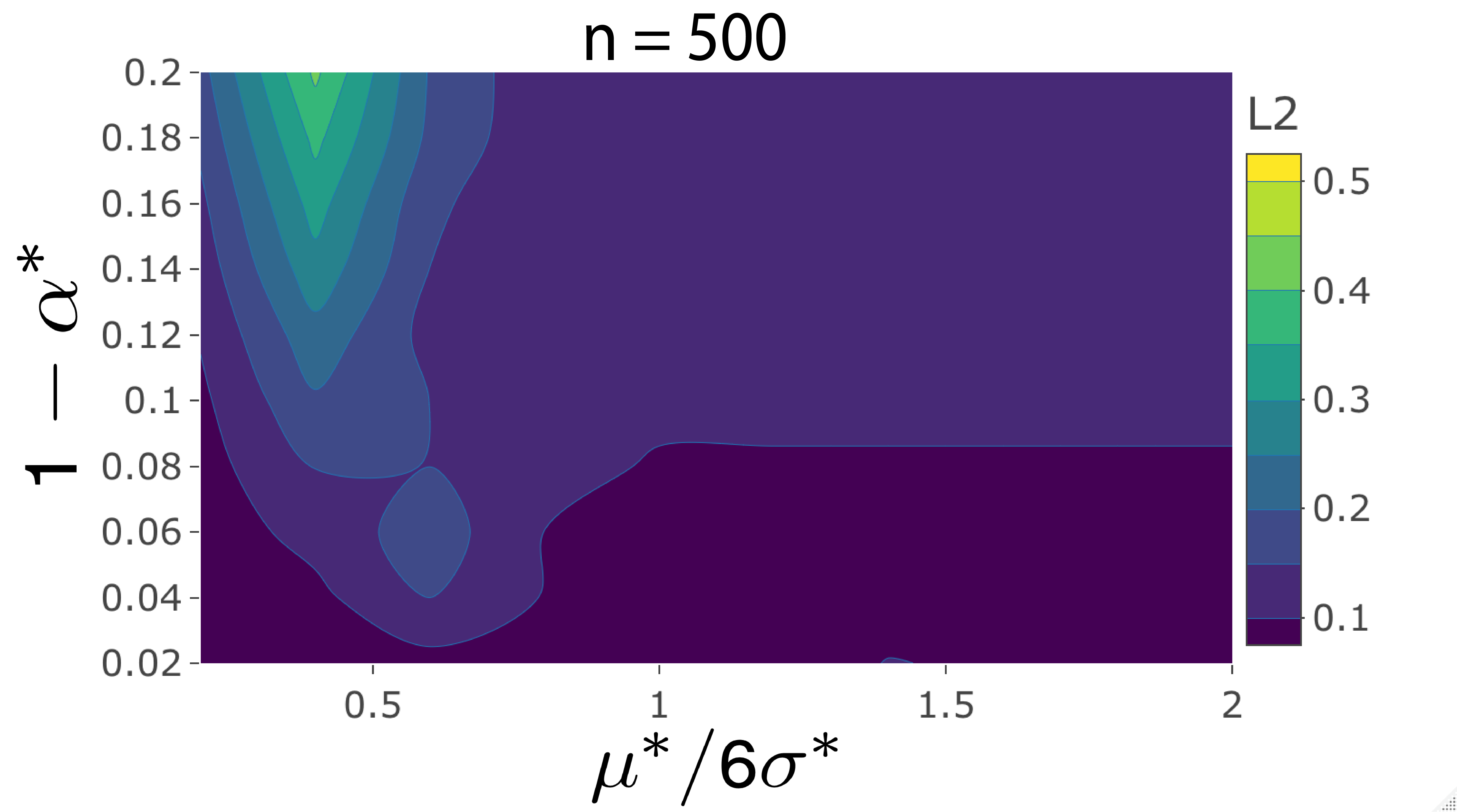}\\
        \includegraphics[width=.45\textwidth]{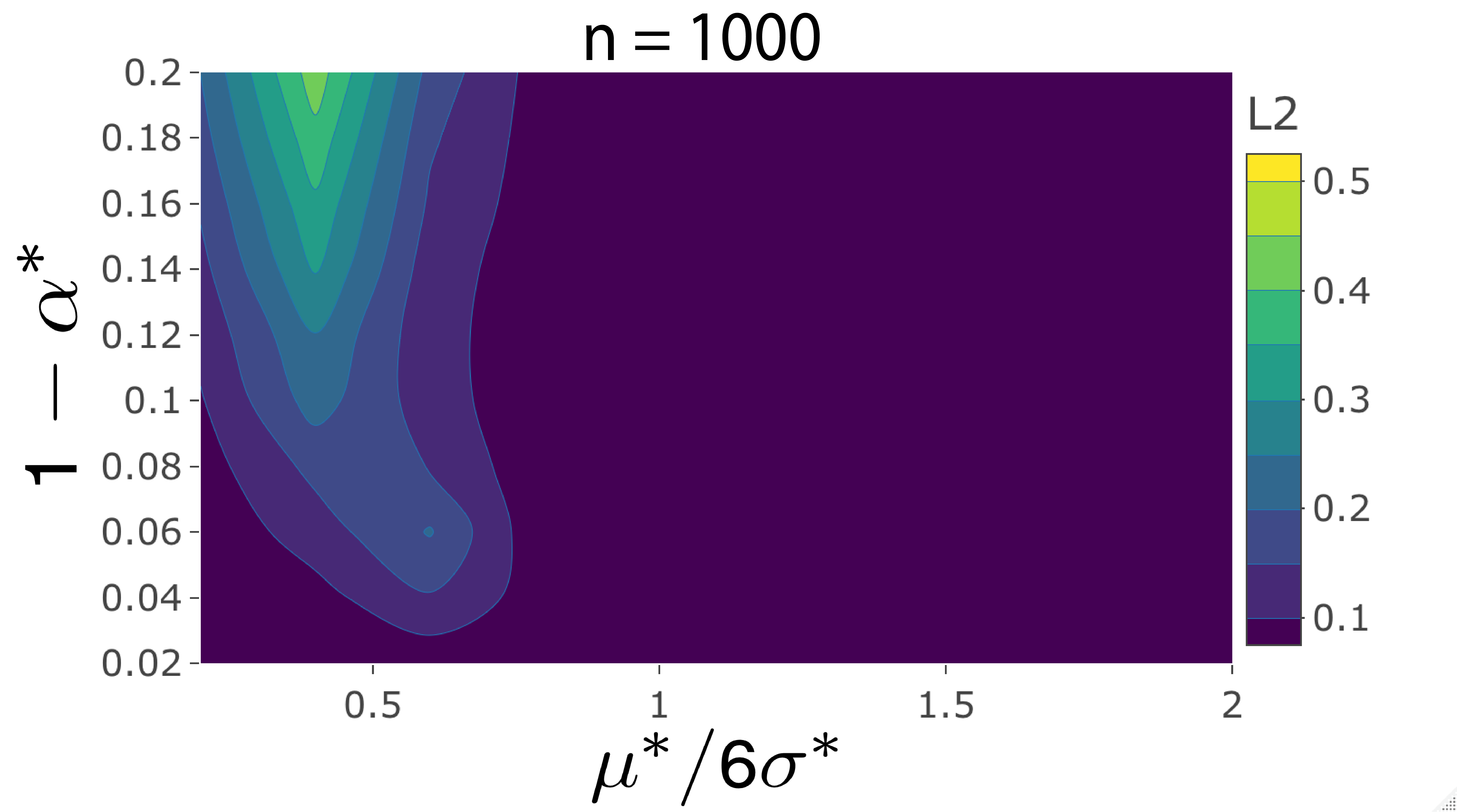}&
        \includegraphics[width=.45\textwidth]{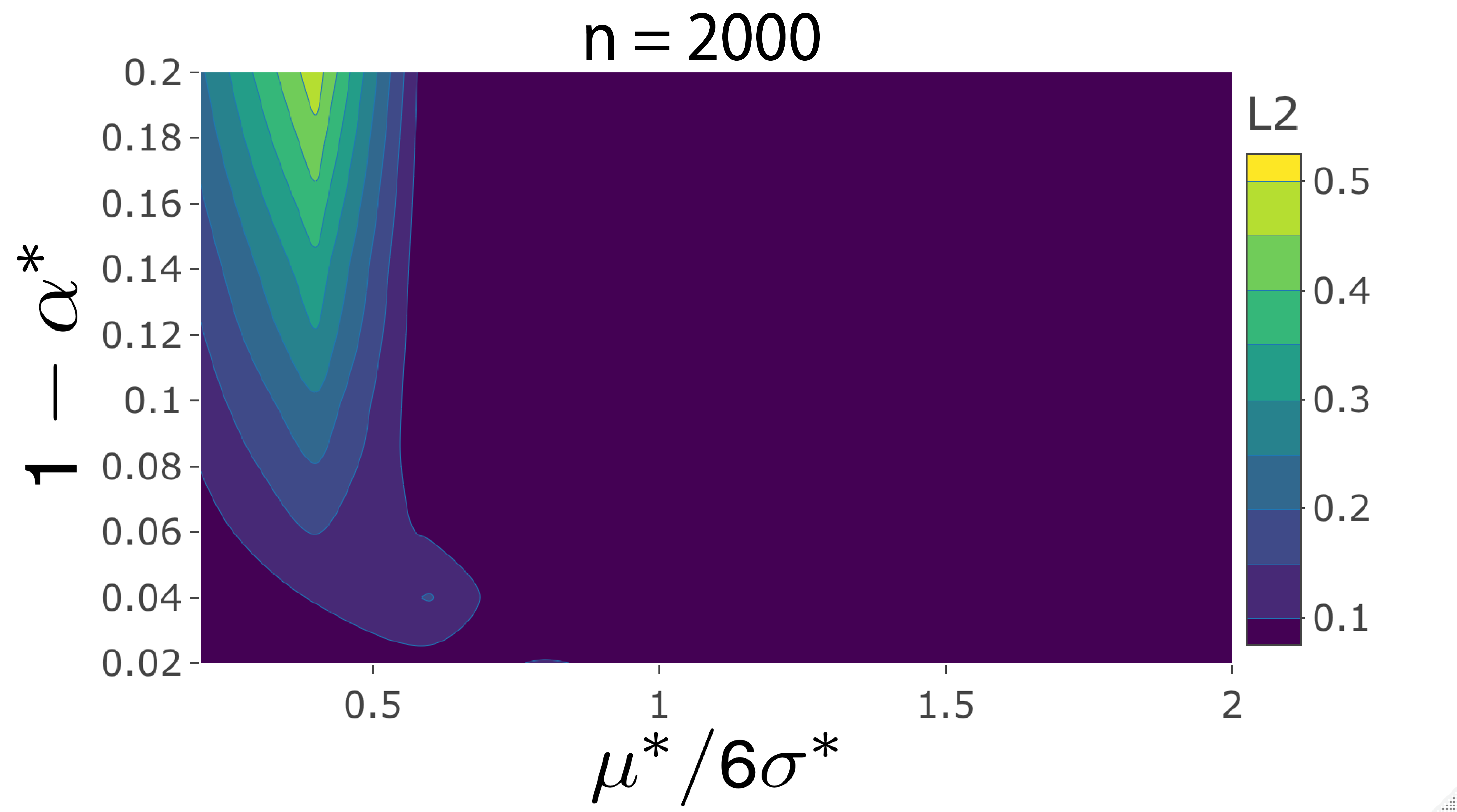}
        \end{tabular}
        \caption{$L_2$ error of the {\em smoothEM} smooth component estimate  
        on simulated data with uniformly distributed spikes. The contour plots show the error (averaged over $20$ simulation replicates) as a function of the spike percentage ($1-\alpha^*$) and the STN $\mu^*/(6\sigma^*)$.
        From left to right, top to bottom, $n = 200, 500, 1000, 2000$.}
        \label{fig:fig8}
    \end{figure}

\begin{figure}[!tbh]
        \centering
        \begin{tabular}{cc}
        \includegraphics[width=.45\textwidth]{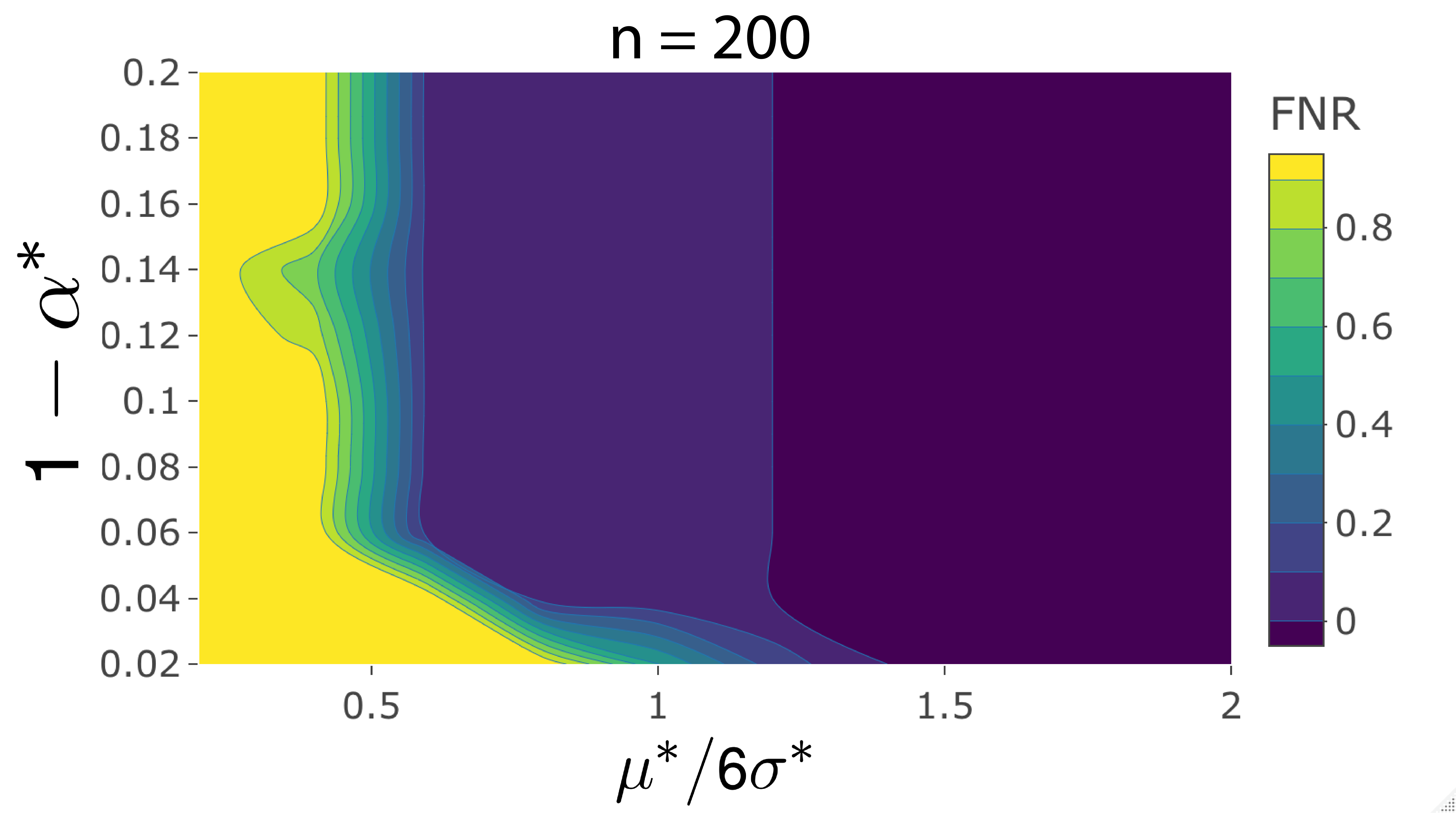}&
        \includegraphics[width=.45\textwidth]{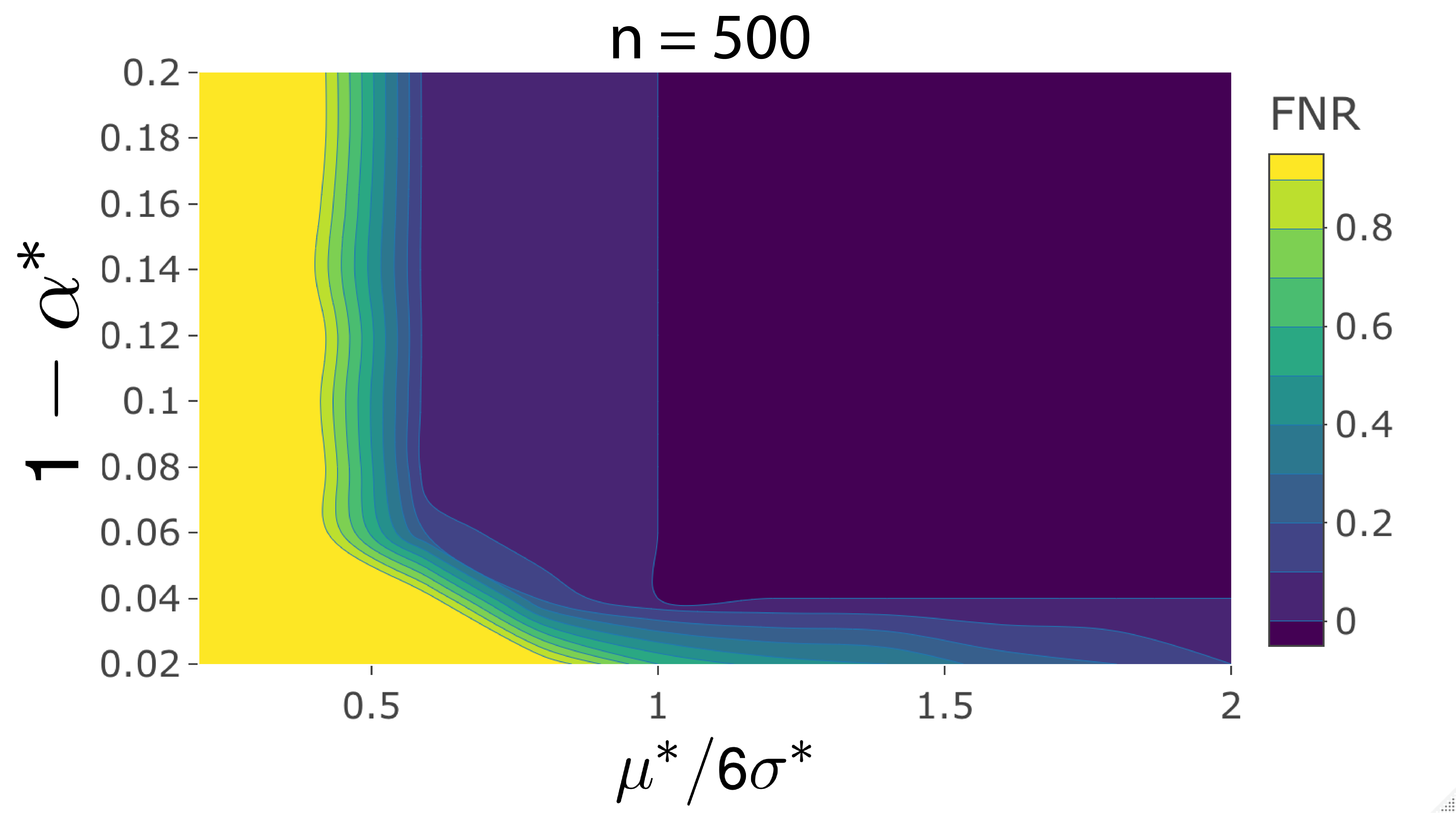}\\
        \includegraphics[width=.45\textwidth]{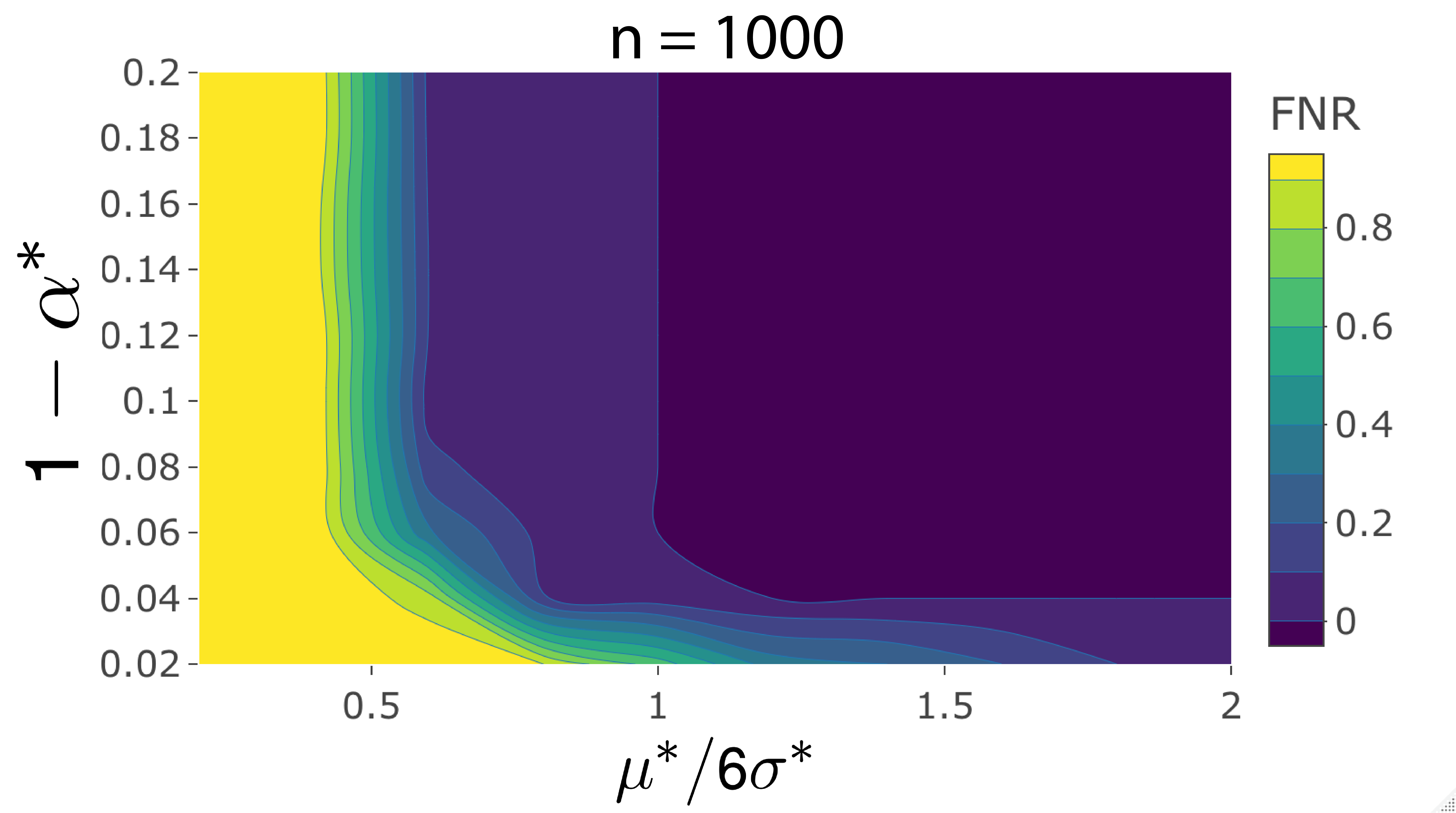}&
        \includegraphics[width=.45\textwidth]{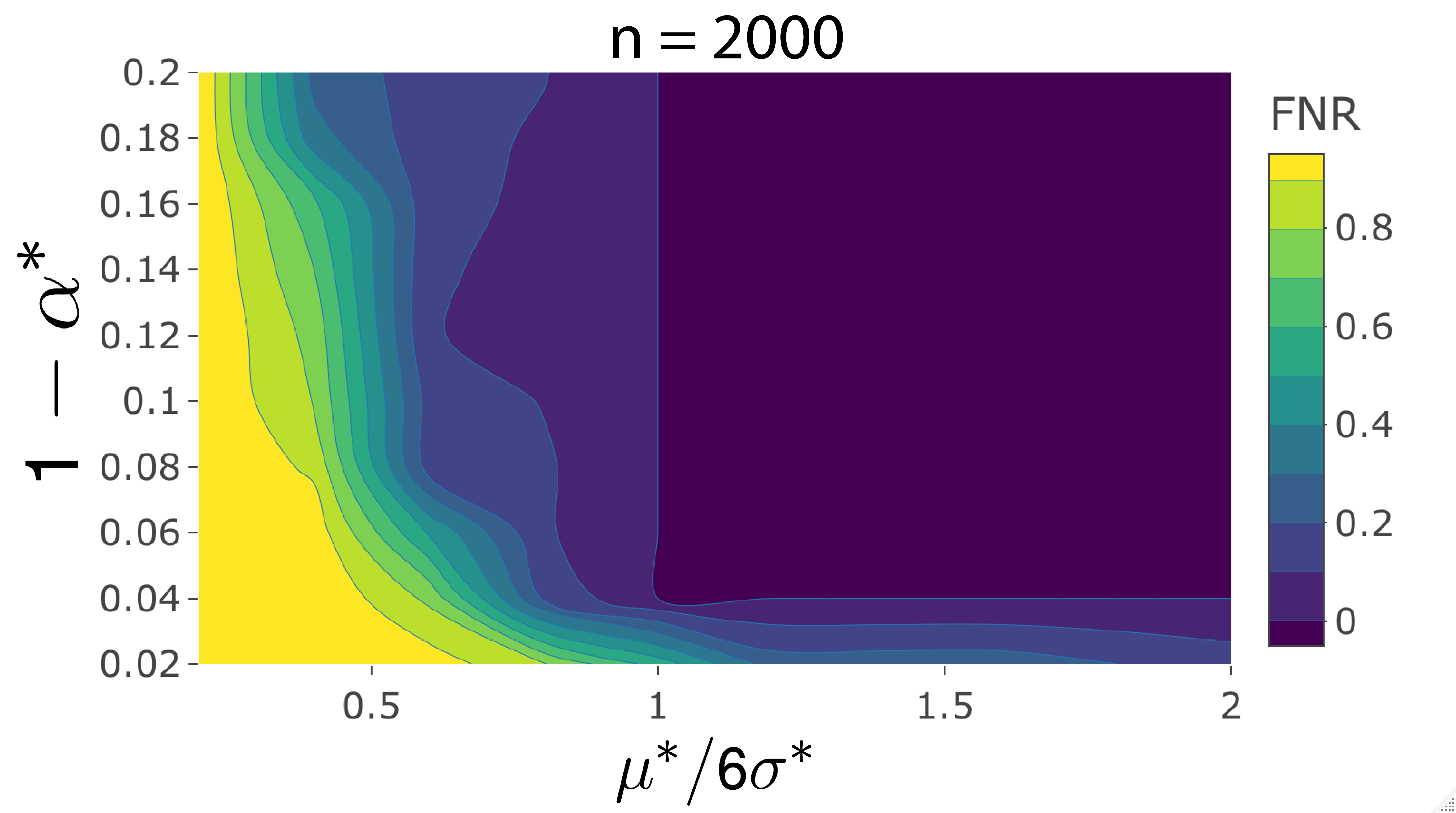}
        \end{tabular}
        \caption{FNR of the {\em smoothEM} spike identification
        on simulated data with uniformly distributed spikes. The contour plots show the FNR (averaged over $20$ simulation replicates) as a function of the spike percentage ($1-\alpha^*$) and the 
        STN $\mu^*/(6\sigma^*)$.
        From left to right, top to bottom, $n = 200, 500, 1000, 2000$.
        }
        \label{fig:fig1} 
    \end{figure}

    Figure~\ref{fig:fig8} contains contour plots of $\small\|  \hat{f}-f\small\|_2$ for the four sample sizes $n$,
    with
    spike percentages ($1-\alpha^*$) and
    STNs $\mu^*/(6\sigma^*)$ on the axes. 
    For each parameter setting, 
    results shown are averages over $20$ simulation replicates.
    Figure~\ref{fig:fig1} contains similar plots for 
    False Negative Rates (FNR) in spike identification.
    False Positive Rates (FPR) 
    are not shown because our procedure has excellent specificity in all
    settings considered; the
    largest FPR is $0.02$.

    We observe that, when $n$ is large, a sufficiently large signal $\mu^*/(6\sigma^*) \geq 1$ corresponds to low FNR (i.e.~good spike classification),
    and thus low error in estimating the smooth component.  When the signal is small, so that spikes are not well separated, our procedure naturally has a harder time recognizing them
    -- but estimation of the smooth component does not suffer much, 
    as smaller spikes do not distort the fit substantially.
    Notably though, both spike identification and smooth component estimation 
    do improve with larger spike size and lower $\alpha^*$, 
    which is in line with our previous discussion in Section \ref{sec:theoretical_guarantees}. 
    Additionally, Figure S2 in the Supplementary Material contains contour plots of the sum of squared error of parameter estimates $\small\|\hat{\theta} -\theta^*\small\|_2$, with the same format as Figures~\ref{fig:fig8} and \ref{fig:fig1}. Here 
    accuracy 
    is consistent with FNR results, which is to be expected since the mis-classification of spikes affects estimation of $\mu^*, \sigma^*$ and $\alpha^*$. 
    
\subsection{Non-homogeneous Poisson spikes}
\label{non-homopp sim}
    The top-right panel in Figure~\ref{fig:2} depicts a scenario where spike locations are generated through a non-homogeneous Poisson, as to be ``clumped'' -- instead of uniformly distributed across the domain. To achieve this, our simulation uses a thinning method by \cite{lewisshedler}. Following their algorithm, the rate function is chosen such that we have spike ``clumps'' of different sizes. We note that by using a non-homogeneous Poisson distribution, at best, we can only approximate the percentage of contamination. As the locations of spikes are now correlated, the mixture Gaussian model specification will not apply. However, simulation results demonstrate excellent robustness for this type of model mis-specification (see Figures S3-S5 in the Supplementary Material, which are the analogs of the figures in the case of uniformly distributed spikes.)
    This 
    suggests that 
    {\em soomthEM} performs similarly well, and thus that it possesses a degree of robustness to the different ways spikes may be distributed across the domain.

\subsection{Comparison to existing methods}
\label{comparison}
    Next, we compare \emph{smoothEM}
    to a recent baseline correction method called \emph{RWSS-GCV} \citep{wei22} and a general adaptive smoothing algorithm, implemented via the \emph{gam} function (with option  \emph{bs = ``ad"})  from the \emph{mgcv} \textsf{R} package \cite{Wood2011-ax} (hereby denoted \emph{mgcv-AS}). Specifically, we
    compare $L_2$ and $L_\infty$ errors in smooth component estimation, i.e.~$\small\|\hat{f}(x) - f(x)\small\|_2$
    and $max_x \small[ \hat{f}(x) - f(x) \small]$, across methods.
    We cannot compare 
    spike identification, since \emph{RWSS-GCV} and \emph{mgcv-AS} do not identify spikes. 
    We consider 
    both a slow-varying 
    and fast-varying underlying smooth curve 
    a $Beta(4,1)$ density
    and $9\pi sin(x)$ in $[0,1]$, respectively (see Figure S6 in the Supplementary Material). Each setting is again run $20$ times, and we report average errors.
        
    Table \ref{benchmark} summarizes the performance of the three methods, and shows that \emph{smoothEM}
    dominates 
    in most cases. 
    Adaptive smoothing (\emph{mgcv-AS}),
    due to its ability to dynamically and locally assign penalty, 
    is more vulnerable to spikes and performs poorly
    -- as it
    assigns lower penalty to spike-affected area. The only scenario in which \emph{mgcv-AS} outperforms \emph{smoothEM} is when both the sample size and the signal-to-noise ratio are low ($n = 200$ and $STN = 0.4$), making
    \emph{smoothEM} 
    more likely to mis-classify spikes and thus deteriorating its
    smooth curve estimation. 
    Both \emph{RWSS-GCV} and \emph{smoothEM} employ iterated penalized smoothing. Whereas \emph{RWSS-GCV} uses flexible weights to down-weight spikes, \emph{smoothEM} uses a hard classification scheme (spike/not spike) and performs classification and estimation simultaneously in a probabilistic framework. When the smooth curve is slow-varying, \emph{RWSS-GCV} trails behind \emph{smoothEM} by a large margin, probably due to the fact that the latter exploits distributional information about the data.
    Moreover, \emph{RWSS-GCV} is highly sensitive to the shape of the smooth curve; its
    performance worsens dramatically when estimating
    the fast-varying curve.
    Lastly, an important observation is that, contrary to \emph{smoothEM}, both \emph{RWSS-GCV} and \emph{mgcv-AS} 
    worsen when the signal-to-noise ratio increase. This suggests that these methods fail to take advantage of the larger separation between spikes and non-spikes to better estimate the smooth curve, and actually include the spikes in this estimation. 
    \begin{table}[!htbp] 
        \caption{\label{benchmark} 
        Comparison of \emph{smoothEM}, \emph{RWSS-GCV} and \emph{mgcv-AS}. 
        Bold entries mark the best performing method in each scenario and for each error.
        } 
    
            \footnotesize  %
            \begin{tabular}{@{}|c|c|c|c|c|c|c|c|c|c|c|} 
            \hline
                \multirow{2}*{
                } & \multirow{2}*{
                $n$} & \multirow{2}*{
                STN} & \multirow{2}*{
                $1-\alpha^*$} & \multicolumn{3}{c|}{ $L_2$} & \multicolumn{3}{c|}{ $L_\infty$}\\ 
                \cline{5-10}
                 &  &  &  & \emph{mgcv-AS} & \emph{RWSS-GCV} & \emph{smoothEM} & \emph{mgcv-AS} & \emph{RWSS-GCV} & \emph{smoothEM}\\ 
                \hline
                & &   2 & 0.1 & 3.4014 & 1.4756 & \textbf{0.0298} & 4.8207 & 2.1527 & \textbf{0.4213}\\ 
                & &   2 & 0.05 & 0.7572 & 1.1956 & \textbf{0.0308} & 2.4117 & 1.9272 & \textbf{0.4297} \\ 
                & &   1 & 0.1 & 0.8475 & 0.8234 & \textbf{0.0298} & 2.3363 & 1.8039 & \textbf{0.4213} \\ 
                & &   1 & 0.05 & 0.1905 & 0.9652 & \textbf{0.0308} & 1.0712 & 1.7230 & \textbf{0.4298}\\ 
                & &  0.4 & 0.1 & \textbf{0.1352} & 0.3609 & 0.1581 & 0.7771 & 1.1084 & \textbf{0.7222}\\ 
                &  \multirow{-6}*{200}&  0.4 & 0.05 & \textbf{0.0453} & 0.1051 & 0.2479 & \textbf{0.4692} & 0.6166 & 0.8674\\ 
                \cline{2-10}
                & &  2 & 0.1 & 2.6524 & 0.7562 & \textbf{0.0095} & 3.8448 & 1.9553 & \textbf{0.2243}\\ 
                & &  2 & 0.05 & 0.5592 & 0.4329 & \textbf{0.0057} & 2.1240 & 1.3041 & \textbf{0.2008}\\ 
                & &  1 & 0.1 & 0.7143 & 0.3901 & \textbf{0.0100} & 2.0590 & 1.2149 & \textbf{0.2259}\\ 
                & &  1 & 0.05 & 0.1698 & 0.4157 & \textbf{0.0060} & 1.1372 & 1.2619 & \textbf{0.2016}\\ 
                & &  0.4 & 0.1 & 0.1437 & 0.1678 & \textbf{0.1249} & 0.9306 & 0.7322 & \textbf{0.6172}\\ 
                 \multirow{-12}*{\makecell{S\\L\\O\\W\\-\\V\\A\\R\\Y\\I\\N\\G}} &  \multirow{-6}*{500} &   0.4 & 0.05 & 0.0519 & 0.3509 & \textbf{0.0419} & 0.4914 & 0.9738 & \textbf{0.3830}\\
                \hline
                & &   2 & 0.1 & 3.4899 & 8.4221 & \textbf{0.0456} & 4.8757 & 4.4467 & \textbf{0.6868}\\ 
                & &   2 & 0.05 & 0.8817 & 7.1128 & \textbf{0.0458} & 2.8577 & 4.0532 & \textbf{0.7235}\\ 
                & &   1 & 0.1 & 0.8803 & 4.8541 & \textbf{0.1064} & 2.3539 & 3.4204 & \textbf{0.8284}\\ 
                & &   1 & 0.05 & 0.2304 & 3.9899 & \textbf{0.0483} & 1.3602 & 2.8410 & \textbf{0.7250}\\ 
                & &  0.4 & 0.1 & \textbf{0.1352} & 0.3609 & 0.1581 & 0.7771 & 1.1084 & \textbf{0.7222} \\ 
                &  \multirow{-6}*{200}&  0.4 & 0.05 & \textbf{0.0453} & 0.1051 & 0.2479 & \textbf{0.4692} & 0.6166 & 0.8674\\ 
                \cline{2-10}
                & &  2 & 0.1 & 2.6753 & 3.6666 & \textbf{0.0225} & 3.9582 & 2.8876 & \textbf{0.6094}\\ 
                & &  2 & 0.05 & 0.5986 & 4.5329 & \textbf{0.0181} & 2.2317 & 2.9951 & \textbf{0.6014}\\ 
                & &  1 & 0.1 & 0.7228 & 3.7041 & \textbf{0.0331} & 2.0649 & 2.6321 & \textbf{0.5704}\\ 
                & &  1 & 0.05 & 0.1818 & 4.5263 & \textbf{0.0273} & 1.1663 & 3.0404 & \textbf{0.5952}\\ 
                & &  0.4 & 0.1 & \textbf{0.1493} & 3.3939 & 0.2239 & \textbf{0.9525} & 2.2901 & 1.1147\\ 
                 \multirow{-12}*{\makecell{F\\A\\S\\T\\-\\V\\A\\R\\Y\\I\\N\\G}} &  \multirow{-6}*{500} &   0.4 & 0.05 & \textbf{0.0569} & 2.3397 & 0.0696 & \textbf{0.5834} & 2.1700 & 0.6490\\
                 \hline
            \end{tabular} 
    \end{table}

\section{Data applications}

    \subsection{Smart meter electricity data}
    \label{SME}
    
    We consider data from the Smart Meter Electricity project of the Irish Commission for Energy Regulation, which collected data on electricity consumption from over 5000 households and businesses during 2009 and 2010.
    Our goal here is to create a meaningful statistical representation of the electricity consumption behaviors, which may be useful to policy makers.
    \begin{figure}[!b]
   \centering
   \includegraphics[width=\textwidth]{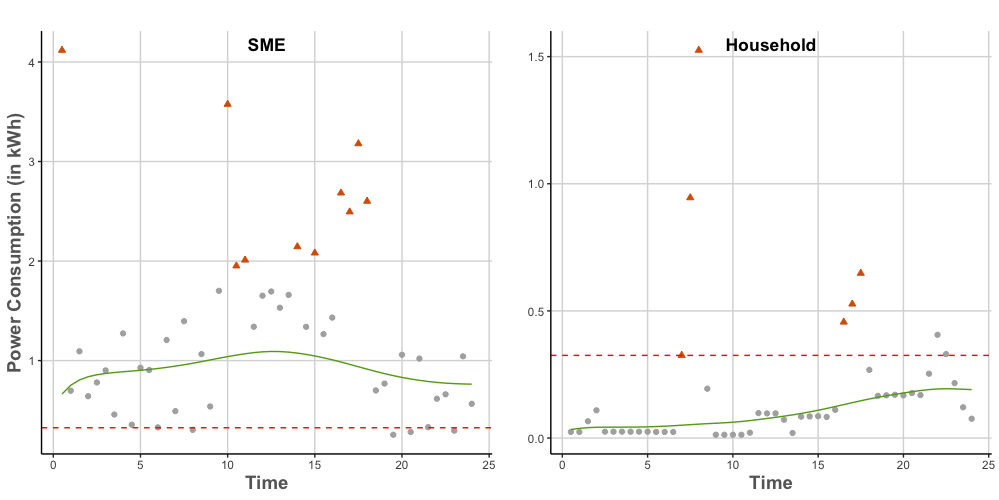}
        \caption{Electricity consumption by an Irish small business (left) and an Irish household (right) on Jan 5, 2010. The estimated smooth components are plotted in green, whereas spikes are identified by red triangles. The red horizontal dashed lines placed at 0.3 on the vertical axes help visualize the different magnitudes of consumption.
        }
        \label{fig:fig9} 
    \end{figure}
    We assume that consumption predominantly follows a smooth pattern with occasional spiked activity -- 
    e.g.~when multiple electrical devices are turned on simultaneously. 
    The data contains daily measurements of electricity consumed, collected at 30 minute intervals in kWh, for   each household and business in the study. 
    As an illustration, we run our procedure on data from one household (meter ID 1976) and one small business enterprise (SME, meter ID 1977) for the months of January and July in 2010. 
    This allows us to highlight differences in patterns of power usage between households and business, and winter and summer months.
    A visual inspection of two exemplar time series, shown in Figure~\ref{fig:fig9}, suggests that assuming the error variances at smooth and spike locations to be the same, as we do in Equation~(\ref{model}), may be too restrictive here. 
    Therefore, we allow errors to have 
    inflated variance at spike locations; that is, we use the model described by Equation (\ref{MSIV_model}). 
    This adds a new variance parameter to be estimated in the EM algorithm, which is not covered in our theoretical treatment in Section~\ref{sec:EMtheory}, but does not cause any convergence slow-down in this application. 
    We also adopt the grid $(10^4, \ldots ,10^{-4})$ for the tuning parameter $\lambda$. 
    \begin{figure}[!b]
        \centering
        \includegraphics[width=\textwidth]{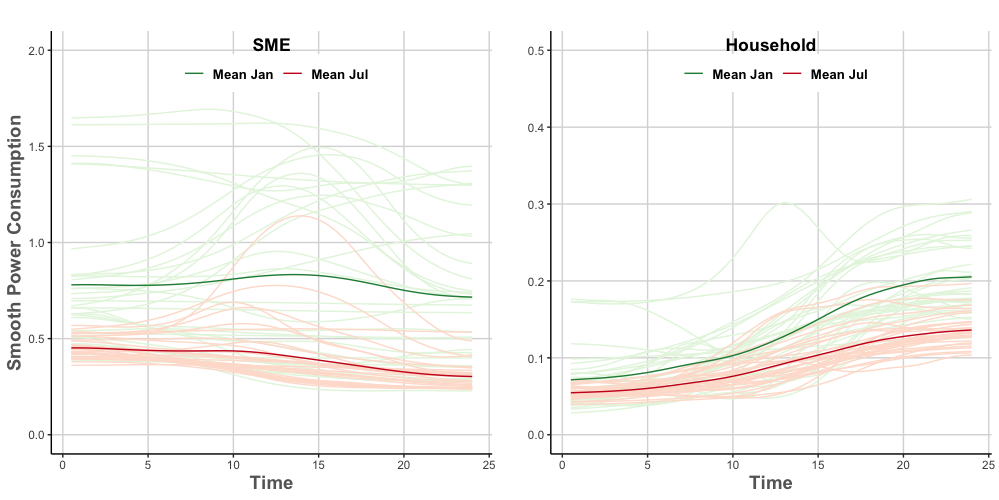}
        \caption{Electricity consumption by an Irish small business (left) and an Irish household (right) for the months of January and July, 2010. Estimated smooth components for each day in January and July are plotted in light green and light red, respectively. The corresponding monthly means are plotted in darker green/red.
        }
        \label{fig:fig10}
    \end{figure}
    \begin{figure}[!b]
        \centering
         \includegraphics[width=\textwidth]{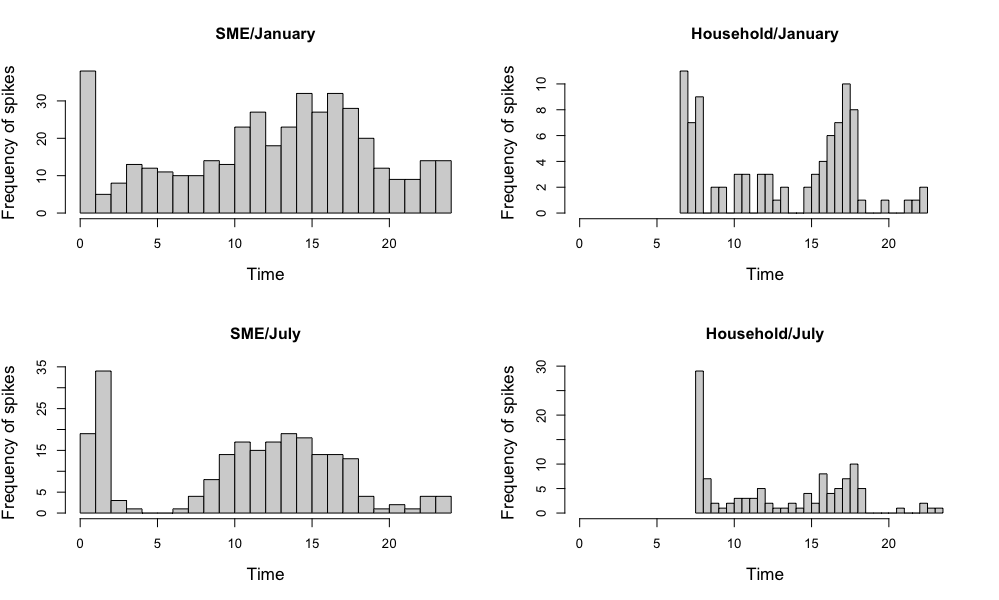}
        \caption{Frequencies of estimated spikes in electricity consumption by an Irish small business (left) and an Irish household (right), for each hour of the day during January, 2010 (top) and July, 2010 (bottom).
        }
        \label{fig:fig12} 
    \end{figure}
    Figure~\ref{fig:fig10} shows estimated smooth curves for the household and small business considered across all January and July days.
    We clearly see that, for both entities, power consumption is higher in January (likely due to heating during cold weather) but follows the same daily pattern in January and July. 
    We also clearly see that such daily pattern is rather different for the small business and the household.
    Power usage tends to peak during the day for the former and at night for the latter. 
    Notably, power usage by the small business is also more variable (from one day to another) than that by the household -- which appears much more consistent.
    Figure~\ref{fig:fig12} shows monthly frequencies of estimated spikes in every hour of the day, plotted again for the household and small business considered and for January and July.
    Notably, the household spikes predominantly occur in the early morning (7:00-8:00am) and, in January, in mid-afternoon (4:00-5:00pm). The small business has numerous spikes in the period of the day when its smooth consumption component is highest (approximately 10:00am to 5:00pm) and, interestingly, right after midnight -- this may correspond to the automatic activation of some appliances.

\subsection{Extreme temperatures in United States}
\begin{figure}[!tbh]
         \centering
         \includegraphics[width=0.9\textwidth]{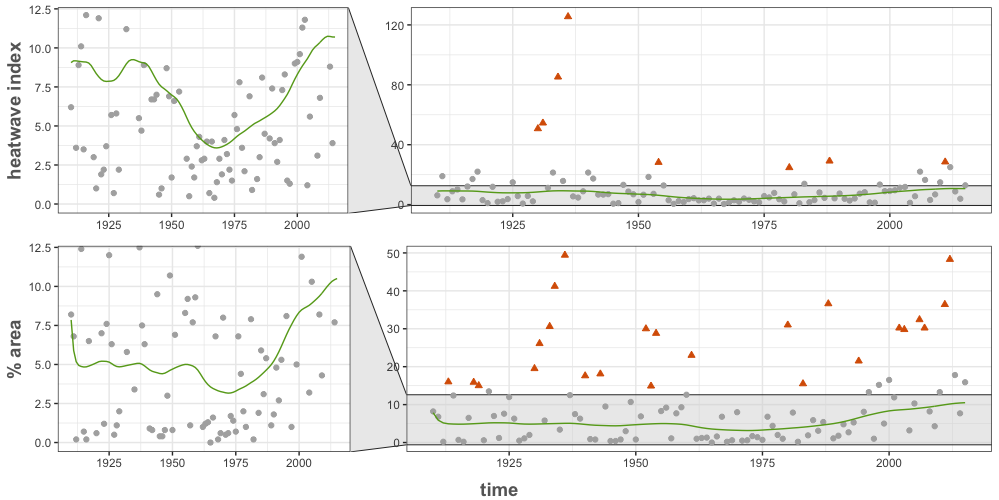}
        \caption{Annual heatwave index in the US (top) and share of its land with unusually high summer temperature (bottom). The estimated smooth component is plotted in green, whereas spikes are identified by the red triangles. 
        For each plot, a vertical zoom (on the left) allows us to visualize the upward trend in the smooth components of both signals in recent decades.
        }
        \label{fig:fig11} 
    \end{figure}

 We consider two temperature-related time series in the US, covering the period from 1910 to 2015. The first is the series of the annual heatwave index. This index treats as a heatwave any period of four or more days with an unusually high average temperature (i.e.~an average temperature that is expected to occur once every 10 years), and takes on values as a function of geographical spread and frequency of heatwaves. 
    The second is the series of the annual percentage of US land area with unusually high summer temperatures. 
    A visual inspection of the two time series again suggests that the error variances at spike locations are inflated. This can be appreciated in Figure~\ref{fig:fig11}. We thus use again the variance inflated model as in the first application.
    Using the grid $(10^4, \ldots ,10^{-4})$ for the tuning parameter $\lambda$, our procedure 
    yields the spike identification and estimated smooth curves shown in red triangles and green lines, respectively, in Figure~\ref{fig:fig11}.
    Probably because geographical spread is part of the definition of the heatwave index, the two time series show rather similar underlying trends.
    {\em smoothEM} detects the 1936 North American heat wave, one of the most intense in modern history, both in terms of heat index and of US area affected by high temperatures -- along with a number of other spikes. 
    Interestingly, only one spike detected in the heatwave index series concerns recent decades -- likely due to the fact that the smooth component estimate exhibits an upward trend; what classified as a spike in the 50's, or even the 80's, is consistent with a standard oscillation around a growing systematic value in more recent times. The picture is different for the US area time series; here recent decades exhibit both an increasing smooth component estimate and an abundance of detected spikes -- suggesting that the geographical dimension of the problem may be yet more concerning. 
    The estimated values for $\alpha^*,\mu^*,\sigma^{*2}_\epsilon, \sigma^{*2}_h$ are respectively $0.08, 30.83, 5.77, 30.79$ for the heatwave index, and $0.28$, $17.88$, $3.96, 10.49$ for the US area.

\section{Discussion}

In this article we propose {\em smoothEM} -- a procedure that, given a signal, simultaneously performs estimation of its smooth component and identification of spikes that may be interspersed within it. {\em smoothEM} uses regularized spline smoothing techniques and the EM algorithm, and is suited for the many applications in which the data comprises discontinuous irregularities superimposed to a noisy curve. We lay out conditions for the procedure to work, and prove asymptotic convergence properties of the EM to a neighborhood of the global optimum under certain restricted conditions. 
We also demonstrate the effectiveness of  {\em smoothEM} under departures from such restricted conditions through simulations and two real data applications, and compare it with a recent algorithm that addresses a similar setting \citep{wei22}, as well as with adaptive smoothing (\emph{mgcv} package in \textsf{R}), which penalizes dynamically spike-affected regions. We demonstrate the superiority of \emph{smoothEM} in a broad range of scenarios. Notably, since it separates spikes and smooth component, 
\emph{smoothEM} could also be used to pre-process functional data prior to the use of other FDA tools. For instance, in a regression context, instead of introducing a functional predictor obtained though traditional spline smoothing of the row data, one could apply our procedure and introduce two distinct predictors; namely, the estimated $\hat{f}$ and, separately, the flagged spike locations. As another example, when performing functional motif discovery or local clustering \citep{cremona22}, \emph{smoothEM} could produce ``de-spiked'' versions of the curves to be searched for recurring smooth patterns, and patterns of detected spikes could be analyzed separately. We shall leave these and other possibilities for future work. An R implementation of \emph{smoothEM} and some examples are provided at \url{https://github.com/hqd1/smoothEM}


\section*{Supplementary Material}
The Supplementary Material includes technical proofs and additional figures.

\bibliographystyle{agsm}
\bibliography{paper-ref}

@book{deboor,
  title = {{A practical guide to splines}},
  author = {de Boor, C.},
publisher = {Springer},
  year = {1978},
}

@book{ramsaysilverman07,
  title = {{Applied functional data analysis: methods and case studies}},
  author = {Ramsay, J. O. and Silverman, B.W.},
publisher = {Springer},
  year = {2007},
}

@book{kozoszka17,
  title = {{Introduction to Functional Data Analysis}},
  author = {Kokoszka, P. and Reimherr, M.},
  year = {2017},
  publisher = {CRC Press}
}

@book{nason08,
  title = {{Wavelet Methods in Statistics with R}},
  author = {Nason, G. P.},
  year = {2008},
  publisher = {Springer}
}

@book{bertsekas,
  title = {{Nonlinear Programming}},
  author = {Bertsekas, D. P.},
  year = {1995},
  publisher = {Athena Scientific}
}

@book{pollard84,
  title = {{Convergence of stochastic processes}},
  author = {Pollard, D.},
  year = {1984},
  publisher = {Springer-Verlag}
}

@book{vershynin18,
  title = {{High-dimensional probability}},
  author = {Vershynin, R.},
  year = {2018},
  publisher = {Cambridge University Press}
}

@article{yaolee06,
  title = {{Penalized spline models for functional
principal component analysis}},
  author = {Yao, F. and Lee, T. C. M.},
  journal = {Journal of the Royal Statistical Society: Series
B (Statistical Methodology)},
  number = {1},
  volume  = 68,
  pages = {3--25},
  year = {2006},
}

@MISC{Wood2011-ax,
title = "Fast stable restricted maximum likelihood and marginal likelihood
estimation of semiparametric generalized linear models",
author = "Wood, S N",
journal = "Journal of the Royal Statistical Society (B)",
volume = 73,
number = 1,
pages = "3--36",
year = 2011
}

@article{goldsmith11,
  title = {{Penalized functional regression}},
  author = {Goldsmith, J. and Bobb, J. and Crainiceanu, C. M. and Caffo, B. and Reich, D.},
  journal = {Journal of Computational and Graphical
Statistics},
  number = {4},
  volume  = 20,
  pages = {830--851},
  year = {2011},
}

@article{descary,
  title = {{Functional data analysis by matrix completion}},
  author = {Descary, M. H. and Panaretos, V. M.},
  journal = {The Annals of Statistics},
  number = {1},
  volume = 47,
  pages = {1--38},
  year = {2019},
}

@article{xiao19,
  title = {{Asymptotic theory of penalized splines}},
  author = {Xiao, L.},
  journal = {Electronic Journal of Statistics},
  volume = 13,
    number = {},
  pages = {747--794},
  year = {2019},
}

@article{zhang10,
  title = {{Baseline correction using adaptive iteratively reweighted penalized least squares}},
  author = {Zhang, Z. M. and Chen, S. and Liang, Y. Z.},
  journal = {Analyst},
  volume = 135,
    number = {5},
  pages = {1138--1146},
  year = {2010},
}

@article{wei22,
  title = {{Two-stage iteratively reweighted smoothing splines for baseline correction}},
  author = {Wei, J. and Zhu, C. and Zhang, Z. M. and He, P.},
  journal = {Chemometrics and Intelligent Laboratory Systems},
  volume = 227,
    number = {},
  pages = {104606},
  year = {2022},
}

@article{pintore06,
  title = {{Spatially adaptive smoothing splines}},
  author = {Pintore, A. and Speckman, P. and Holmes, C. C.},
  journal = {Biometrika},
  volume = 93,
    number = {1},
  pages = {113-125},
  year = {2006},
}

@article{luo97,
  title = {{Hybrid Adaptive Splines}},
  author = {Luo, Z. and Wahba, G.},
  journal = {Journal of the American Statistical Association},
  volume = 92,
    number = {},
  pages = {107-116},
  year = {1997},
}

@unpublished{wu17,
  author       = {Wu, C. and Yang, C. and Zhao, H. and Zhu, J.}, 
  title        = {On the Convergence of the EM Algorithm:
A Data-Adaptive Analysis},
  note         = {arXiv:1611.00519v2},
  year         = 2017
}

@article{balakrishnan,
  title = {{Statistical guarantees for the EM algorithm: from population to sample-based analysis}},
  author = {Balakrishnan, S. and Wainwright, M. J. and Yu, B.},
  journal = {The Annals of Statistics},
  volume = 45,
    number = {1},
  pages = {77--120},
  year = {2017},
}

@article{osullivan,
  title = {{A statistical perspective on ill-posed inverse problems}},
  author = {O'Sullivan, F.},
  journal = {Statistical Science},
  volume = 1,
    number = {4},
  pages = {502--518},
  year = {1986},
}

@article{lewisshedler,
  title = {{Simulation of Nonhomogeneous Poisson Processes with Degree-Two Exponential Polynomial Rate Function}},
  author = {Lewis, P. A. W. and Shedler, G. S.},
  journal = {Operations Research},
  volume = 27,
    number = {5},
  pages = {1026--1040},
  year = {1979},
}

@article{cremona22,
author = {Marzia A. Cremona and Francesca Chiaromonte},
title = {Probabilistic K-means with local alignment for clustering and motif discovery in functional data},
journal = {Journal of Computational and Graphical Statistics},
volume = {0},
number = {ja},
pages = {1-17},
year  = {2022},
publisher = {Taylor & Francis},
}

\end{document}


\def\spacingset#1{\renewcommand{\baselinestretch}%
{#1}\small\normalsize} \spacingset{1}

\if1\blind
{
  \title{\bf \textit{smoothEM}: a new approach for the simultaneous assessment of smooth patterns and spikes - supplementary materials}
  \author{Huy Dang \\
    Dept.~of Statistics, Pennsylvania State University, University Park, USA\\
    and \\
    Marzia A. Cremona \\
    Dept.~of Operations and Decision Systems, Université Laval, Québec, Canada\\
    CHU de Québec – Université Laval Research Center, Québec, Canada\\
    and \\
    Francesca Chiaromonte\\
    Dept.~of Statistics, Pennsylvania State University, University Park, USA\\
    Inst.~of Economics and L'EMbeDS, Sant'Anna School of Advanced Studies, Pisa, Italy}
  \maketitle
} \fi

\if0\blind
{
  \bigskip
  \bigskip
  \bigskip
  \begin{center}
    {\LARGE\bf Title}
\end{center}
  \medskip
} \fi

\bigskip
\spacingset{1.9} 

\section{Proofs and technical details}
\subsection{Proof of Theorem 1}
 We start by stating three conditions that are needed to guarantee good properties for the EM algorithm:
    
    \begin{enumerate}[label = C\arabic*]
        \item
            ($\nu$-strong concavity) There is some $\nu>0$ such that
            $$
            q(\theta_1) - q(\theta_2) - \langle \nabla q(\theta_2),\theta_1-\theta_2\rangle \leq -\frac{\nu}{2} \left\Vert  \theta_1-\theta_2\right\Vert_2^2, \text{ for all } \theta_1, \theta_2 \in   \mathbb{B}_2(r;\theta^*);
            $$
        \item
            (Lipschitz-smoothness) There is some $L > 0$ such that
            $$
            q(\theta_1) - q(\theta_2) - \langle \nabla q(\theta_2),\theta_1-\theta_2\rangle \geq -\frac{L}{2} \left\Vert  \theta_1-\theta_2\right\Vert_2^2,  \text{ for all } \theta_1, \theta_2 \in   \mathbb{B}_2(r;\theta^*);
            $$
        \item
            (Gradient smoothness) For an appropriately small $\gamma >0$
            $$
            \left\Vert  \nabla q(\theta) - \nabla Q(\theta|\theta)\right\Vert_2 \leq \gamma \left\Vert  \theta-\theta^*\right\Vert_2, \text{ for all } \theta \in \mathbb{B}_2(r;\theta^*).
            $$
    \end{enumerate}
    The Theorem below, proved in \cite{balakrishnan}, utilizes these
    conditions to formulate guarantees for the population level EM. 
    
    \begin{theorem}[\textbf{Balakrishnan, Wainwright \& Yu, 2017}]
    \label{Balak1}
        For some radius \mbox{$r>0$} and a triplet $(\gamma,\nu,L)$ such that $0 \leq \gamma < \nu \leq L$, 
        assume that 
        the conditions 
        C1, C2 and C3 hold, and 
        that the stepsize is chosen as $s = 2/(L + \nu)$.  Then given any initialization $\theta_0 \in  \mathbb{B}_2(r;\theta^*)$, with probability $1- \delta$ the population first order EM iterates satisfy the bound
        $$
            \left\Vert  \theta_k - \theta^*\right\Vert_2 \leq \bigg(1-\frac{2\nu-\gamma}{L+\nu}\bigg)^k\left\Vert  \theta_0 - \theta^*\right\Vert_2 \quad \text{ for all } k = 1,2, \ldots
        $$
    \end{theorem}
    For the mixtured Gaussian model with all unknown parameters $\theta^* = (\alpha^*, \mu^*, \sigma^{*2})$, to prove 
    convergence of the EM at the population level, we need to ensure that the 
    conditions C1, C2 and C3 are satisfied. \\
    
    \noindent
    \textbf{C1)}
    The first condition is $\nu$-strong concavity, i.e. 
    existence of some $\nu>0$ such that 
    $$
        q(\theta_1) - q(\theta_2) - \langle \nabla q(\theta_2),\theta_1-\theta_2\rangle \leq -\frac{\nu}{2} \left\Vert  \theta_1-\theta_2\right\Vert_2^2.
    $$
    We have
    $$
        q(\theta) = Q(\theta|\theta^*)
        =  \mathbb{E}_{\theta^*}\left[\frac{1}{A^*+B^*}(A^*\log{A} + B^*\log{B})\right]
    $$
    where $A = \alpha\phi(\xi;0,\sigma^2)$ and $B = (1-\alpha)\phi(\xi; \mu, \sigma^2)$.
    Let $f = -q$;
    strong concavity of $q$ is equivalent to 
    strong convexity of $f$. 
    Since $f$ is twice continuously differentiable, this 
    is equivalent to 
    positive semi-definiteness of $\nabla^2f(\theta) - \nu I$ for every $\theta \in   \mathbb{B}_2(r;\theta^*)$, where $I$ is the identity matrix, see Proposition B.5 in \cite{bertsekas}.
    We have
    \begin{align*}
        \nabla f(\theta) =  \mathbb{E}\bigg\{ & \frac{1}{A^*+B^*}\left(\frac{B^*}{1-\alpha}-\frac{A^*}{\alpha}\right),\\
        & \frac{-B^*(\xi-\mu)}{(A^*+B^*)\sigma^2},\\
        & \frac{A^*}{A^*+B^*}\left(\frac{1}{2\sigma^2} - \frac{\xi^2}{2\sigma^4} \right) + \frac{B^*}{A^*+B^*}\left(\frac{1}{2\sigma^2} - \frac{(\xi-\mu)^2}{2\sigma^4} \right)
        \bigg\} .
    \end{align*}
    Differentiating one more time, we get
    \begin{align*}
        \nabla^2 f(\theta) =  &\mathbb{E}
        \begin{bmatrix}
            \frac{1}{A^*+B^*}\left(\frac{B^*}{(1-\alpha)^2}+\frac{A^*}{\alpha^2}\right) & 0 & 0 \\
            0 & \frac{B^*}{(A^*+B^*)\sigma^2} & \frac{B^*}{A^*+B^*}\frac{(\xi-\mu)}{\sigma^4}\\
            0 & \frac{B^*}{A^*+B^*}\frac{(\xi-\mu)}{\sigma^4} & \frac{A^*}{A^*+B^*}\left(-\frac{1}{2\sigma^4} + \frac{\xi^2}{\sigma^6}\right) + \frac{B^*}{A^*+B^*}\left(-\frac{1}{2\sigma^4} + \frac{(\xi-\mu)^2}{\sigma^6}\right)
        \end{bmatrix}\\
        =& 
        \begin{bmatrix}
            \frac{\alpha^*}{\alpha^2} + \frac{1-\alpha^*}{(1-\alpha)^2} & 0 & 0 \\
            0 & \frac{1-\alpha^*}{\sigma^2} & \frac{(1-\alpha^*)(\mu^*-\mu)}{\sigma^4}\\
            0 & \frac{(1-\alpha^*)(\mu^*-\mu)}{\sigma^4} & \alpha^*\left(-\frac{1}{2\sigma^4} + \frac{\sigma^{*2}}{\sigma^6}\right) + (1-\alpha^*)\left(-\frac{1}{2\sigma^4} + \frac{\sigma^{*2}}{\sigma^6}\right)
        \end{bmatrix}\\
        =& 
        \begin{bmatrix}
            \frac{\alpha^*}{\alpha^2} + \frac{1-\alpha^*}{(1-\alpha)^2} & 0 & 0 \\
            0 & \frac{1-\alpha^*}{\sigma^2} & \frac{(1-\alpha^*)(\mu^*-\mu)}{\sigma^4}\\
            0 & \frac{(1-\alpha^*)(\mu^*-\mu)}{\sigma^4} & \left(-\frac{1}{2\sigma^4} + \frac{\sigma^{*2}}{\sigma^6}\right)
        \end{bmatrix}.
    \end{align*}
    Thus,
    \begin{align*}
        \nabla^2 f(\theta)- \nu I &=  
        \begin{bmatrix}
            \frac{\alpha^*}{\alpha^2} + \frac{1-\alpha^*}{(1-\alpha)^2} -\nu & 0 & 0 \\
            0 & \frac{1-\alpha^*}{\sigma^2} -\nu & \frac{(1-\alpha^*)(\mu^*-\mu)}{\sigma^4}\\
            0 & \frac{(1-\alpha^*)(\mu^*-\mu)}{\sigma^4} & \left(-\frac{1}{2\sigma^4} + \frac{\sigma^{*2}}{\sigma^6}\right) - \nu
        \end{bmatrix}
        = 
        \begin{bmatrix}
            a & 0 & 0 \\
            0 & b & c\\
            0 & c & d
        \end{bmatrix}.
    \end{align*}
    The task becomes finding a $\nu >0$ such that the matrix above is positive semi-definite, or equivalently, such that all principal minors are non-negative. This means we need to ensure:
    \begin{enumerate}
        \item
            $a \geq 0$.
            We note that for $\theta \in   \mathbb{B}_2(r;\theta^*)$ and $\alpha > 0.5$, $\frac{\alpha^*}{\alpha^2} + \frac{1-\alpha^*}{(1-\alpha)^2} \geq \frac{\alpha^*}{\alpha^2} + \frac{1-\alpha^*}{\alpha^2} = \frac{1}{\alpha^2} > \frac{1}{(\alpha^*+r\ \wedge \ 1)^2}$. Thus this condition is satisfied for $\nu <\frac{1}{(\alpha^*+r\ \wedge \ 1)^2} $;
        \item
            $b \geq 0$.
            This condition is satisfied
            for $\nu < \frac{1-\alpha^*}{\sigma^{*2}+r}$;
        \item 
            $d \geq 0$.
            We note that 
            $
                \inf_{\sigma^2 \in   \mathbb{B}_2(r;\sigma^{*2})} -\frac{1}{2\sigma^4} + \frac{\sigma^{*2}}{\sigma^6} = \frac{\sigma^{*2} - r}{2(\sigma^{*2}+r)^3}, \text{ if $r < \sigma^{*2}$}.
             $   
            Thus
            this condition is satisfied
            for $\nu < \frac{\sigma^{*2} - r}{2(\sigma^{*2}+r)^3}$ and $r < \sigma^{*2}$;
        \item
            $bd - c^2 \geq 0$.
            If $\frac{1-\alpha^*}{\sigma^{*2} +r} \leq \frac{\sigma^{*2} - r}{2(\sigma^{*2}+r)^3}$, we combine with $ \frac{(1-\alpha^*)(\mu^*-\mu)}{\sigma^4} \leq \frac{(1-\alpha^*)r}{(\sigma^{*2} -r)^2} $ to reduce the last condition to a stricter one of $\frac{\sigma^{*2} - r}{2(\sigma^{*2}+r)^3}-\nu \geq \frac{(1-\alpha^*)r}{(\sigma^{*2} -r)^2}$, or equivalently, $\nu \leq \frac{\sigma^{*2} - r}{2(\sigma^{*2}+r)^3}-\frac{(1-\alpha^*)r}{(\sigma^{*2} -r)^2}$.
            If for some values of $\sigma^{*2}$, $\frac{1-\alpha^*}{\sigma^{*2} +r} \geq \frac{\sigma^{*2} - r}{2(\sigma^{*2}+r)^3}$, then by similar reasoning, we require $\nu \leq \frac{1-\alpha^*}{\sigma^{*2}+r}-\frac{(1-\alpha^*)r}{(\sigma^{*2} -r)^2}$. The right hand side is positive for $r < \frac{\sigma^{*2}}{3}$.
    \end{enumerate}
    Putting the pieces together, $\nu = \min \left\{\left(\frac{1}{(\alpha^*+r)^2} \vee 1 \right),  \frac{\sigma^{*2}-r}{2(\sigma^{*2}+r)^3}-\frac{(1-\alpha^*)r}{(\sigma^{*2} -r)^2}, \frac{1-\alpha^*}{\sigma^{*2}+r}-\frac{(1-\alpha^*)r}{(\sigma^{*2} -r)^2}\right\}$.\\

    \noindent
    \textbf{C2)}
    The second condition is Lipschitz smoothness, which is stated in terms of $f = -q$ as follows:
    $$
        f(\theta_1) - f(\theta_2) - \langle \nabla f(\theta_2),\theta_1-\theta_2\rangle \leq \frac{L}{2} \left\Vert  \theta_1-\theta_2\right\Vert_2^2.
    $$
    In order to prove this condition, we start by introducing and demonstrating the following Lemma. 
    \begin{lemma}
        \label{lemma1}
        If f is twice continuously differentiable, and $- \nabla^2 f + LI$ is positive semi-definite, 
        where $I$ is the identity matrix, then $f$ 
        satisfies the Lipschitz smoothness condition.
    \end{lemma}
    \begin{proof}[Proof of Lemma \ref{lemma1}]
    Since the Lipschitz smoothness condition is equivalent to:
    $$
        (\nabla f(\theta_1) - \nabla f(\theta_2))^\top(\theta_1 - \theta_2) \leq L  \left\Vert  \theta_1-\theta_2\right\Vert_2^2
    $$
    the proof of the lemma follows as an adaptation of 
    Proposition B.5 in \cite{bertsekas}. 
    Assume that $- \nabla^2 f(\theta)+ LI$ is positive sem-idefinite, then for all $a \in   \mathbb{R}^d$, $a^\top(\nabla^2 f(\theta)- LI)a \leq 0$. Let $g:   \mathbb{R} \rightarrow \mathbb{R}$ be given as
    $$
        g(t) = \nabla f(t\theta_1+ (1-t)\theta_2)^\top(\theta_1-\theta_2).
    $$
    Using the mean value theorem, we have
    $$
        (\nabla f(\theta_1) - \nabla f(\theta_2))^\top(\theta_1 - \theta_2) = g(1) - g(0) = \frac{\partial g}{\partial t} (t^*) 
    $$
    for some $t^* \in [0,1]$. Finally, we have
    $$
        \frac{\partial g}{\partial t} (t^*) = (\theta_1-\theta_2)^\top\nabla^2 f(t\theta_1+ (1-t)\theta_2)^\top(\theta_1-\theta_2) \leq L  \left\Vert  \theta_1-\theta_2\right\Vert_2^2
    $$
    where the last inequality follows from the positive semi-definiteness of  $- \nabla^2 f(\theta)+ LI$.
    \end{proof}

    \noindent
    By Lemma \ref{lemma1}, if $- \nabla^2 f + LI$ is positive semi-definite, 
    the Lipschitz smoothness condition is met. Now, 
    \begin{align*}
        -\nabla^2 f(\theta)+ L I &=  
        \begin{bmatrix}
            -(\frac{\alpha^*}{\alpha^2} + \frac{1-\alpha^*}{(1-\alpha)^2}) +L & 0 & 0 \\
            0 & -\frac{1-\alpha^*}{\sigma^2} +L & -\frac{(1-\alpha^*)(\mu^*-\mu)}{\sigma^4}\\
            0 & -\frac{(1-\alpha^*)(\mu^*-\mu)}{\sigma^4} & -\left(-\frac{1}{2\sigma^4} + \frac{\sigma^{*2}}{\sigma^6}\right) +L
        \end{bmatrix}\\
        &= 
        \begin{bmatrix}
            m & 0 & 0 \\
            0 & n & p\\
            0 & p & q
        \end{bmatrix}.
    \end{align*}
    As before, we need to seek an $L > 0$ such that all of the principal minors are no-nnegative, i.e. we need to ensure:
    \begin{enumerate}
        \item
            $m \geq 0$. 
            This condition is satisfied
            for $L \geq \frac{\alpha^*}{(\alpha^*-r \ \vee \ 0.5)^2} + \frac{1-\alpha^*}{(1-\alpha^*-r \ \vee \ \omega)^2}$, where $\omega$ is a small positive constant;
        \item
            $n \geq 0$. 
            This condition is satisfied
            for $L \geq \frac{1-\alpha^*}{\sigma^{*2} -r}$;
        \item
            $q \geq 0$.
            We note that 
            $    
            \sup_{\sigma^2 \in \mathbb{B}_2(r;\sigma^{*2})} \left[-\frac{1}{2\sigma^4} + \frac{\sigma^{*2}}{\sigma^6} \right]= \frac{\sigma^{*2} + r}{2(\sigma^{*2}-r)^3}, \text{ if $r < \sigma^{*2}$}.
            $
            Thus, the condition is satisfied 
            for $L \geq \frac{\sigma^{*2}+r}{2(\sigma^{*2}-r)^3}$.
        \item
            $nq \geq p^2$.
            If $\frac{\sigma^{*2}+r}{2(\sigma^{*2}-r)^3} <  \frac{1-\alpha^*}{\sigma^{*2} -r}$, the left hand side 
            is bounded from below by $\left(L - \frac{1-\alpha^*}{\sigma^{*2}-r}\right)^2$,
            and the right hand side is bounded from above by $\frac{(1-\alpha^*)^2r^2}{(\sigma^{*2}-r)^4}$. Thus, the condition is satisfied for $L \geq \frac{(1-\alpha^*)\sigma^{*2}}{(\sigma^{*2} -r)^2}$. 
            On the other hand, if $\frac{\sigma^{*2}+r}{2(\sigma^{*2}-r)^3} >  \frac{1-\alpha^*}{\sigma^{*2} -r}$, by a similar reasoning, we can take $L \geq \frac{\sigma^{*2}+r}{2(\sigma^{*2}-r)^3} +  \frac{1-\alpha^*}{\sigma^{*2} -r}$.
    \end{enumerate}
     Putting 
    the pieces together, $L = \max\left\{\frac{\alpha^*}{(\alpha^*-r \ \vee \ 0.7)^2} + \frac{1-\alpha^*}{(1-\alpha^*-r \ \vee \ \omega)^2},\frac{(1-\alpha^*)\sigma^{*2}}{(\sigma^{*2} -r)^2},\frac{\sigma^{*2}+r}{2(\sigma^{*2}-r)^3} +  \frac{1-\alpha^*}{\sigma^{*2} -r}\right\}$.\\

    \noindent
    \textbf{C3)}
    The third condition is gradient smoothness: there exists an appropriately small 
    $\gamma \geq 0$, such that:
    $$
        \left\Vert  \nabla q(\theta) - \nabla Q(\theta|\theta)\right\Vert_2 \leq \gamma \left\Vert  \theta-\theta^*\right\Vert_2 \ .
    $$
    Recall that the data is generated according to 
    $
    g_\theta(\xi) = \alpha\phi(\xi; 0, \sigma^2) + (1-\alpha)\phi(\xi; \mu, \sigma^2)
    $
    with $\theta = (\alpha, \mu, \sigma^2)^\top$.
    Denote as before $A = A_\theta := \alpha\phi(\xi;0,\sigma^2)$ and $B = B_\theta := (1-\alpha)\phi(\xi; \mu, \sigma^2)$. 
    We have 
    \begin{align*}
        \nabla q(\theta) = \frac{\partial q(\theta)}{\partial \theta} = \frac{\partial Q(\theta|\theta^*)}{\partial \theta} = 
        \mathbb{E} \bigg\{&\frac{1}{A^* + B^*}\left(A^* \frac{1}{\alpha} - B^* \frac{1}{1-\alpha}\right),\\
        & \frac{B^*}{A^* + B^*} \frac{\xi - \mu}{\sigma^2},\\
        & \frac{A^*}{A^* + B^*}\left(-\frac{1}{2\sigma^2} + \frac{\xi^2}{2\sigma^4}\right) + \frac{B^*}{A^* + B^*}\left(-\frac{1}{2\sigma^2} + \frac{(\xi-\mu)^2}{2\sigma^4}\right)\bigg\} \ .
    \end{align*}
    Similarly, for $Q(\theta|\theta)$ we have
    \begin{align*}
        \nabla q(\theta) - \nabla Q(\theta|\theta) =  \mathbb{E}\bigg\{&\left(\frac{A^*}{A^* + B^*} - \frac{A}{A+B}\right)\frac{1}{\alpha} - \left(\frac{B^*}{A^* + B^*} - \frac{B}{A+B}\right)\frac{1}{1-\alpha},\\
         &\left(\frac{B^*}{A^* + B^*} - \frac{B}{A+B}\right)\frac{\xi - \mu}{\sigma^2},\\
         & \left(\frac{A^*}{A^* + B^*} - \frac{A}{A+B}\right)\left(-\frac{1}{2\sigma^2} + \frac{\xi^2}{2\sigma^4}\right) \\
         &+ \left(\frac{B^*}{A^* + B^*} - \frac{B}{A+B}\right) \left(-\frac{1}{2\sigma^2} + \frac{(\xi-\mu)^2}{2\sigma^4}\right)\bigg\} \ .
    \end{align*}
    Let $w = \frac{A^*}{A^* + B^*} - \frac{A}{A+B}$ and note that $-w = \frac{B^*}{A^* + B^*} - \frac{B}{A+B}$. Then we obtain
    \begin{align*}
        \nabla q(\theta) - \nabla Q(\theta|\theta) &= \mathbb{E}\left\{w \frac{1}{\alpha(1-\alpha)}, w\frac{\mu-\xi}{\sigma^2}, w\frac{2\xi\mu - \mu^2}{2\sigma^4}\right\}^\top\\
        &=  \mathbb{E} \left\{-w \cdot 
        \begin{bmatrix}
            0 & -\frac{1}{\alpha(1-\alpha)}\\
            \frac{1}{\sigma^2} & -\frac{\mu}{\sigma^2}\\
            -\frac{\mu}{\sigma^4} & \frac{\mu^2}{2\sigma^4}
        \end{bmatrix} \cdot 
        \begin{bmatrix}
            \xi\\
            1
        \end{bmatrix}
        \right\} \ .
    \end{align*}
    Writing $-w = \frac{A}{A+B} - \frac{A^*}{A^* + B^*} = w(\theta;\xi) - w(\theta^*;\xi)$,
    by Taylor's Theorem for multivariate function we have
    \begin{align}
        \label{w1}
        -w = w(\theta;\xi) - w(\theta^*;\xi) = \sum_{i = 1}^{3} \left(\int_0^1\nabla_i w(\theta_u; \xi)du\right) (\theta - \theta^*)_i 
    \end{align}
    where $\nabla_i w(\theta_u; \xi) = \frac{\partial w(\theta;\xi)}{\partial \theta_i}|_{ \theta = \theta_u}$ and $\theta_u = \theta^* + u(\theta - \theta^*)$ for $u \in [0,1]$.
    Now, 
    \begin{align}
        \label{w2}
        \nabla w(\theta_u;\xi) &= \left\{\frac{A_uB_u}{(A_u+B_u)^2} \frac{1}{\alpha_u(1-\alpha_u)}, \frac{A_uB_u}{(A_u+B_u)^2} \frac{\xi-\mu_u}{\sigma_u^2}, \frac{A_uB_u}{(A_u+B_u)^2} \frac{2\xi\mu_u -\mu_u^2}{2\sigma_u^4}\right\}^\top \nonumber \\
        &= -\frac{A_uB_u}{(A_u+B_u)^2} 
        \begin{bmatrix}
            0 & -\frac{1}{\alpha_u(1-\alpha_u)}\\
            \frac{1}{\sigma_u^2} & -\frac{\mu_u}{\sigma_u^2}\\
            -\frac{\mu_u}{\sigma_u^4} & \frac{\mu_u^2}{2\sigma_u^4}
        \end{bmatrix} 
        \begin{bmatrix}
            \xi\\
            1
        \end{bmatrix} \ .
    \end{align}
    Thus
    \begin{align*}
        &\nabla q(\theta) - \nabla Q(\theta|\theta) \\
        &= \mathbb{E} \left\{
        \begin{bmatrix}
            0 & -\frac{1}{\alpha(1-\alpha)}\\
            \frac{1}{\sigma^2} & -\frac{\mu}{\sigma^2}\\
            -\frac{\mu}{\sigma^4} & \frac{\mu^2}{2\sigma^4}
        \end{bmatrix} 
        \begin{bmatrix}
            \xi\\
            1
        \end{bmatrix} 
        \begin{bmatrix}
            \xi & 1
        \end{bmatrix} 
        \int_0^1 \frac{-A_uB_u}{(A_u+B_u)^2} 
        \begin{bmatrix}
            0 & -\frac{1}{\alpha_u(1-\alpha_u)}\\
            \frac{1}{\sigma_u^2} & -\frac{\mu_u}{\sigma_u^2}\\
            -\frac{\mu_u}{\sigma_u^4} & \frac{\mu_u^2}{2\sigma_u^4}
        \end{bmatrix}^\top du \ (\theta-\theta^*)\right\}\\
        &= \mathbb{E} \left\{M 
        \begin{bmatrix}
            \xi\\
            1
        \end{bmatrix} 
        \begin{bmatrix}
            \xi & 1
        \end{bmatrix} \int_0^1 \frac{-A_uB_u}{(A_u+B_u)^2} M_u^\top du \ (\theta-\theta^*)\right\}
    \end{align*}
    and
    \begin{align}
        \label{huy8}
        \left\Vert  \nabla q(\theta) - \nabla Q(\theta|\theta)\right\Vert_2 &= \left\Vert   \mathbb{E} \bigg\{M \begin{bmatrix}
        \xi\\
        1
        \end{bmatrix} \begin{bmatrix}
        \xi & 1
        \end{bmatrix} \int_0^1 \frac{-A_uB_u}{(A_u+B_u)^2} M_u^\top du \ (\theta-\theta^*)\bigg\}
        \right\Vert_2\nonumber\\
        &= \left\Vert   \int_0^1 \mathbb{E} \bigg\{M \begin{bmatrix}
        \xi\\
        1
        \end{bmatrix} \begin{bmatrix}
        \xi & 1
        \end{bmatrix} \frac{-A_uB_u}{(A_u+B_u)^2} M_u^\top du \bigg\} (\theta-\theta^*)
        \right\Vert_2\nonumber\\
        &\leq \sup_{u \in [0,1]} \left\Vert  \mathbb{E} \bigg\{M\frac{-A_uB_u}{(A_u+B_u)^2}  \begin{bmatrix}
        \xi\\
        1
        \end{bmatrix} \begin{bmatrix}
        \xi & 1
        \end{bmatrix}  M_u^\top \bigg\}\right\Vert_{op} \left\Vert  \theta-\theta^*\right\Vert_2 \nonumber\\
        &\leq \sup_{u \in [0,1]} \left\Vert M \ \mathbb{E} \bigg\{ \frac{-A_uB_u}{(A_u+B_u)^2} \begin{bmatrix}
        \xi\\
        1
        \end{bmatrix} \begin{bmatrix}
        \xi & 1
        \end{bmatrix}   \bigg\}M_u^\top\right\Vert_{F} \left\Vert  \theta-\theta^*\right\Vert_2\nonumber\\
        &\leq \left\Vert  M\right\Vert_F \sup_{u \in [0,1]} \left\Vert  \mathbb{E}  \bigg\{\frac{-A_uB_u}{(A_u+B_u)^2}\begin{bmatrix}
        \xi^2 & \xi\\
        \xi &1
        \end{bmatrix} \bigg\}\right\Vert_{F} \left\Vert  M_u\right\Vert_F \left\Vert  \theta-\theta^*\right\Vert_2 \ .
    \end{align}
    The 
    idea is to bound $ \mathbb{E}\bigg\{\frac{A_uB_u \xi^2}{(A_u+B_u)^2}\bigg\}$, $\mathbb{E}\bigg\{\frac{A_uB_u \xi}{(A_u+B_u)^2}\bigg\}$
    and $\mathbb{E}\bigg\{\frac{A_uB_u}{(A_u+B_u)^2}\bigg\}$ appropriately, so that the terms in front of $\left\Vert  \theta-\theta^*\right\Vert_2$ 
    are appropriately small, or 
    go to $0$ quickly as the signal (in a sense to be clarified later) becomes large.
    First, we bound $\mathbb{E}\bigg\{\frac{A_uB_u}{(A_u+B_u)^2}\bigg\}$. It 
    is easy to verify that
    \begin{align*}
        &\sup_{\xi \in   \mathbb{R}} \frac{A_uB_u}{(A_u+B_u)^2} = \frac{A_uB_u}{(A_u+B_u)^2}\bigg|_{\xi = \frac{\mu_u}{2} + \frac{\sigma_u^2}{\mu_u}\log(\frac{\alpha_u}{1-\alpha_u})} = \frac{1}{4}\\
        &\sup_{\xi \leq t_1} \frac{A_uB_u}{(A_u+B_u)^2} = \frac{A_uB_u}{(A_u+B_u)^2}\bigg|_{\xi = t_1}, \text{ for all } t_1 < \frac{\mu_u}{2} \\
        &\sup_{\xi \geq t_2} \frac{A_uB_u}{(A_u+B_u)^2} = \frac{A_uB_u}{(A_u+B_u)^2}\bigg|_{\xi = t_2}, \text{ for all } t_2 > \frac{\mu_u}{2} + \frac{\sigma_u^2}{\mu_u}\log\left(\frac{\alpha_u}{1-\alpha_u}\right).
    \end{align*}
    Now, 
    \begin{align*}
        \mathbb{E}\left\{\frac{A_uB_u}{(A_u+B_u)^2}\bigg| \xi \geq t_2 \right\} &\leq  \mathbb{E}\left\{\frac{1}{\frac{A_u}{B_u} + \frac{B_u}{A_u} + 2} \bigg|_{\xi = t_2}\right\}\\
        &= \frac{1}{\frac{A_u}{B_u} + \frac{B_u}{A_u} + 2} \bigg|_{\xi = t_2}\\
        &= \frac{1}{\frac{\alpha_u}{1-\alpha_u}\exp(-\frac{t_2^2}{2\sigma_u^2} + \frac{(t_2-\mu_u)^2}{2\sigma_u^2}) + \frac{1-\alpha_u}{\alpha_u}\exp(\frac{t_2^2}{2\sigma_u^2} - \frac{(t_2-\mu_u)^2}{2\sigma_u^2}) + 2}\\
        & \leq \frac{1}{\frac{1-\alpha_u}{\alpha_u}\exp(\frac{t_2^2}{2\sigma_u^2} - \frac{(t_2-\mu_u)^2}{2\sigma_u^2})} = \frac{\alpha_u}{1-\alpha_u} \exp\left[\frac{\mu_u(\frac{\mu_u}{2}-t_2)}{\sigma_u^2}\right].
    \end{align*}
    Similarly,
    \begin{align*}
        \mathbb{E}\left\{\frac{A_uB_u}{(A_u+B_u)^2}\bigg| \xi \leq t_1 \right\} &\leq \mathbb{E}\left\{\frac{1}{\frac{A_u}{B_u} + \frac{B_u}{A_u} + 2} \bigg|_{\xi = t_1}\right\}\\
        &= \frac{1}{\frac{A_u}{B_u} + \frac{B_u}{A_u} + 2} \bigg|_{\xi = t_1}\\
        &= \frac{1}{\frac{\alpha_u}{1-\alpha_u}\exp\left(-\frac{t_1^2}{2\sigma_u^2} + \frac{(t_1-\mu_u)^2}{2\sigma_u^2}\right) + \frac{1-\alpha_u}{\alpha_u}\exp\left(\frac{t_1^2}{2\sigma_u^2} - \frac{(t_1-\mu_u)^2}{2\sigma_u^2}\right) + 2}\\
        & \leq \frac{1}{\frac{\alpha_u}{1-\alpha_u}\exp\left(-\frac{t_1^2}{2\sigma_u^2} + \frac{(t_1-\mu_u)^2}{2\sigma_u^2}\right)} = \frac{1-\alpha_u}{\alpha_u} \exp\left[\frac{\mu_u(t_1-\frac{\mu_u}{2})}{\sigma_u^2}\right].
    \end{align*}
    Thus, we can bound $\mathbb{E}\left\{\frac{A_uB_u}{(A_u+B_u)^2}\right\}$ as follows:
    \begin{align}
        \label{huy1}
        &\mathbb{E}\left\{\frac{A_uB_u}{(A_u+B_u)^2}\right\} \nonumber \\
        &\leq \mathbb{E}\left\{\frac{A_uB_u}{(A_u+B_u)^2}\bigg| \xi \leq t_1 \right\}  +\mathbb{E}\left\{\frac{A_uB_u}{(A_u+B_u)^2}\bigg| \xi \geq t_2 \right\} + \mathbb{E}\left\{\frac{A_uB_u}{(A_u+B_u)^2}\bigg| t_1 \leq \xi \leq t_2 \right\}  \mathbb{P}(t_1 \leq \xi \leq t_2) \nonumber\\
        &\leq \frac{1-\alpha_u}{\alpha_u} \exp\left[\frac{\mu_u(t_1-\frac{\mu_u}{2})}{\sigma_u^2}\right] + \frac{\alpha_u}{1-\alpha_u} \exp\left[\frac{\mu_u(\frac{\mu_u}{2}-t_2)}{\sigma_u^2}\right] + \frac{1}{4}  \mathbb{P} (t_1 \leq \xi \leq t_2) \ .
    \end{align}
    Let $t_2 = \frac{\mu_u}{2} + \frac{\sigma_u^2}{\mu_u}log\bigg(\frac{\alpha_u}{1-\alpha_u}\bigg) + \omega_0 = \frac{\mu_u}{2} + \omega$, where $\omega_0$ is a fixed small constant
    and $t_1 = \frac{\mu_u}{2}- \omega$. Then, assuming $\omega < \frac{\mu_u}{2}$, $\frac{\mu_u}{2} < \mu^* -\omega$ and $r < \mu^* - 2\omega$, we have
    \begin{align}
    \label{huy2}
        \mathbb{P}(t_1 \leq \xi \leq t_2) &= 
        \alpha^*\mathbb{P}(t_1 \leq \xi_0 \leq t_2) + (1-\alpha^*)\mathbb{P}(t_1 \leq \xi_1 \leq t_2)\nonumber\\
        & \qquad \qquad \qquad \qquad \text{where $\xi_0 \sim N(0, \sigma^{*2})$ and $\xi_1 \sim N(\mu^*, \sigma^{*2})$}\nonumber\\
        &\leq \alpha^*\mathbb{P}\left(\xi_0 > \frac{\mu_u}{2} - \omega\right) +   (1-\alpha^*)\mathbb{P}\left(\xi_1 < \frac{\mu_u}{2} + \omega \right)\nonumber\\
        &\leq \alpha^*\phi^c\left(\frac{\frac{\mu_u}{2} - \omega}{\sigma^*}\right) + (1-\alpha^*)\phi^c\left(\frac{\mu^*-\frac{\mu_u}{2} - \omega}{\sigma^*}\right)\nonumber\\
        &\leq \alpha^*\frac{1}{\sqrt{2\pi}}\frac{\sigma^*}{\frac{\mu_u}{2}-\omega} \exp\left[-\frac{1}{2\sigma^{*2}}\left(\frac{\mu_u}{2}-\omega\right)^2\right]\nonumber \\
        &  \quad \quad + (1-\alpha^*)\frac{1}{\sqrt{2\pi}}\frac{\sigma^*}{\mu^*-\frac{\mu_u}{2}-\omega} \exp\left[-\frac{1}{2\sigma^{*2}}\left(\mu^*-\frac{\mu_u}{2}-\omega\right)^2\right]\nonumber\\ 
        &\leq \alpha^*\frac{1}{\sqrt{2\pi}}\frac{\sigma^*}{\frac{\mu^*-r}{2}-\omega} \exp\left[-\frac{1}{2\sigma^{*2}}\left(\frac{\mu^*-r}{2}-\omega\right)^2\right]\nonumber\\
        &  \quad \quad + (1-\alpha^*)\frac{1}{\sqrt{2\pi}}\frac{\sigma^*}{\mu^*-\frac{\mu^*+r}{2}-\omega} \exp\left[-\frac{1}{2\sigma^{*2}}\left(\mu^*-\frac{\mu^*+r}{2}-\omega\right)^2\right]\nonumber\\ 
        &= \frac{1}{\sqrt{2\pi}}\frac{\sigma^*}{\frac{\mu^*}{2}-\frac{r}{2}-\omega} \exp\left[-\frac{1}{2\sigma^{*2}}\left(\frac{\mu^*}{2}-\frac{r}{2}-\omega\right)^2\right] \nonumber\\
        &= \frac{1}{\sqrt{2\pi}}\exp\left[-\log\left(\frac{\frac{\mu^*}{2}-\frac{r}{2}-\omega}{\sigma^*}\right)- \frac{1}{2}\left(\frac{\frac{\mu^*}{2}-\frac{r}{2}-\omega}{\sigma^*}\right)^2\right].
    \end{align}
    Similarly,
    \begin{align}
        \label{huy3}
        \frac{1-\alpha_u}{\alpha_u} \exp\left[\frac{\mu_u(t_1-\frac{\mu_u}{2})}{\sigma_u^2}\right] &\leq \frac{1-\alpha_u}{\alpha_u} \exp\left[\frac{\mu_u(-\frac{\sigma^2_u}{\mu_u}\log(\frac{\alpha_u}{1-\alpha_u})-\omega_0)}{\sigma_u^2}\right] \nonumber\\
        &= \frac{1-\alpha_u}{\alpha_u} \exp\left[-\log\bigg(\frac{\alpha_u}{1-\alpha_u}\bigg) -\frac{\mu_u}{\sigma_u^2}\omega_0\right] \nonumber\\
        &=  \left(\frac{1-\alpha_u}{\alpha_u}\right)^2\exp\left(-\frac{\mu_u}{\sigma^2_u} \omega_0\right) \nonumber\\
        &< \exp\left(-\frac{\mu^* -r}{\sigma^{*2}+r}\omega_0\right)
    \end{align}
    and
    \begin{align}
        \label{huy4}
        \frac{\alpha_u}{1-\alpha_u} \exp\left[\frac{\mu_u\left(\frac{\mu_u}{2}-t_2\right)}{\sigma_u^2}\right] < \exp\left(-\frac{\mu^* -r}{\sigma^{*2}+r}\omega_0\right). \qquad \qquad \qquad \qquad
    \end{align}
    Combining equations (\ref{huy1}), (\ref{huy2}), (\ref{huy3}) and (\ref{huy4}), we have that for appropriately large $\mu^*$
    \begin{align}
        \label{huy5}
        \mathbb{E}\left\{\frac{A_uB_u}{(A_u+B_u)^2}\right\} &\leq 2\exp\left(-\frac{\mu^* -r}{\sigma^{*2}+r}\omega_0\right)+ \frac{1}{\sqrt{2\pi}}\exp\left[-\log\left(\frac{\frac{\mu^*}{2}-\frac{r}{2}-\omega}{\sigma^*}\right)- \frac{1}{2}\left(\frac{\frac{\mu^*}{2}-\frac{r}{2}-\omega}{\sigma^*}\right)^2\right] \nonumber\\
        &< C \exp\left(-\frac{\mu^* -r}{\sigma^{*2}+r}\right) \ .
    \end{align}
    Now, we bound $ \mathbb{E}\left\{\frac{A_uB_u}{(A_u+B_u)^2}\xi^2\right\}$ as follows:
    \begin{align}
        \label{huy6}
        \mathbb{E}\left\{\frac{A_uB_u}{(A_u+B_u)^2}\xi^2\right\} &\leq \mathbb{E}\left\{\frac{A_uB_u}{(A_u+B_u)^2}\xi^2\bigg|\xi \leq t_1\right\} + \mathbb{E}\left\{\frac{A_uB_u}{(A_u+B_u)^2}\xi^2\bigg|\xi \geq t_2\right\} \nonumber\\
        & \qquad + \mathbb{E}\left\{\frac{A_uB_u}{(A_u+B_u)^2}\xi^2\bigg|t_1 \leq \xi \leq t_2\right\}  \mathbb{P}(t_1 \leq \xi \leq t_2)\nonumber\\
        &\leq \sup_{\xi \leq t_1} \frac{A_uB_u}{(A_u+B_u)^2}  \mathbb{E}(\xi^2) + \sup_{\xi \geq t_2} \frac{A_uB_u}{(A_u+B_u)^2}  \mathbb{E}(\xi^2) \nonumber\\
        & \qquad + \frac{1}{4} \mathbb{E}(\xi^2) \mathbb{P}(t_1 \leq \xi \leq t_2) \nonumber\\
        &< C \exp\left(-\frac{\mu^* -r}{\sigma^{*2}+r}\right) \left(\sigma^{*2}+ (1-\alpha^*)\mu^*\right) \ .
    \end{align}
    Lastly, from (\ref{huy5}) and (\ref{huy6}) we have
    \begin{align}
        \label{huy7}
        \mathbb{E}\left\{\frac{A_uB_u\xi}{(A_u+B_u)^2}\right\} &= \mathbb{E}\left\{\frac{\sqrt{A_uB_u}}{A_u+B_u} \frac{\sqrt{A_uB_u}\xi}{A_u+B_u} \right\} \leq \sqrt{\mathbb{E}\left\{\frac{A_uB_u}{(A_u+B_u)^2}\right\}} \sqrt{\mathbb{E}\left\{\frac{A_uB_u\xi^2}{(A_u+B_u)^2}\right\}} \nonumber\\
        &< C \exp\left(-\frac{\mu^* -r}{\sigma^{*2}+r}\right) \sqrt{\sigma^{*2}+ (1-\alpha^*)\mu^*}
    \end{align}
    where the first inequality follows from the classic Cauchy-Schwarz inequality.
    Putting everything together, and noting that the terms $\left\Vert  M\right\Vert_2$ and $\left\Vert  M_u\right\Vert_2$ in (\ref{huy8}) are polynomial in $\mu^*$ and $\sigma^{*2}$, we can conclude that
    \begin{align}
        \left\Vert  \nabla q(\theta) - \nabla Q(\theta|\theta)\right\Vert_2 \leq \gamma(\alpha^*,\mu^*, \sigma^{*2}) \left\Vert  \theta-\theta^*\right\Vert_2
    \end{align}
    where $\gamma(\alpha^*,\mu^*, \sigma^{*2}) \sim O\left(\frac{\mu^{*5}}{\sigma^{*8}}\exp(-\frac{\mu^* -r}{\sigma^{*2}+r})\right)$ goes to 0 exponentially fast with large $\frac{\mu^*}{\sigma^{*2}}$.
    %
    %
    
 \subsection{Proof of Theorem 2}
  We start considering a result for the sample EM proved in \cite{balakrishnan}.
    \begin{theorem}[\textbf{Balakrishnan, Wainwright \& Yu, 2017}]
        \label{Balak2}
        For a given size $n$ and tolerance parameter $\delta \in (0,1)$, let $\epsilon_Q^{unif}(n, \delta)$ be the smallest scalar such that with probability at least $1-\delta$
        $$
            \sup_{\theta \in   \mathbb{B}_2(r;\theta^*)} \left\Vert  \nabla Q_n(\theta|\theta) - \nabla Q(\theta|\theta)\right\Vert_2 \leq \epsilon_Q^{unif}(n, \delta) \ .
        $$
        Suppose that, in addition to the conditions of Theorem \ref{Balak1}, the sample size $n$ is large enough to ensure that $\epsilon_Q^{unif}(n, \delta) \leq (\nu-\gamma)r$.
        Then with probability at least $1- \delta$, given any initial vector $\theta_0 \in \mathbb{B}_2(r;\theta^*)$, the finite sample EM iterates $\{\theta_k\}_{k=0}^\infty$ satisfy the bound
        $$
            \left\Vert  \theta_t - \theta^*\right\Vert_2 \leq \left(1- \frac{2\nu-2\gamma}{L + \nu}\right)^t\left\Vert  \theta_0- \theta^*\right\Vert_2 + \frac{\epsilon_Q^{unif}(n, \delta)}{\nu-\gamma} \ .
        $$
    \end{theorem}
    The proof of Theorem 2 follows from Theorem \ref{Balak2} if we can show that $\epsilon_Q^{unif}(n, \delta) \rightarrow 0$ almost surely.
    Let $w(\theta;\xi) = \frac{A_\theta}{A_\theta+B_\theta}$, where $A_\theta := \alpha\phi(\xi;0,\sigma^2)$ and $B_\theta := (1-\alpha)\phi(\xi; \mu, \sigma^2)$. We have 
    \begin{align*}
        \sup_{\theta \in   \mathbb{B}_2(r;\theta^*)} \left\Vert  \nabla Q_n(\theta|\theta) - \nabla Q(\theta|\theta)\right\Vert_2 = \sup_{\theta \in   \mathbb{B}_2(r;\theta^*)} \left\Vert  \boldsymbol{R}(\theta)\boldsymbol{x}\right\Vert_2 \leq \sup_{\theta \in   \mathbb{B}_2(r;\theta^*)} \left\Vert  \boldsymbol{R}(\theta)\right\Vert_{op} \left\Vert  \boldsymbol{x}\right\Vert_2 
    \end{align*}
    where 
    \begin{align*}
        \boldsymbol{R}(\theta) = \begin{bmatrix}
        0&0&\frac{1}{\alpha(1-\alpha)}&0\\
        \frac{1}{\sigma^2}&0&\frac{\mu}{\sigma^2}&\frac{-1}{\sigma^2}\\
        \frac{-\mu}{\sigma^4}&\frac{1}{2\sigma^4}&\frac{\mu^2}{2\sigma^4}&\frac{\mu}{\sigma^4}
        \end{bmatrix} \text{ and } \boldsymbol{x} = \begin{pmatrix*}[l]
        \frac{1}{n}\sum_{i = 1}^{n}\xi_i -  \mathbb{E}(\xi)\\
        \frac{1}{n}\sum_{i = 1}^{n}\xi^2_i -  \mathbb{E}(\xi^2)\\
        \frac{1}{n}\sum_{i = 1}^{n}w(\theta;\xi_i) -  \mathbb{E}(w(\theta;\xi))\\
        \frac{1}{n}\sum_{i = 1}^{n}w(\theta;\xi_i)\xi_i -  \mathbb{E}(w(\theta;\xi)\xi)\\
        \end{pmatrix*} \ .
    \end{align*}
    The elements of the column vector $\boldsymbol{x}$ can be regarded as empirical processes and can be bound separately. The first two
    do not involve $\theta$.
    Thus, by SLLN, 
    $
        P\left(\left|\frac{1}{n}\sum_{i = 1}^{n}\xi_i -  \mathbb{E}(\xi)\right| > \omega\right) \rightarrow 0  \text{ almost surely, as } n \rightarrow \infty,
    $
    and 
    $
        P\left(\left|\frac{1}{n}\sum_{i = 1}^{n}\xi^2_i -  \mathbb{E}(\xi^2)\right| > \omega\right) \rightarrow 0 \text{ almost surely, as } n \rightarrow \infty.
    $
    To bound the remaining 
    two elements, we use covering numbers. The following 
    correspond to Definition 23 and Theorem 24 from \cite{pollard84}.
    \begin{definition}[\citeauthor{pollard84}, \citeyear{pollard84}]
        \label{coveringnumber}
        Let $P$ be a probability measure on $\xi$, and $\mathscr{F}$ be a class of functions in $\mathscr{L}^1(P)$. For each $\omega > 0$, define the covering number $N(\omega, P, \mathscr{F})$ as the smallest value of $m$ for which there
        exist functions $g_1, \ldots, g_m$ such that $\min_j   \mathbb{E}_P|f-g_j| \leq \omega$ for each $f \in \mathscr{F}$. Set $N(\omega, P, \mathscr{F}) = \infty$ if no such $m$ exists.
    \end{definition}
    \begin{theorem} [\citeauthor{pollard84}, \citeyear{pollard84}]
        \label{Pollard}
        Let $\mathscr{F}$ be a class of functions 
        whose envelope $F$ is integrable with respect to $P$. If $P_n$ is obtained by independent sampling from the probability measure $P$ and if $\log N(\omega,P_n, \mathscr{F}) = o_p(n)$ for each fixed $\omega>0$, then $\sup_\mathscr{F} | \mathbb{E}_{P_n} f -  \mathbb{E}_Pf| \rightarrow 0$ almost surely, as $n \rightarrow \infty$. 
    \end{theorem}
    Now, let $\mathscr{F}_1 = \{w(\theta;\xi): \theta \in  \mathbb{B}_2(r;\theta^*)\}$, and $\mathscr{F}_2 = \{w(\theta;\xi)\xi: \theta \in  \mathbb{B}_2(r;\theta^*)\}$. Then $F_1(\xi) = 1$ is an envelope of $\mathscr{F}_1$, $F_2(\xi) = |\xi|$ is an envelope of $\mathscr{F}_2$,
    and both $F_1$
    and $F_2$ are integrable. We set out to prove that $\log N(\omega,P_n, \mathscr{F}_1) = o_p(n)$ and $\log N(\omega,P_n, \mathscr{F}_2) = o_p(n)$.
    Similar to (\ref{w1}) and (\ref{w2}), letting $\theta_u = \theta_2 + u(\theta_1 - \theta_2)$, we have
    \begin{align*}
        P_n |w(\theta_1;\xi) - w(\theta_2;\xi)| &= \frac{1}{n}\sum_{i =1}^{n} \left|w(\theta_1;\xi_i) - w(\theta_2;\xi_i)\right| \\
        &= \frac{1}{n}\sum_{i=1}^{n} \left|\int_0^1 -\frac{A_uB_u}{(A_u+B_u)^2} [\xi_i \ 1]
        \begin{bmatrix}
            0 & -\frac{1}{\alpha_u(1-\alpha_u)}\\
            \frac{1}{\sigma_u^2} & -\frac{\mu_u}{\sigma_u^2}\\
            -\frac{\mu_u}{\sigma_u^4} & \frac{\mu_u^2}{2\sigma_u^4}
        \end{bmatrix}^\top du (\theta_1-\theta_2) \right|\\
        &\leq \frac{1}{n}\sum_{i=1}^{n} \left|\sup_{u \in [0,1]} -\frac{A_uB_u}{(A_u+B_u)^2} [\xi_i \ 1]\ M_u^\top (\theta_1-\theta_2) \right| \text{ where } M_u \text{ is the above matrix}\\
        & \leq \frac{1}{n}\sum_{i=1}^{n} \sup_{u \in [0,1]} \left\Vert  \frac{A_uB_u}{(A_u+B_u)^2} [\xi_i \ 1]\right\Vert_2 \left\Vert  M_u\right\Vert_{op} \left\Vert  \theta_1-\theta_2\right\Vert_2\\
        & \leq \frac{1}{n}\sum_{i=1}^{n} \sup_{u \in [0,1]} \left\Vert  \frac{A_uB_u}{(A_u+B_u)^2} [\xi_i \ 1]\right\Vert_2 \left\Vert  M_u\right\Vert_{F} \left\Vert  \theta_1-\theta_2\right\Vert_2.
    \end{align*}
    We note that $\theta_1, \theta_2$ and $\theta_u$ are in $\mathbb{B}_2(r;\theta^*)$.
    Thus, $\left\Vert  M_u\right\Vert_F$ can be bounded by a constant $M(\theta^*; r) \in   \mathbb{R}$, and this constant is of the order of a polynomial 
    in the elements 
    of $\theta^*$.
    Let $D = \frac{1}{n}\sum_{i=1}^{n} \sup_{u \in [0,1]} \left\Vert  \frac{A_uB_u}{(A_u+B_u)^2} [\xi_i \ 1]\right\Vert_2 M(\theta^*;r)$. Then 
    $\left\Vert  \theta_1-\theta_2\right\Vert_2 < \frac{\omega}{D}$ implies $P_n |w(\theta_1;\xi) - w(\theta_2;\xi)| < \omega$. This means that $N(\omega,P_n, \mathscr{F}_1) \leq N(\frac{\omega}{D}, \left\Vert  \cdot\right\Vert_2,  \mathbb{B}_2(r;\theta^*)) \leq (1+\frac{2rD}{\omega})^3$, where the last inequality 
    follows from Proposition 4.2.12 in \cite{vershynin18}
    (or can be proved directly by comparing volumes of the corresponding balls in $  \mathbb{R}^3$). Thus, our task now becomes proving that $3\log(1+\frac{2rD}{\omega}) = o_p(n)$. Let $\delta$ be any positive constant, then
    \begin{align}
        \label{huy9}
        P\left\{3\log\left(1+\frac{2rD}{\omega}\right) > n\delta \right\} &= P\left\{D> \frac{(e^{n\delta/3}-1)\omega}{2r} \right\}\nonumber\\
        &= P\bigg\{\frac{1}{n}\sum_{i=1}^{n} \sup_{u \in [0,1]} \left\Vert  \frac{A_uB_u}{(A_u+B_u)^2} [\xi_i \ 1]\right\Vert_2 M(\theta^*;r)> \frac{(e^{n\delta/3}-1)\omega}{2r} \bigg\}\nonumber\\
        &\leq P\bigg\{\frac{1}{n}\sum_{i=1}^{n} \sup_{u \in [0,1]} \left\Vert  \frac{A_uB_u}{(A_u+B_u)^2} [\xi_i \ 1]\right\Vert_2 > \frac{(e^{n\delta/3}-1)\omega}{2rM(\theta^*;r)} \bigg\}\nonumber\\
        &\leq P\bigg\{\max_{\xi_i}\sup_{u \in [0,1]} \left\Vert  \frac{A_uB_u}{(A_u+B_u)^2} [\xi_i \ 1]\right\Vert_2 > \frac{(e^{n\delta/3}-1)\omega}{2rM(\theta^*;r)} \bigg\}\nonumber\\
        &\leq nP\bigg\{\sup_{u \in [0,1]} \left\Vert  \frac{A_uB_u}{(A_u+B_u)^2} [\xi\ 1]\right\Vert_2 > \frac{(e^{n\delta/3}-1)\omega}{2rM(\theta^*;r)} \bigg\}\nonumber\\
        &\leq nP\bigg\{\sup_{u \in [0,1]} \frac{A_uB_u|\xi|}{(A_u+B_u)^2} + \frac{A_uB_u}{(A_u+B_u)^2} > \frac{(e^{n\delta/3}-1)\omega}{2rM(\theta^*;r)} \bigg\}\nonumber\\
        &\leq nP\bigg\{\sup_{u \in [0,1]} \frac{A_uB_u|\xi|}{(A_u+B_u)^2} > \frac{(e^{n\delta/3}-1)\omega}{4rM(\theta^*;r)} \bigg\}\nonumber\\
        &\quad + nP\bigg\{\sup_{u \in [0,1]} \frac{A_uB_u}{(A_u+B_u)^2} > \frac{(e^{n\delta/3}-1)\omega}{4rM(\theta^*;r)} \bigg\}\nonumber\\
        &\leq n  \mathbb{E}\bigg\{\sup_{u \in [0,1]} \frac{A_uB_u|\xi|}{(A_u+B_u)^2}\bigg\} \frac{4rM(\theta^*;r)}{(e^{n\delta/3}-1)\omega}\\
        &\quad + n \mathbb{E}\bigg\{\sup_{u \in [0,1]} \frac{A_uB_u}{(A_u+B_u)^2}\bigg\} \frac{4rM(\theta^*;r)}{(e^{n\delta/3}-1)\omega}
    \end{align}
    where the last inequality 
    follows from a Markov inequality. Note that in (\ref{huy5}) and (\ref{huy7}), the bounds for $\mathbb{E}\left\{ \frac{A_uB_u}{(A_u+B_u)^2}\right\}$ and $\mathbb{E}\left\{\frac{A_uB_u\xi}{(A_u+B_u)^2}\right\}$ work the same as $\mathbb{E}\left\{\sup_{u \in [0,1]}  \frac{A_uB_u}{(A_u+B_u)^2}\right\}$ and $\mathbb{E}\left\{\sup_{u \in [0,1]} \frac{A_uB_u|\xi|}{(A_u+B_u)^2}\right\}$.
    Thus, (\ref{huy9}) can be bounded as follows:
    \begin{align}
        \label{huy10}
        P\left\{3\log\left(1+\frac{2rD}{\omega}\right) > n\delta \right\} &\leq C \exp\left(-\frac{\mu^* -r}{\sigma^{*2}+r}\omega_0\right) \sqrt{\sigma^{*2}+ (1-\alpha^*)\mu^*} \frac{4nrM(\theta^*;r)}{(e^{n\delta/3}-1)\omega}\nonumber\\
        & \quad + C \exp\left(-\frac{\mu^* -r}{\sigma^{*2}+r}\omega_0\right)\frac{4nrM(\theta^*;r)}{(e^{n\delta/3}-1)\omega}\quad  \rightarrow 0 \text{ as } n \rightarrow \infty 
    \end{align}
    proving the case for $N(\omega,P_n, \mathscr{F}_1)$. 
    For $N(\omega,P_n, \mathscr{F}_2)$, we proceed similarly:
    \begin{align*}
        P_n |w(\theta_1;\xi)\xi - w(\theta_2;\xi)\xi| &= \frac{1}{n}\sum_{i =1}^{n} \left|w(\theta_1;\xi_i)\xi_i - w(\theta_2;\xi_i)\xi_i\right| \\
        &= \frac{1}{n}\sum_{i=1}^{n} \left|\int_0^1 -\frac{A_uB_u}{(A_u+B_u)^2} [\xi_i^2 \ \xi_i]
        \begin{bmatrix}
        0 & -\frac{1}{\alpha_u(1-\alpha_u)}\\
            \frac{1}{\sigma_u^2} & -\frac{\mu_u}{\sigma_u^2}\\
            -\frac{\mu_u}{\sigma_u^4} & \frac{\mu_u^2}{2\sigma_u^4}
        \end{bmatrix}^\top du (\theta_1-\theta_2) \right|\\
        & \leq \frac{1}{n}\sum_{i=1}^{n} \sup_{u \in [0,1]} \left\Vert  \frac{A_uB_u}{(A_u+B_u)^2} [\xi_i^2 \ \xi_i]\right\Vert_2 \left\Vert  M_u\right\Vert_{F} \left\Vert  \theta_1-\theta_2\right\Vert_2 \ .
    \end{align*}
    Replacing $D$ with $ \frac{1}{n}\sum_{i=1}^{n} \sup_{u \in [0,1]} \left\Vert  \frac{A_uB_u}{(A_u+B_u)^2} [\xi_i^2 \ \xi_i]\right\Vert_2 M(\theta^*;r)$, 
    we have $N(\omega,P_n, \mathscr{F}_2) \leq N(\frac{\omega}{D}, \left\Vert  \cdot\right\Vert_2,  \mathbb{B}_2(r;\theta^*)) \leq (1+\frac{2rD}{\omega})^3$, and as before
    \begin{align}
        \label{huy11}
        P\left\{3\log\left(1+\frac{2rD}{\omega}\right) > n\delta \right\} &= P\left\{D> \frac{(e^{n\delta/3}-1)\omega}{2r} \right\}\nonumber\\
        &\leq n  \mathbb{E}\bigg\{\sup_{u \in [0,1]} \frac{A_uB_u \xi^2}{(A_u+B_u)^2}\bigg\} \frac{4rM(\theta^*;r)}{(e^{n\delta/3}-1)\omega} \\
        &\quad + n \mathbb{E}\bigg\{\sup_{u \in [0,1]} \frac{A_uB_u |\xi|}{(A_u+B_u)^2}\bigg\} \frac{4rM(\theta^*;r)}{(e^{n\delta/3}-1)\omega} \nonumber \\
        &\leq C \exp\bigg(-\frac{\mu^* -r}{\sigma^{*2}+r}\omega_0\bigg) \bigg(\sigma^{*2}+ (1-\alpha^*)\mu^*\bigg)\frac{4nrM(\theta^*;r)}{(e^{n\delta/3}-1)\omega} \nonumber\\
        &\quad + C \exp\bigg(-\frac{\mu^* -r}{\sigma^{*2}+r}\omega_0\bigg) \sqrt{\sigma^{*2}+ (1-\alpha^*)\mu^*} \frac{4nrM(\theta^*;r)}{(e^{n\delta/3}-1)\omega}
        \rightarrow 0 \text{ as } n \rightarrow \infty 
    \end{align}
    
\clearpage
\section{Additional Figures}
\begin{figure}[!hbt]
        \centering
\begin{tikzpicture}[node distance=3cm]
		\node (data) [box, align = center] {Data \\ $(x_i, y_i)_{i=1}^n$}; 
		\node (resid) [box, align = center, right of = data, xshift = 1.7cm] {Residuals \\ $\xi_i(\lambda)$};
		\node (spikes) [box, align = center, above right of = resid, xshift = 1.3cm, yshift = -0.5cm] {``spikes'' \\ $M_i(\lambda)= 1$};
		\node (nonspikes) [box, align = center, below right of = resid, xshift = 1.3cm, yshift = 0.5cm] {``smooth'' \\ $M_i(\lambda) = 0$};
		\node (newresid) [box, align = center, right of = nonspikes, xshift = 1.9cm] {Updated \\ residuals \\ $\xi'_i(\lambda)$};
		\node (EM) [box, align = center, below left of = newresid, xshift = 0.5cm, yshift = -1.4cm] {$\hat\theta_\lambda = (\hat{\alpha}, \hat{\mu}, \hat{\sigma}^2)_\lambda$ \\ 
			$\ell(\lambda; \hat{\alpha}, \hat{\mu}, \hat{\sigma}^2)$ \\ 
			$\boldsymbol{M'}(\lambda)$};   
		\node (optim) [box, align = center, left of = EM, xshift = -6.7cm] {$\hat\theta_{\lambda^*} = (\hat{\alpha}, \hat{\mu}, \hat{\sigma}^2)_{\lambda^*}$ \\
			$\boldsymbol{M'}(\lambda^*)$};  
		\node (fest) [box, align = center, below of = optim, yshift = -0.5cm] {$\hat f$}; 
		\draw [arrow] (data) -- node[text width=1.5cm,align=center,above] {smoothing splines} node[below]{$\lambda$} (resid);
		\draw [arrow] (resid) -- node[sloped, above] {large} (spikes);
		\draw [arrow] (resid) -- node[sloped, below] {small} (nonspikes);
		\draw [arrow] (nonspikes) -- node[text width=1.5cm,align=center,above] {smoothing splines} node[below]{$\lambda$} (newresid);
		\draw [arrow] (newresid) -- node[sloped, above] {EM} (EM);
		\draw [arrow] (EM) -- node[sloped, above] {$\lambda^* = \arg \max_\lambda \allowbreak \left[\ell(\lambda;\hat{\alpha}, \hat{\mu}, \hat{\sigma}^2) + F(\lambda) \right]$} (optim);
		\draw [arrow] (optim) -- node[text width=2cm,align=right,left] {smoothing splines} node[text width=2cm,align=left,right] {$\lambda^*$ \mbox{$M'_i(\lambda^*)=0$}} (fest);
	\end{tikzpicture}
	\caption{Flowchart of the \emph{smoothEM} algorithm}
\end{figure}
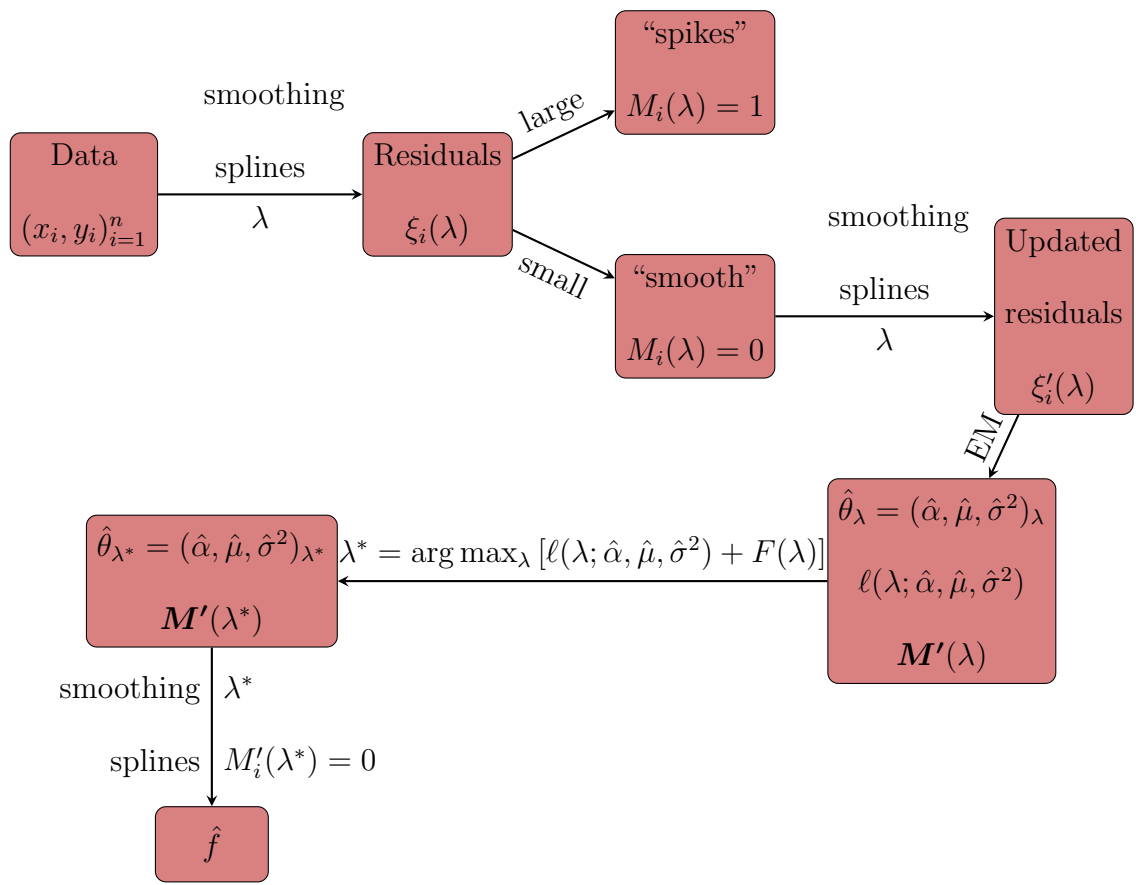 

\begin{figure}[!ht]
        \begin{tabular}{cc}
        \includegraphics[width=.45\textwidth]{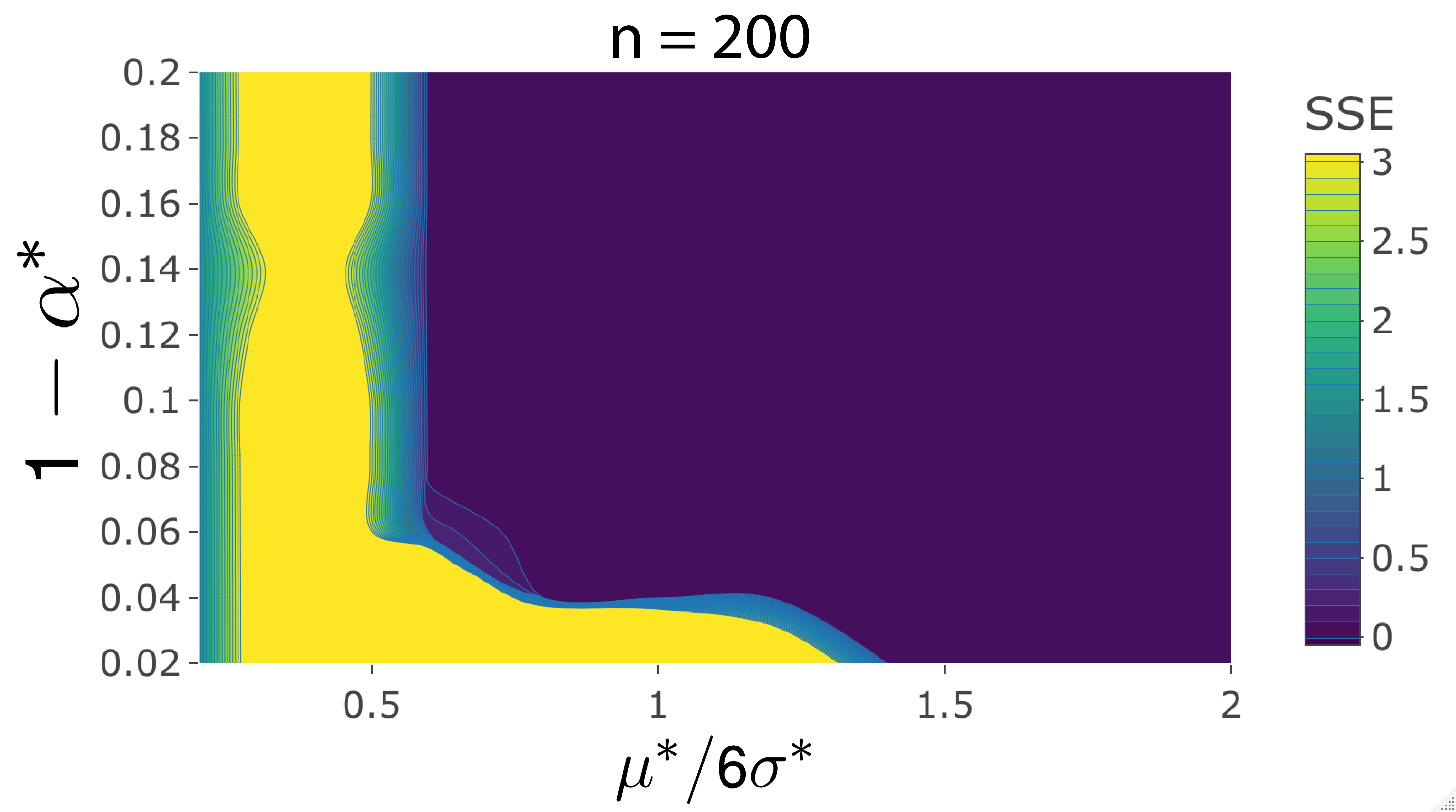}&
        \includegraphics[width=.45\textwidth]{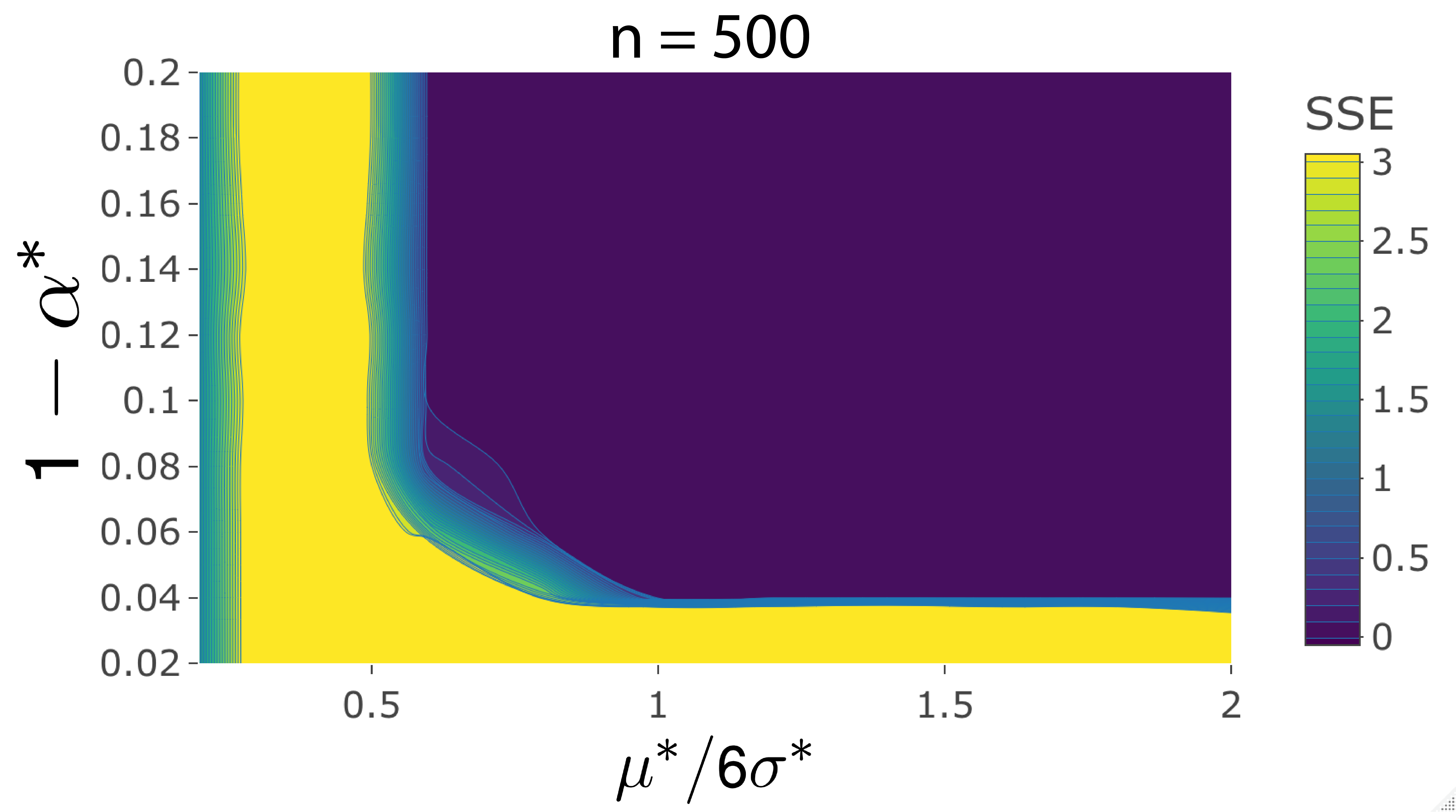}\\
        \includegraphics[width=.45\textwidth]{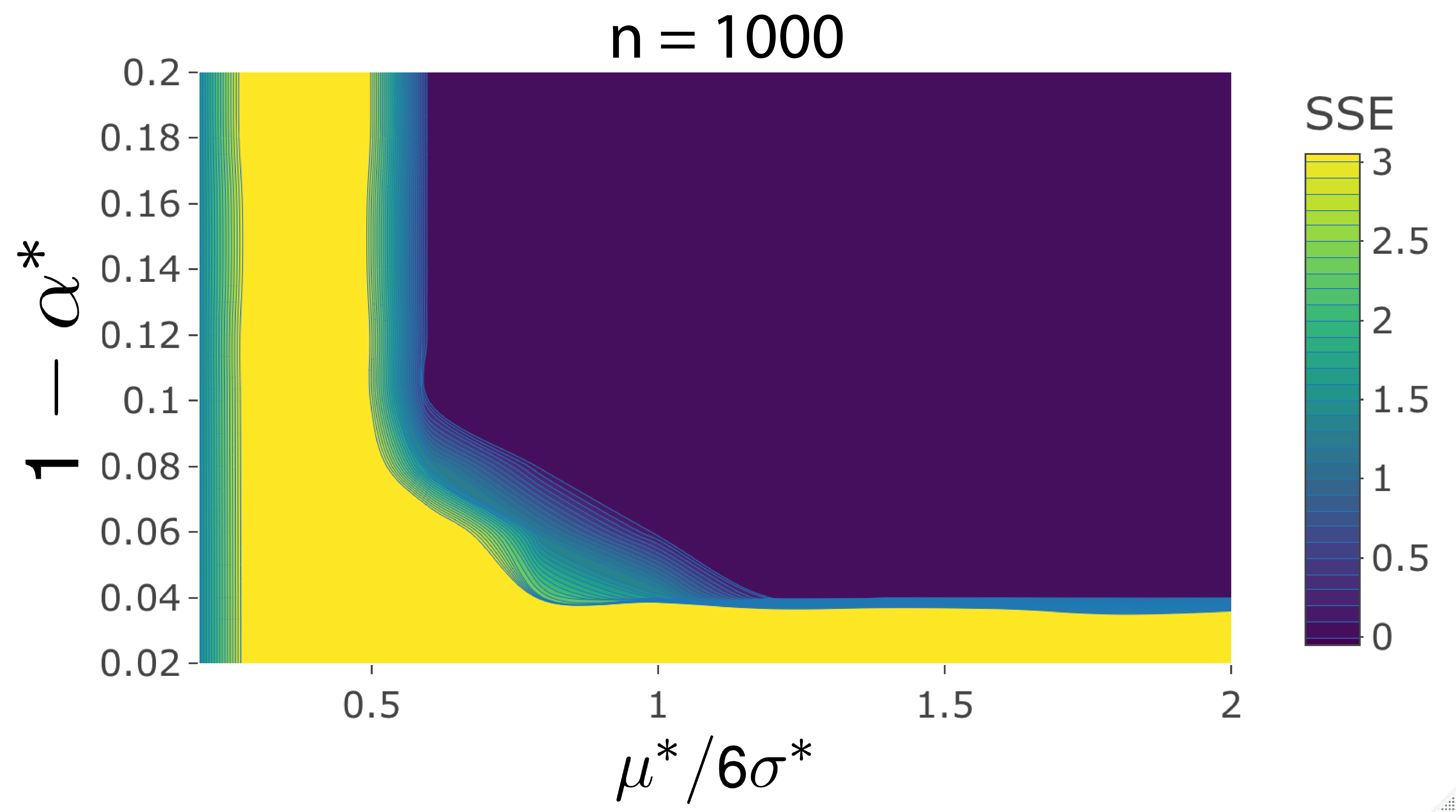}&
        \includegraphics[width=.45\textwidth]{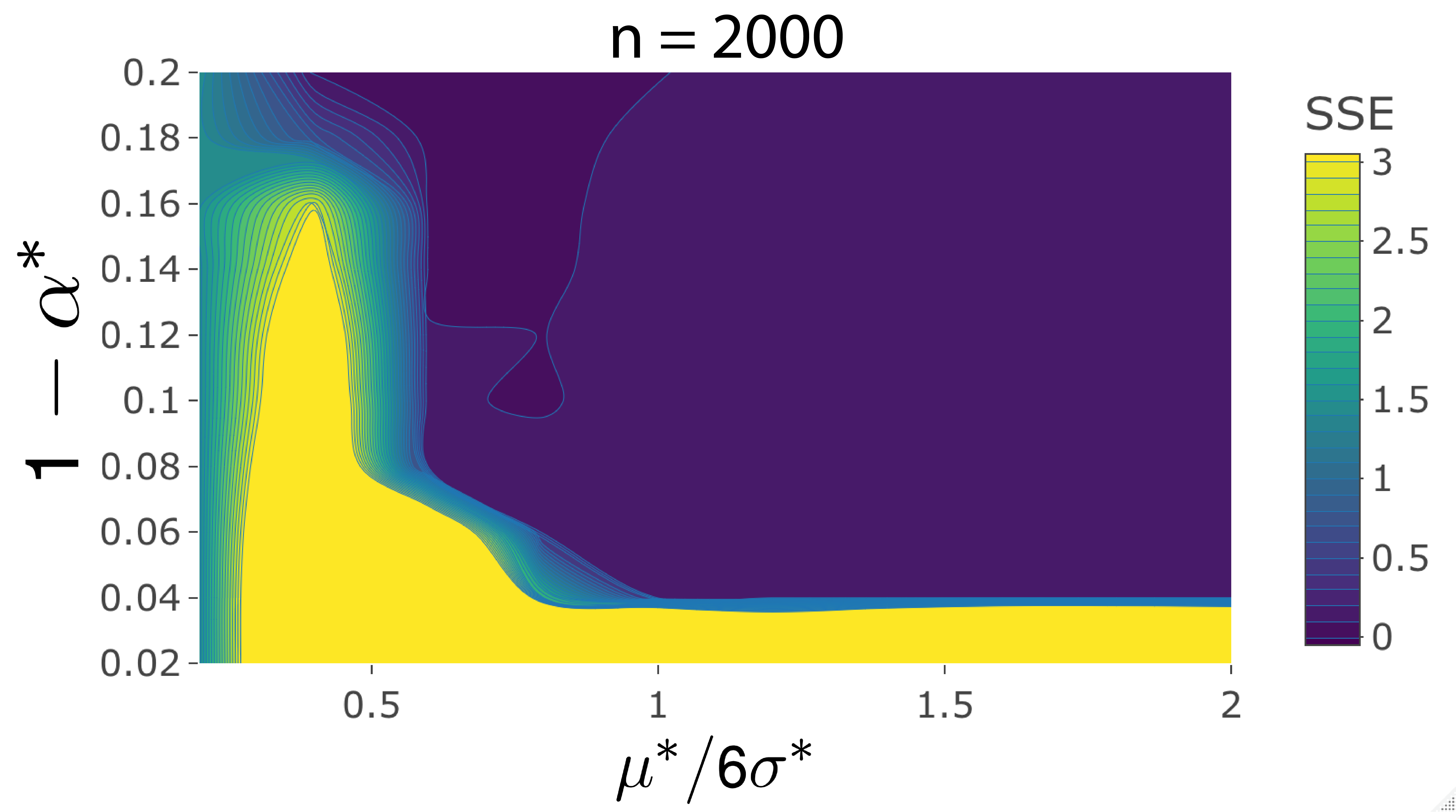}
        \end{tabular}
        \caption{ SSE of the {\em smoothEM} parameter estimates on simulated data with uniformly distributed spikes. The contour plots show the SSE (averaged over $20$ simulation replicates) as a function of the spike percentage ($\alpha^*$) and the STN ($\frac{\mu^*}{6\sigma^*}$).
        From left to right, top to bottom, $n = 200, 500, 1000, 2000$.
        }
        \label{fig:fig14} 
    \end{figure}
    
    \begin{figure}[hbt!]
        \centering
        \begin{tabular}{cc}
        \includegraphics[width=.45\textwidth]{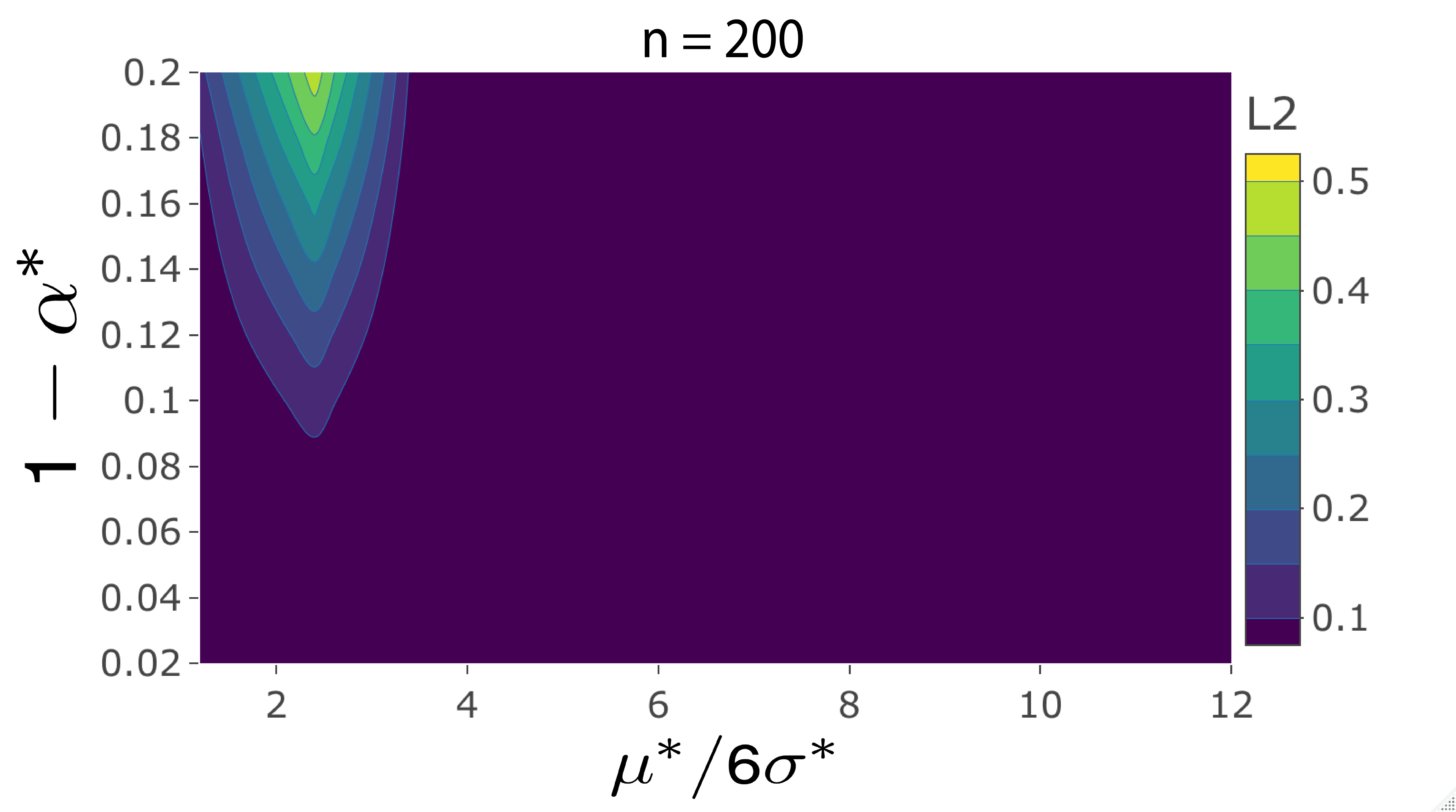}&
        \includegraphics[width=.45\textwidth]{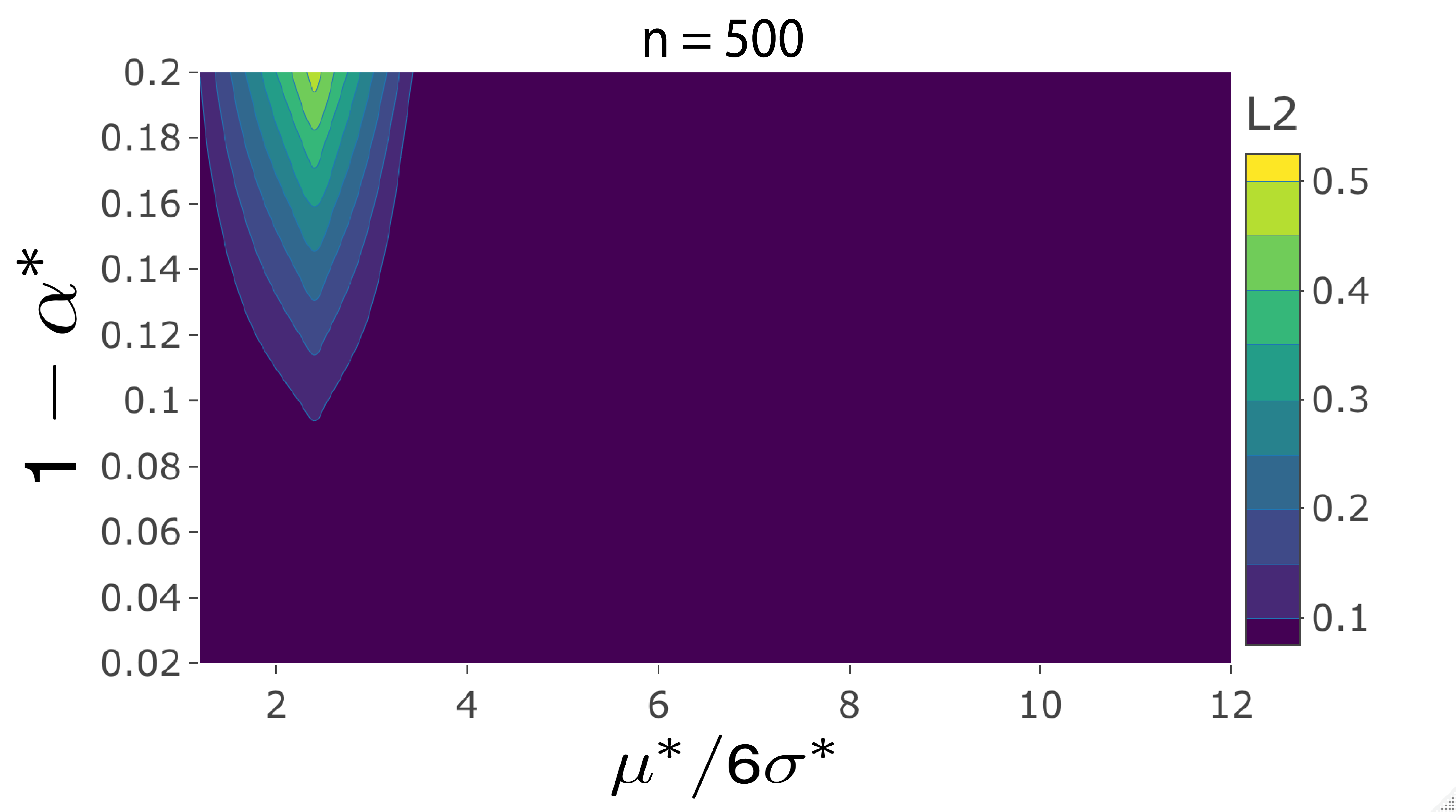}\\
        \includegraphics[width=.45\textwidth]{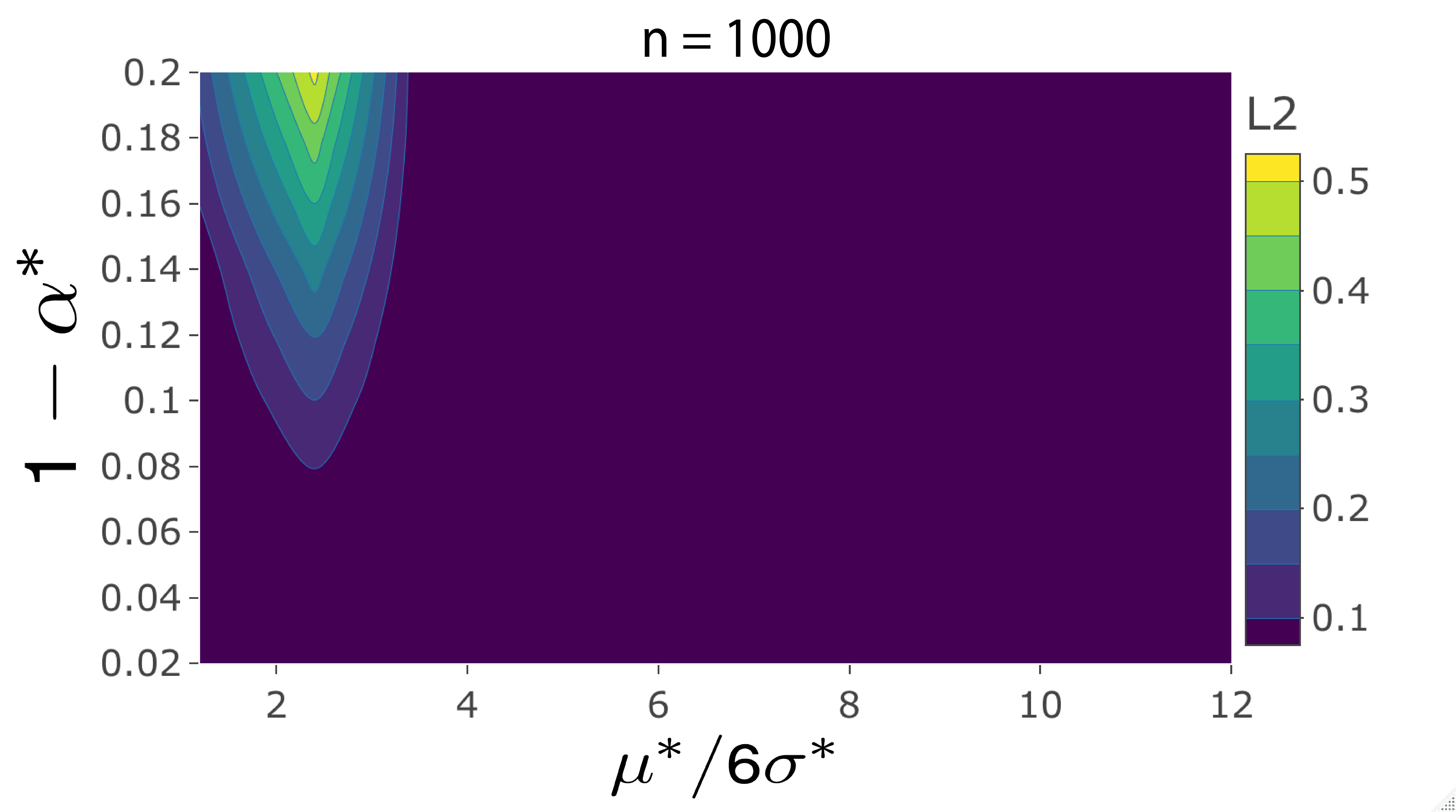}&
        \includegraphics[width=.45\textwidth]{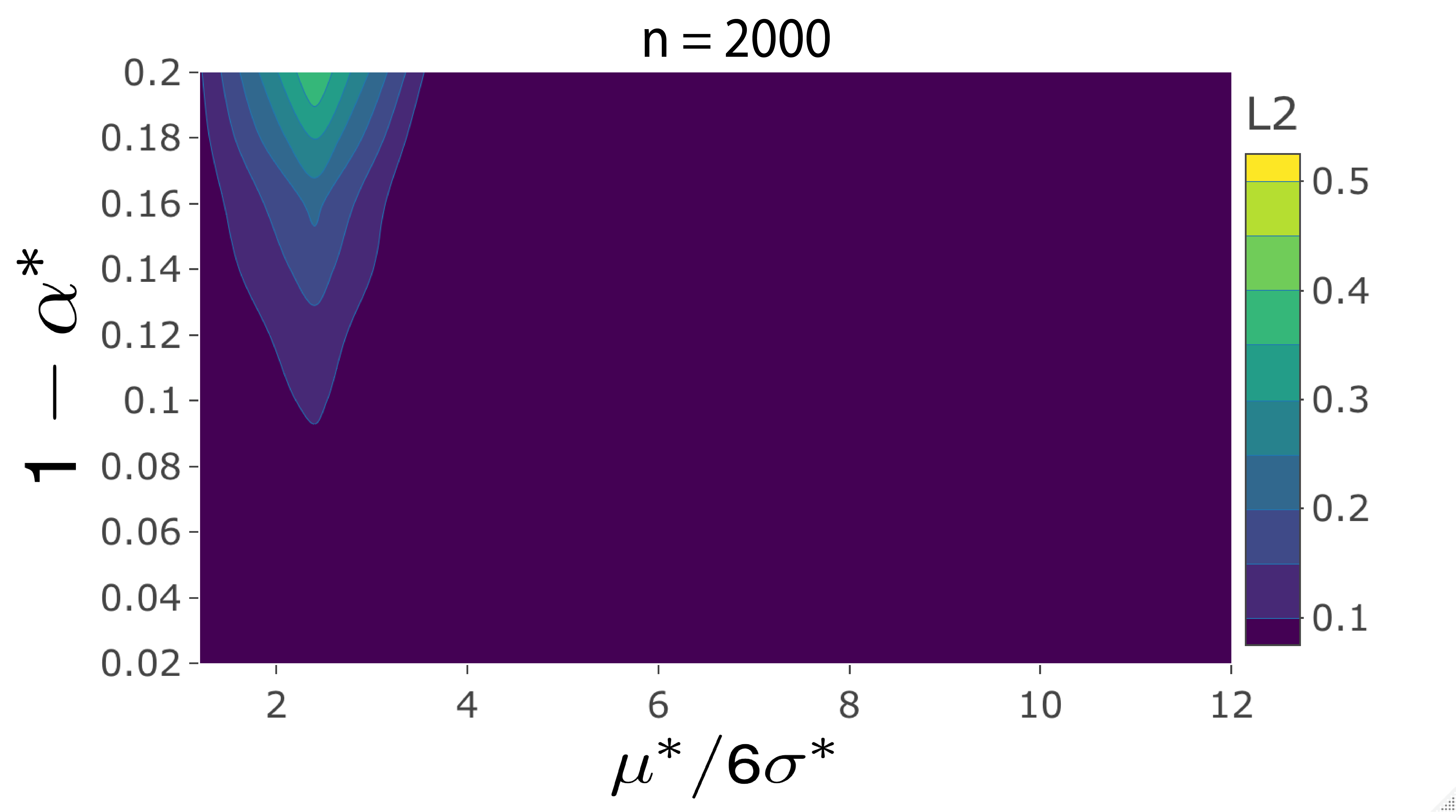}
        \end{tabular}
       \caption{$L_2$ error of the {\em smoothEM} smooth component estimate on simulated data with 'clumped' spikes. The contour plots show the error (averaged over $20$ simulation replicates) as a function of the spike percentage ($1-\alpha^*$) and the STN ($\frac{\mu^*}{6\sigma^*}$).
        From left to right, top to bottom, $n = 200, 500, 1000, 2000$.}
       \label{fig:fig21}
    \end{figure}
    \begin{figure}[hbt!]
        \centering
        \begin{tabular}{cc}
        \includegraphics[width=.45\textwidth]{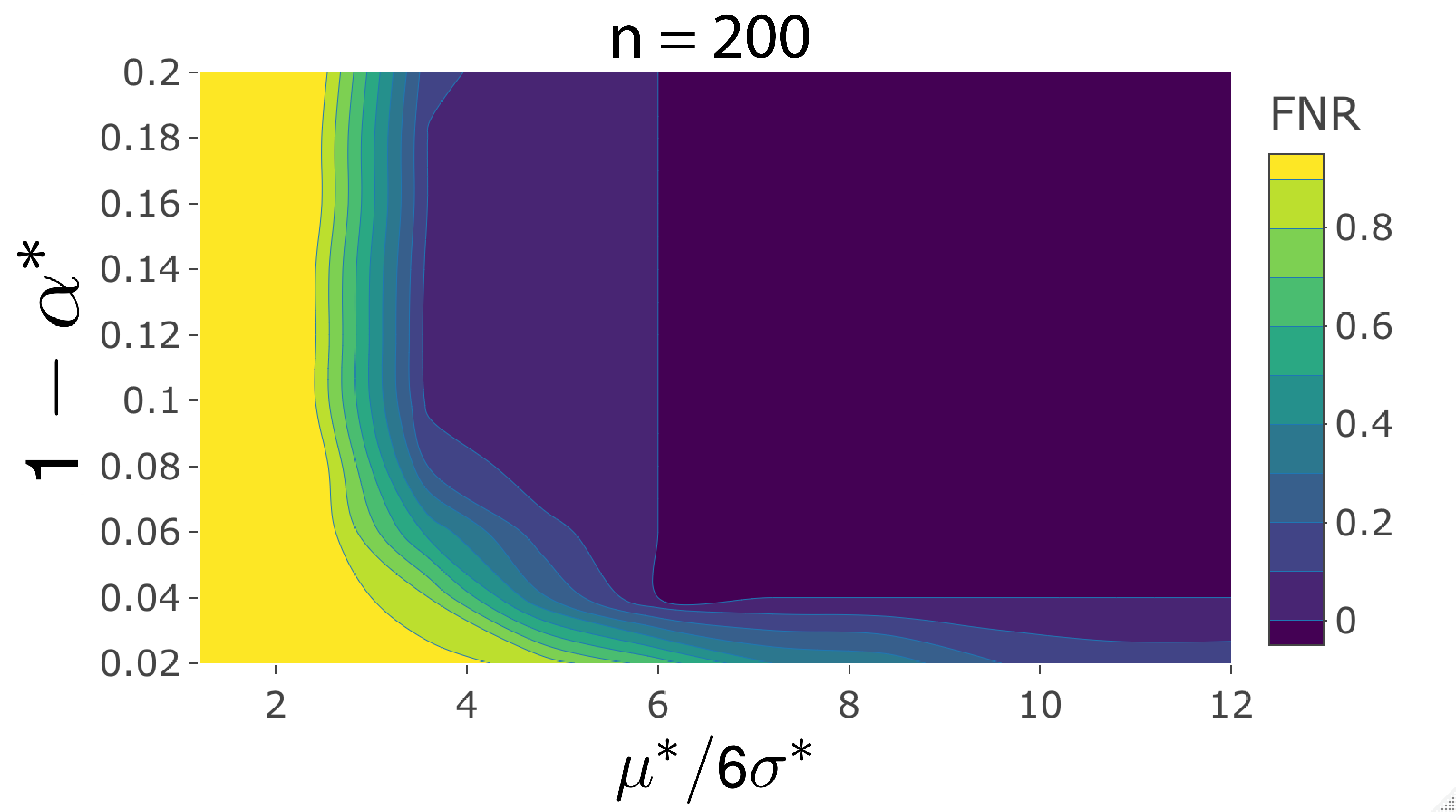}&
        \includegraphics[width=.45\textwidth]{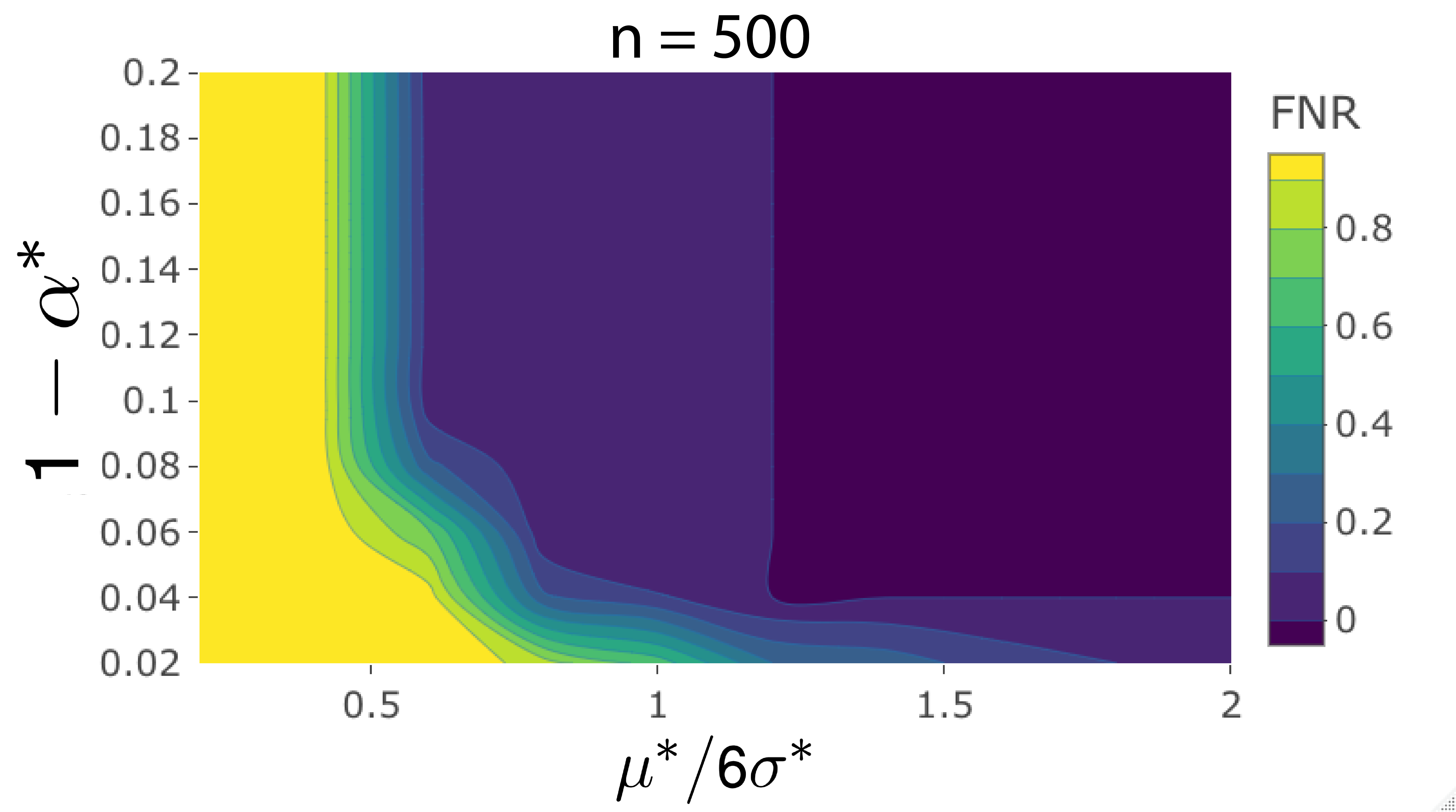}\\
        \includegraphics[width=.45\textwidth]{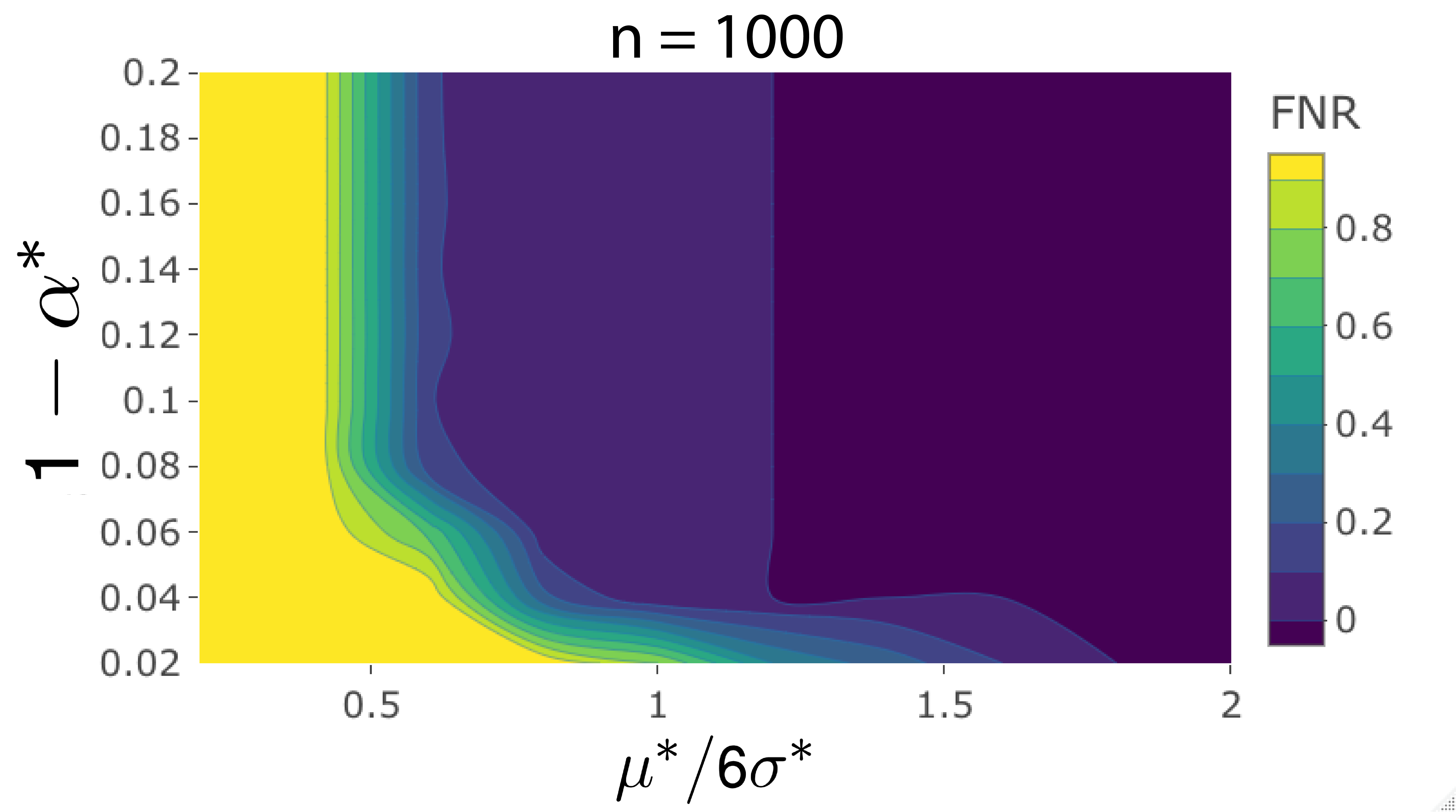}&
        \includegraphics[width=.45\textwidth]{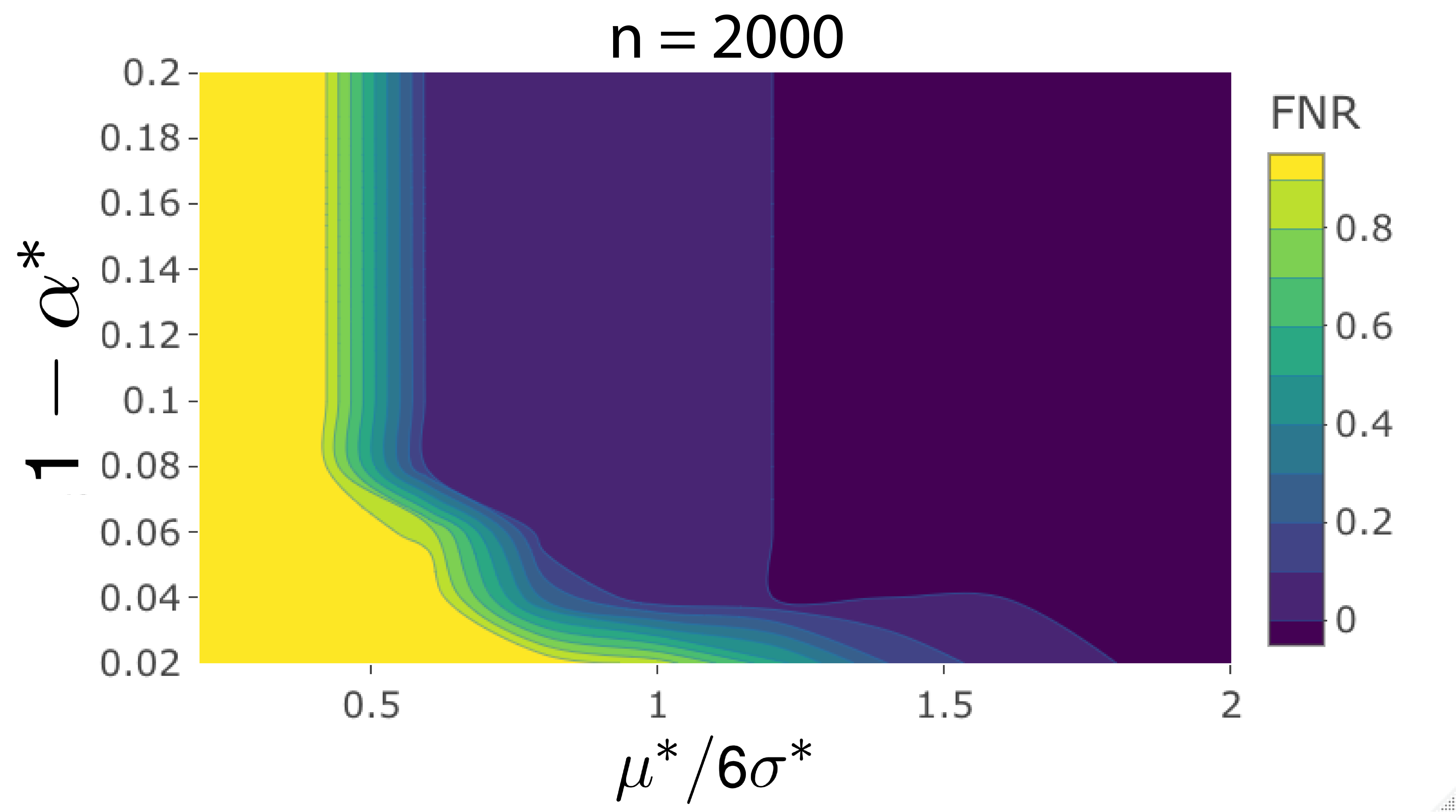}
        \end{tabular}
        \caption{FNR of the {\em smoothEM} spike identification on simulated data with 'clumped' spikes. The contour plots show the FNR (averaged over $20$ simulation replicates) as a function of the spike percentage ($1-\alpha^*$) and the SNT ($\frac{\mu^*}{6\sigma^*}$).
        From left to right, top to bottom, $n = 200, 500, 1000, 2000$.}
        \label{fig:fig22} 
    \end{figure}
    \begin{figure}
        \centering
        \begin{tabular}{cc}
        \includegraphics[width=.45\textwidth]{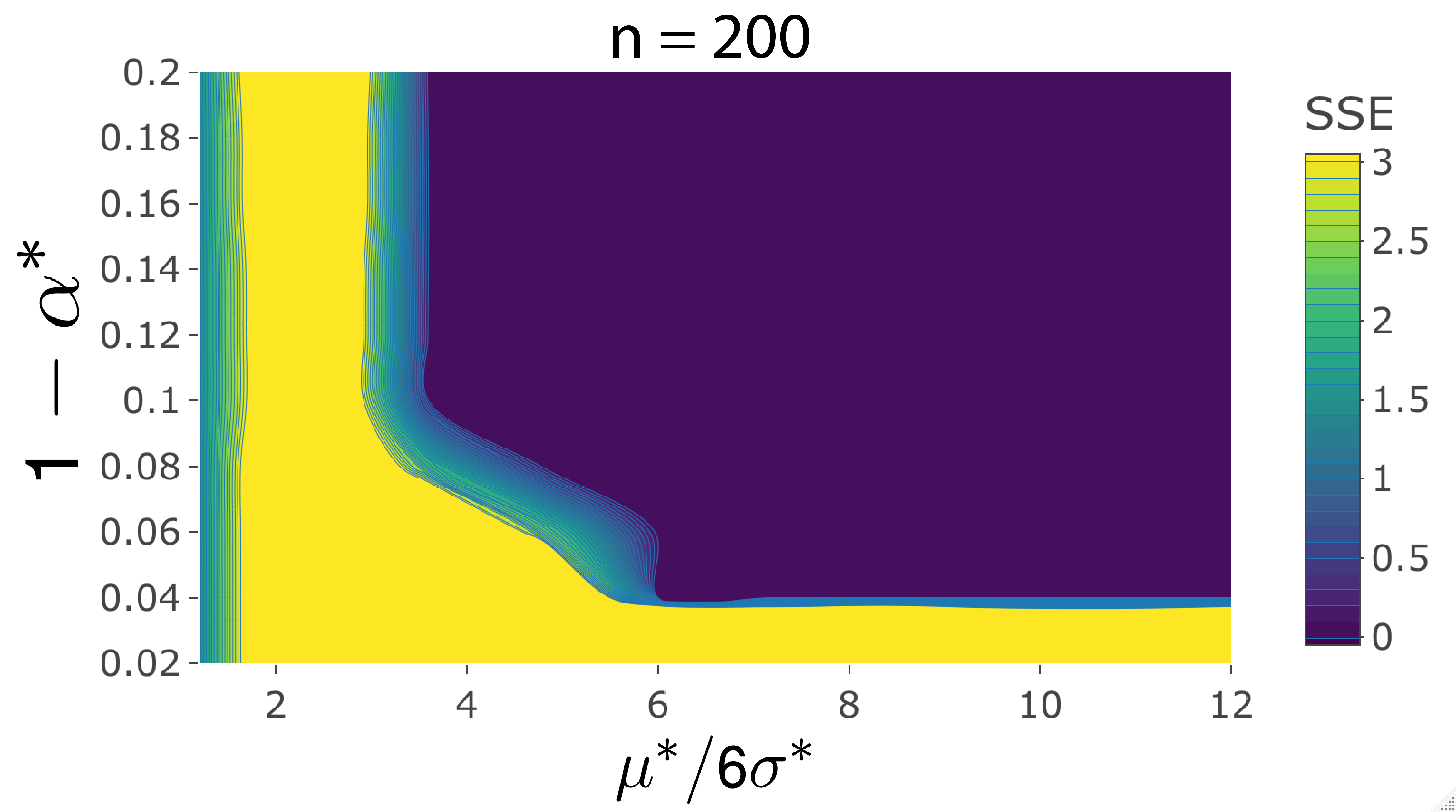}&
        \includegraphics[width=.45\textwidth]{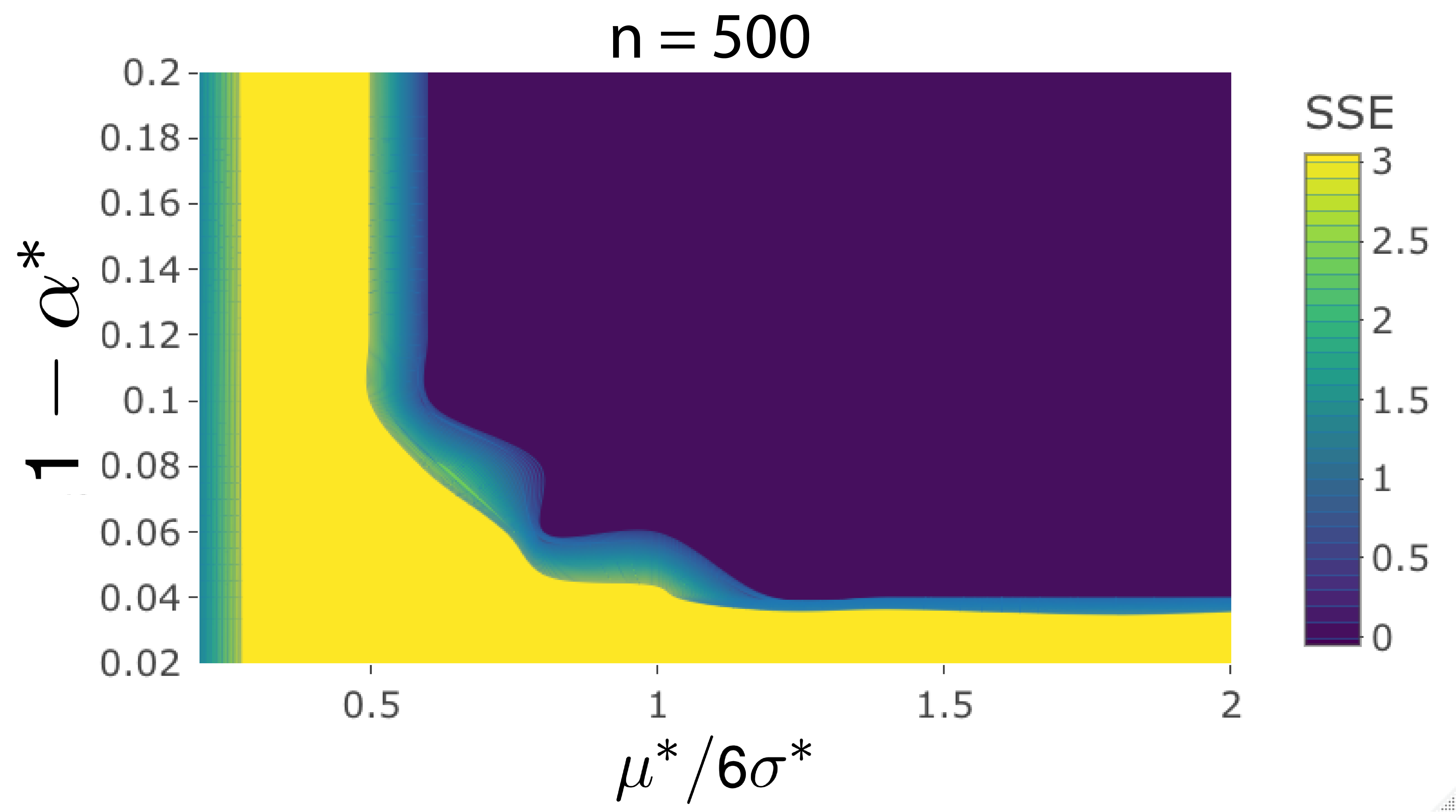}\\
        \includegraphics[width=.45\textwidth]{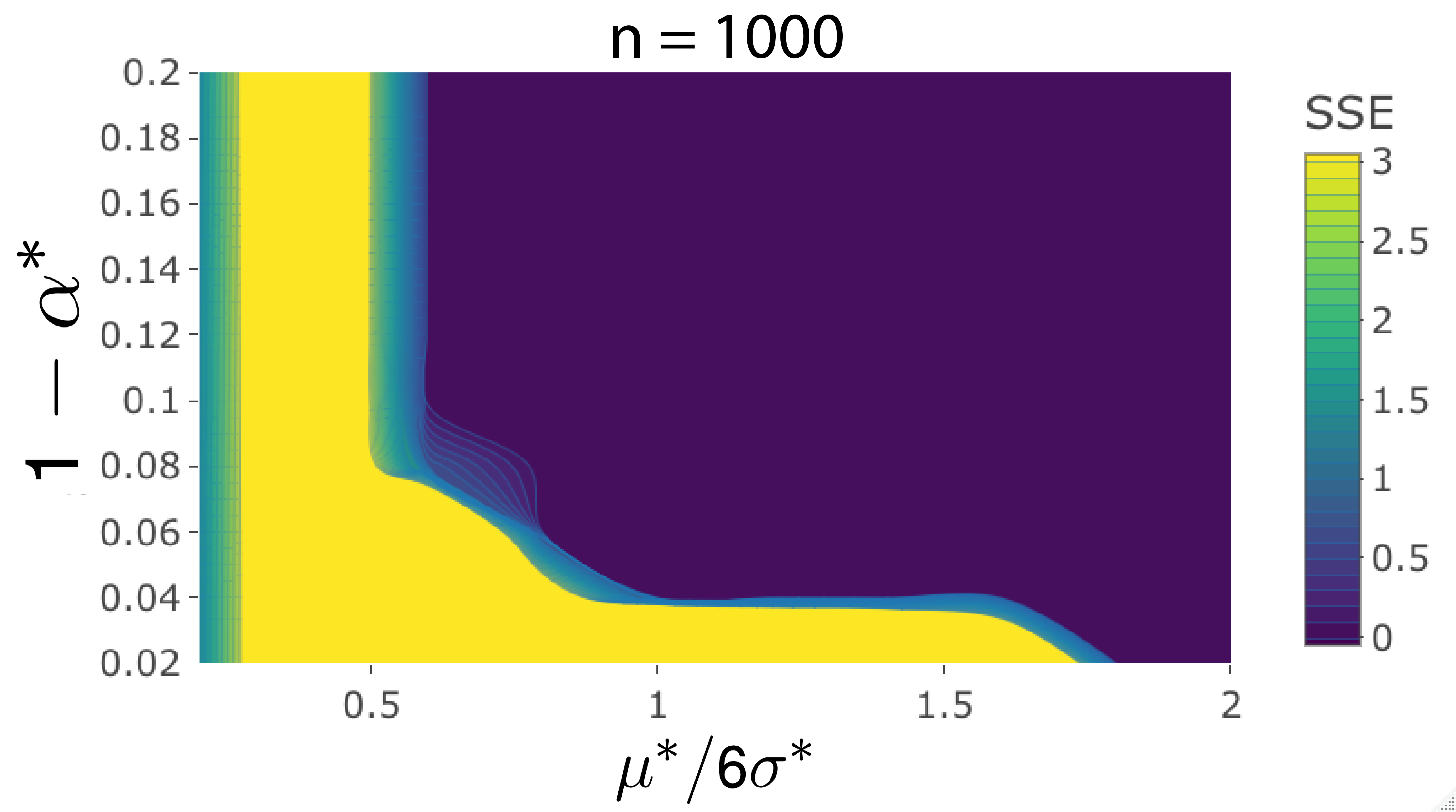}&
        \includegraphics[width=.45\textwidth]{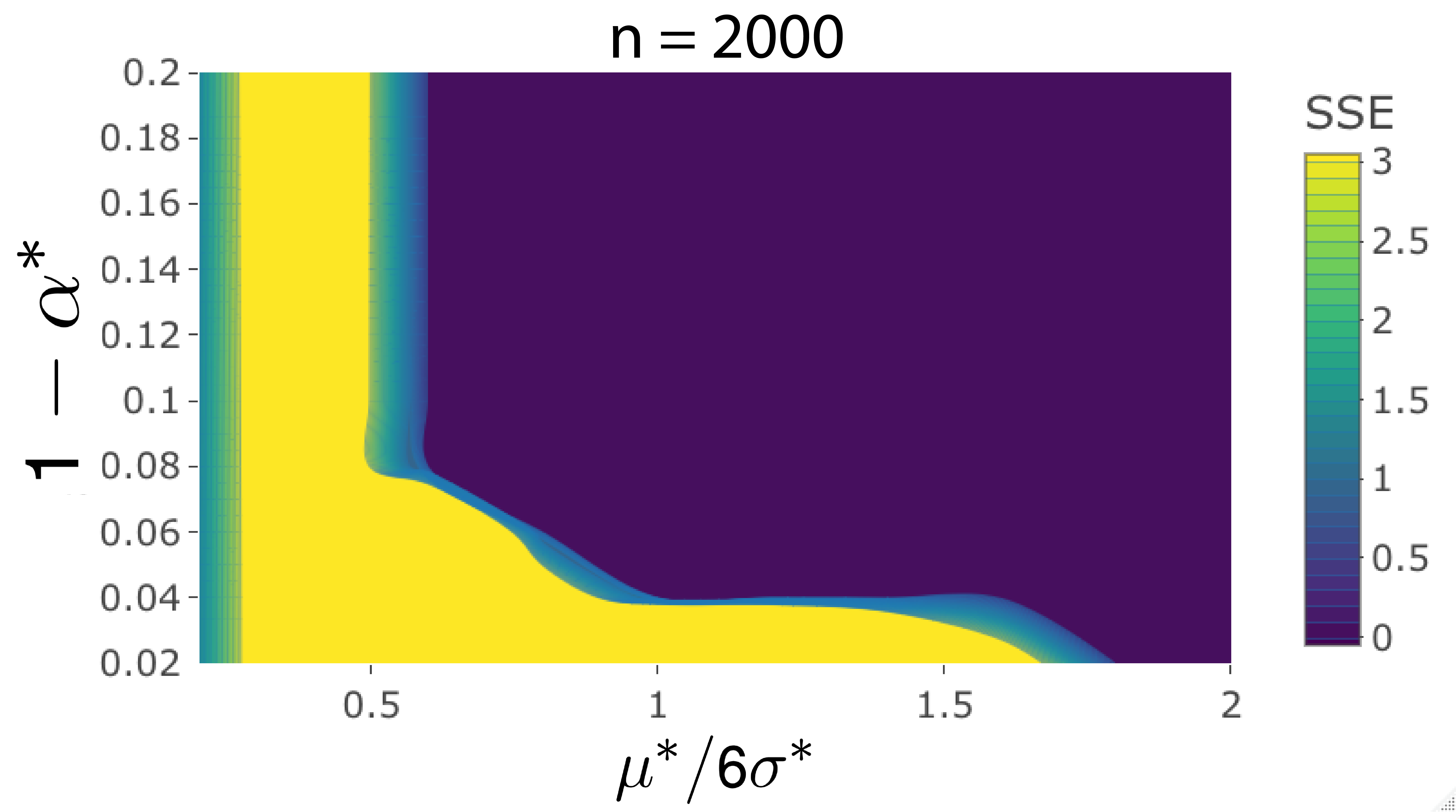}
        \end{tabular}
        \caption{SSE of the {\em smoothEM} parameter estimates on simulated data with 'clumped' spikes. The contour plots show the SSE (averaged over $20$ simulation replicates) as a function of the spike percentage ($\alpha^*$) and the STN ($\frac{\mu^*}{6\sigma^*}$).
        From left to right, top to bottom, $n = 200, 500, 1000, 2000$.}
        \label{fig:fig23} 
    \end{figure}

 \begin{figure}
        \centering
        \includegraphics[width=\textwidth]{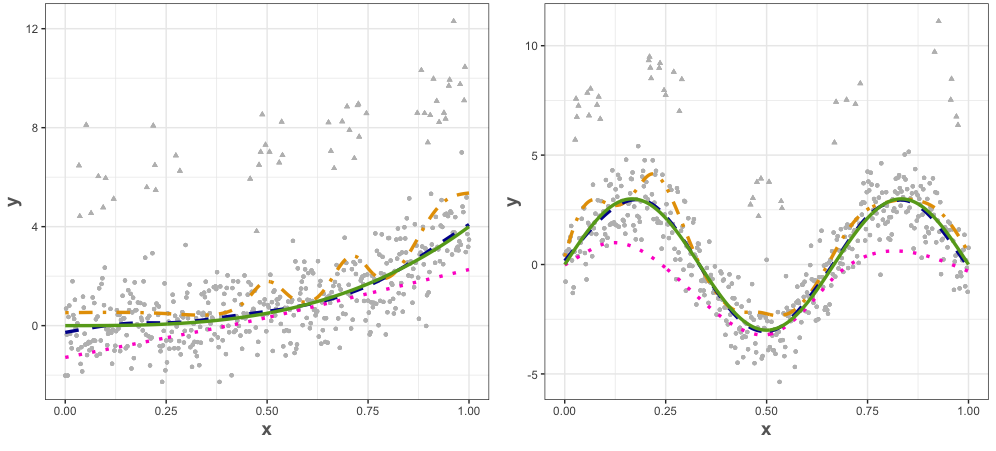}
        \caption{Comparison of \emph{smoothEM}, \emph{RWSS-GCV} and \emph{mgcv-AS} 
        on simulated data with slow-varying (left) and fast-varying (right) smooth components, $n = 500$, STN $\mu^*/(6\sigma^*) = 1$, and $1-\alpha^* \approx 10\%$. True curves (solid green) and {\em smoothEM} fits (dashed blue) are almost indistinguishable, while \emph{RWSS-GCV} fits (dotted red) and \emph{mgcv-AS} fits (dashed gold) depart markedly from the truth. True spikes and non-spikes are plotted as triangles and circles, respectively (the competing methods do not perform spike identifications)
        }
        \label{fig:benchmark_sim} 
    \end{figure}

\clearpage
\bibliographystyle{agsm}
\bibliography{paper-ref}